\documentclass{svmult}

\usepackage{color}
\usepackage{graphicx}
\usepackage{journals}
\usepackage{natbib}
\begin{document}

\title*{Theory and Applications of Non-Relativistic and Relativistic Turbulent Reconnection}
\author{A. Lazarian, G. Kowal, M. Takamoto, E. M. de Gouveia Dal Pino, and J. Cho}
\titlerunning{Turbulent reconnection}
\authorrunning{Lazarian et al.}
\institute{A. Lazarian \at Department of Astronomy, University of Wisconsin, Madison, Wisconsin, USA, \email{lazarian@astro.wisc.edu}
\and G. Kowal \at Escola de Artes, Ci\^{e}ncias e Humanidades, Universidade de S\~{a}o Paulo, S\~{a}o Paulo, SP, Brazil, \email{grzegorz.kowal@usp.br}
\and M. Takamoto \at Department of Earth and Planetary Science, The University of Tokyo, Tokyo, Japan, \email{mtakamoto@eps.s.u-tokyo.ac.jp}
\and E. M. de Gouveia Dal Pino \at Instituto de Astronomia, Geof\'isica e Ciencias Atmosfericas, Universidade de Sao Paulo, S\~{a}o Paulo, SP, Brazil, \email{dalpino@astro.iag.usp.br}
\and J. Cho \at Department of Astronomy and Space Science, Chungnam National University, Daejeon, Korea, \email{jcho@cnu.ac.kr}}

\maketitle

\abstract{Realistic astrophysical environments are turbulent due to the
extremely high Reynolds numbers of the flows. Therefore, the theories of
reconnection intended for describing astrophysical reconnection should not
ignore the effects of turbulence on magnetic reconnection. Turbulence is known
to change the nature of many physical processes dramatically and in this review
we claim that magnetic reconnection is not an exception. We stress that not only
astrophysical turbulence is ubiquitous, but also magnetic reconnection itself induces
turbulence. Thus turbulence must be accounted for in any realistic astrophysical reconnection
setup. We
argue that due to the similarities of MHD turbulence in relativistic and
non-relativistic cases the theory of magnetic reconnection developed for the
non-relativistic case can be extended to the relativistic case and we provide
numerical simulations that support this conjecture. We also provide quantitative
comparisons of the theoretical predictions and results of numerical experiments,
including the situations when turbulent reconnection is self-driven, i.e. the
turbulence in the system is generated by the reconnection process itself. We show
how turbulent reconnection entails the violation of magnetic flux freezing, the conclusion that has really far reaching consequences for many realistically turbulent astrophysical environments.
 In addition, we consider observational testing of turbulent reconnection as well as
numerous implications of the theory. The former includes the Sun and solar wind
reconnection, while the latter include the process of reconnection diffusion
induced by turbulent reconnection, the acceleration of energetic particles,
bursts of turbulent reconnection related to black hole sources as well as gamma ray
bursts. Finally, we explain why turbulent reconnection cannot be explained by
turbulent resistivity or derived through the mean field approach. We also argue
that the tearing reconnection transfers to fully turbulent reconnection in 3D
astrophysically relevant settings with realistically high Reynolds numbers.}


\section{Problem of reconnection as we see it}

This is a chapter that deals with magnetic reconnection in astrophysical
environments that are generically turbulent. We discuss how turbulence makes
reconnection fast and what this means for many astrophysical systems.

This is a contribution to the book on magnetic reconnection and therefore it is
not particularly productive to repeat that magnetic reconnection is important
for variety of processes from solar flares to gamma ray bursts. What we would
like to stress here is that magnetic reconnection is not some exotic process
that may be taking place occasionally in astrophysical environments, but it is
bread and butter of most processes taking place in magnetized plasmas. The
key to understanding of omnipresence of magnetic reconnection is the
ubiquity of turbulence in astrophysical environments.

Turbulence is a feature of high Reynolds number flows and most of magnetized
flows have extremely high Reynolds numbers. We show that even if the initial
astrophysical setup is not turbulent or ``not sufficiently turbulent'', the
development of reconnection, e.g. outflow, is bound to transfer the process of
reconnection to fully turbulent regime. Therefore we view the laminar models
with plasma instabilities, e.g. tearing instability, as transient states
to the fully turbulent reconnection.

What is the speed of reconnection? It is important to stress that turbulent
reconnection can address the apparent dichotomy suggested by observations, e.g.
reconnection is sometimes fast and sometimes slow. The theory of turbulent
reconnection relates this to the dependence of magnetic reconnection rate on the
level of turbulence in the system. As the intensity of turbulence changes,
the reconnection rate also changes.

As we will discuss in the review, the theory of turbulent reconnection predicts
reconnection rates that do not depend on the details of plasma microphysics, but
only on the level of MHD turbulence. The plasma physics related to the local
reconnection events may still be important at small scales e.g. for the acceleration
of particles from the thermal pool. At the same time, for understanding of
particle acceleration at large energies we will claim that the MHD description of turbulent
reconnection is sufficient. We note, however, that the turbulent reconnection
theory that we describe in the review is based on MHD and therefore it is not
applicable to the Earth magnetosphere where the current sheets are comparable to
the ion inertial scale.

The theory of turbulent reconnection has been covered in a number of reviews
that include
\cite{Lazarian_etal:2015a,Lazarian_etal:2015b,BrowningLazarian:2013}. In a
review by \cite{KarimabadiLazarian:2013} there was also an attempt to present
side by side both the theories of turbulent reconnection based on plasma turbulence
and on MHD approach, that we discuss here. We warn our readers, however, that the
statement in the latter review that the MHD approach has problems with describing reconnection
events in Solar wind was shown to be incorrect in
\citep{Lalescu_etal:2015}.

Within the present review we also discuss the implications of turbulent
reconnection, the  latter study becoming more important as the interconnection between
turbulence and astrophysical reconnection is appearing more evident to the
community. However, in terms of implications, we are just scratching the surface
of a very rich subject. Some implications are rather dramatic and has far reaching
consequences. For instance,  it is generally believed that magnetic
fields embedded in a highly conductive fluid retain their topology for all time
due to the magnetic fields being frozen-in \citep{Alfven:1943, Parker:1979}.
This concept of frozen-in magnetic fields is at the foundation of many theories,
e.g. of the theory of star formation in magnetized interstellar medium. At the
same time, in the review we discuss that this concept is not correct in the presence of
turbulence. As a result, serious revisions are necessary for the theoretical description
of a large number of astrophysical systems.

In what follows we briefly discuss modern ideas on non-relativistic and
relativistic MHD turbulence in \S 2 and \S 3 respectively, introduce the basic
concepts of turbulent non-relativistic reconnection theory in \S 4, provide
numerical testing of turbulent reconnection in \S 5. In \S 6 we discuss how
non-relativistic turbulent reconnection theory can be generalized for the case
of relativistic reconnection and provide numerical testing of the idea,
while \S 7 deals with the case of turbulent reconnection where turbulence is
injected by the reconnection process itself. The observational testing of
turbulent reconnection is discussed in \S 8 and the implications of the theory
of turbulent reconnection are summarized in \S 9. A comparison of turbulent
reconnection to other popular ideas can be found in \S 10 and the final remarks
are given in \S 11. There we also discuss the relation of the material in this chapter to
that in other chapters of the book.

\section{Non-Relativistic MHD turbulence}

Non-relativistic MHD turbulence is the best explored case with a lot of
observational and numerical data available to test the theory.

\subsection{Astrophysical turbulence: expectations and evidence}

Magnetized astrophysical fluids have huge Reynolds numbers $Re\equiv LV/\nu$ as
magnetic field limits the diffusion of charged particles perpendicular to its
local direction making viscosity $\nu$ small\footnote{In addition, the mean free
path of particles can also be constrained by the instabilities developed in turbulent plasmas
below the scale determined by Coulomb collisions \cite[see][]{Schekochihin_etal:2009,
LazarianBeresnyak:2006, BrunettiLazarian:2011}. The resulting scattering arising from the
ion interactions with the perturbed magnetic field ensures that compressible
motions are much more resilient to the collisionless damping compared to the textbook results
obtained with the Spitzer cross sections for ion collisions.} while the scales of the flow
$L$ are astrophysically huge. High $Re$ number flows are prey to numerous linear
and finite-amplitude instabilities, from which turbulent motions readily
develop. The plasma turbulence is can be driven by external energy sources,
such as supernovae in the ISM \citep{NormanFerrara:1996, Ferriere:2001}, merger
events and active galactic nuclei outflows in the intracluster medium (ICM)
\citep{Subramanian_etal:2006, EnsslinVogt:2006, Chandran:2005}, and baroclinic
forcing behind shock waves in interstellar clouds.  In other cases, the
turbulence is spontaneous, with available energy released by a rich array of
instabilities, such as magneto-rotational instability (MRI) in accretion disks
\citep{BalbusHawley:1998}, kink instability of twisted flux tubes in the solar
corona \citep{GalsgaardNordlund:1997, GerrardHood:2003}, etc. Finally, as we
discuss in the review, magnetic reconnection can also be a source of turbulence.

\begin{figure}[t]
\centering
\includegraphics[width=0.45\textwidth]{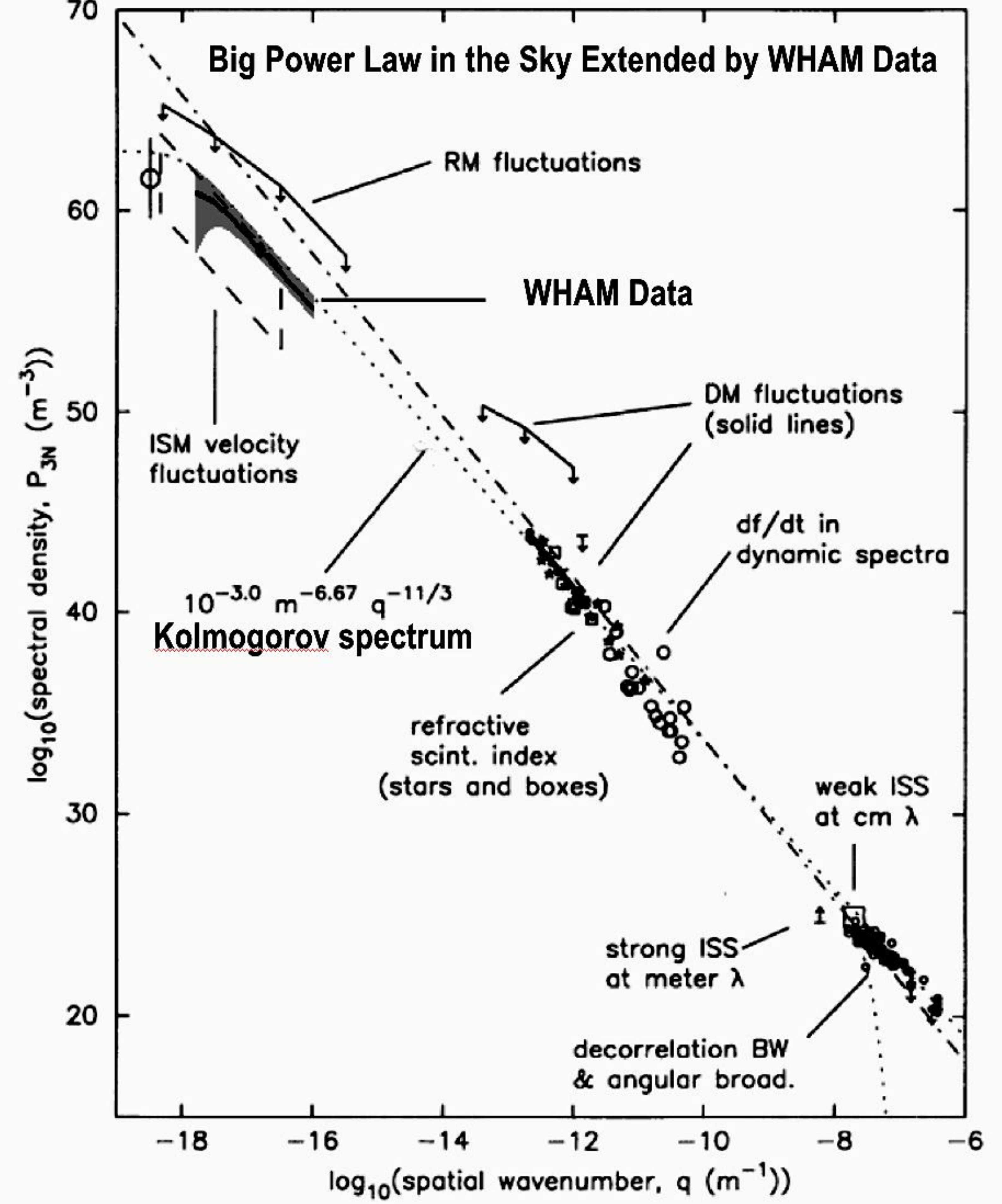}
\includegraphics[width=0.45\textwidth]{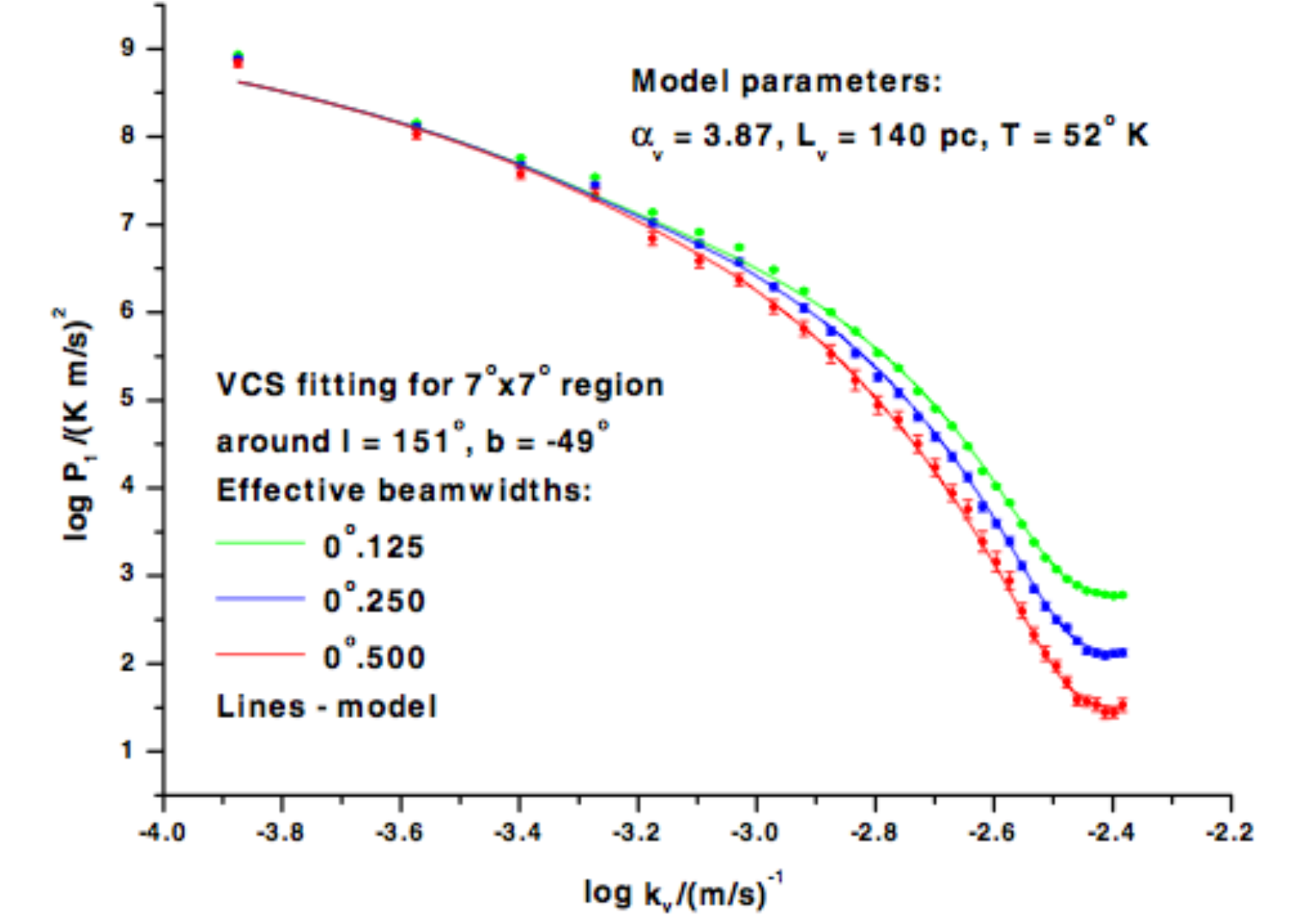}
\caption{{\it Left Panel} Big power law in the sky from
\cite{Armstrong_etal:1995} extended to scale of parsecs using the Wisconsin
H-Alpha Mapper (WHAM) data.
Reproduced from \cite{ChepurnovLazarian:2010} by permission of the AAS. {\it
Right Panel}. Properties of turbulence in HI obtained with VCS techniques.
Reproduced from \cite{Chepurnov_etal:2010} by permission of the AAS. \label{f1}}
\end{figure}

Observations confirm that astrophysical environments are indeed turbulent. The
spectrum of electron density fluctuations in the Milky Way is presented in
Figure~\ref{f1}, but similar examples are discussed in \cite{Leamon_etal:1998},
\cite{Bale_etal:2005} for solar wind, and \cite{VogtEnsslin:2005} for the
intracluster medium. As new techniques for studying turbulence are being applied to
observational data, the evidence of the turbulent nature of astrophysical media
becomes really undeniable. For instance, the Velocity Channel Analysis (VCA) and
Velocity Coordinate Spectrum (VCS) techniques \citep{LazarianPogosyan:2000,
LazarianPogosyan:2004, LazarianPogosyan:2006} provided unique insight into the
velocity spectra of turbulence in molecular clouds
\cite[see][]{Padoan_etal:2006,Padoan_etal:2010}, galactic and extragalactic
atomic hydrogen
(\cite{StanimirovicLazarian:2001, Chepurnov_etal:2010,
Chepurnov_etal:2015}, see also the review by \cite{Lazarian:2009}, where a
compilation of velocity and density spectra obtained with contemporary HI and CO
data is presented). We expect new flow of information on magnetic field spectra
to come from the new techniques that treat synchrotron fluctuations
\citep{LazarianPogosyan:2012}.

\subsection{Theory of weak and strong MHD turbulence}

MHD theory is applicable to astrophysical plasmas at sufficiently large scales
and for many astrophysical situations the Alfv\'enic turbulence, which is the
most important for turbulent reconnection and is applicable at the scales
substantially larger than the ion gyroradius $\rho_i$ \cite[see a discussion
in][]{Eyink_etal:2011, Lazarian_etal:2015a}.

The history of the theory of MHD turbulence can be traced back to the pioneering
studies by \cite{Iroshnikov:1964} and \cite{Kraichnan:1965}.  A good account for
the state of the field could be found in \cite{Biskamp:2003}.  Usually
turbulence is subdivided into weak and strong regimes, depending on the strength of
nonlinear interaction.  While weak MHD turbulence allows for analytical perturbative
treatment \citep{NgBhattacharjee:1996, Galtier_etal:2002, Chandran:2005}, the
progress in understanding strong turbulence came primarily from
phenomenological and closure models that were tested by comparison with results
of numerical simulations.  Important theoretical works on strong MHD turbulence
include \cite{MontgomeryTurner:1981}, \cite{Matthaeus_etal:1983},
\cite{Shebalin_etal:1983}, and \cite{Higdon:1984}.  Those clarified the
anisotropic nature of the energy cascade and paved the way for further
advancement in the field. The study by \cite{GoldreichSridhar:1995} identified
the balance between perturbations parallel and perpendicular to the local
direction of magnetic field, i.e., ``critical balance'', as the key component of
dynamics in strong magnetic turbulence.  For detailed recent reviews on MHD
turbulence, see \cite{BrandenburgLazarian:2013} and
\cite{BeresnyakLazarian:2015}.  Below we provide a simplified derivation of the
Alfv\'enic turbulence scaling that we employ later to understand turbulent
reconnection \cite[see also][]{Cho_etal:2003}.  Other ways of obtaining the same
relations may be found in e.g. \cite{GoldreichSridhar:1995},
\cite{LazarianVishniac:1999}, and \cite{Galtier_etal:2000}. Other fundamental
modes, i.e. slow and fast modes are of relatively marginal importance for the
theory of turbulent reconnection and we do not discuss them in the review\footnote{These
modes may play decisive role other processes, e.g. for cosmic ray scattering (see Yan \& Lazarian 2002, 2003a), acceleration of charged dust (see Yan \& Lazarian 2003b), star formation (see McKee \& Ostriker 2007).}.

If all the Alfv\'enic wave packets are moving in one direction, then they are
stable to nonlinear order. Therefore, in order to initiate turbulence, there
must be opposite-traveling wave packets of similar dimensions and the energy
cascade occurs only when they collide. It is also natural to assume that the
wave packets are anisotropic and therefore to distinguish between the parallel
$l_{\|}$ and the perpendicular $l_{\bot}$ scales of the wave-packets. The change
of energy per collision is
\begin{equation}
 \Delta E \sim (du^2/dt) \Delta t
 \sim {\bf u}_l \cdot \dot{\bf u}_l\Delta t
 \sim  (u_l^3/l_{\perp}) (l_{\|}/V_A),
\end{equation}
where we take into account that Alfv\'enic motions perpendicular to magnetic field
providing $\dot{\bf u}_l\sim u_l^2/l_{\bot}$, while the time of interactions is
determined by the time of wave packets interacting with each other, i.e. $\Delta
t \sim l_{\|}/V_A$, where $V_A$ is the Alfv\'en velocity.

The fractional energy change per collision is the ratio of $\Delta E$ to $E$,
\begin{equation}
  \zeta_l \equiv \frac{\Delta E}{u^2_l}
                           \sim \frac{ u_l l_{\|} }{ V_A l_{\perp} },
                         \label{zeta}
\end{equation}
which characterizes the strength of the nonlinear interaction. The cascading is
a random walk process in such a description with
\begin{equation}
t_{cas}\sim \zeta_l^{-2} \Delta t .
\label{tcas}
\end{equation}

The Alfv\'enic 3-wave resonant interactions are characterized by
\begin{eqnarray}
  {\bf k}_1 + {\bf k}_2 & = & {\bf k}_3, \label{k123} \\
  \omega_1 +  \omega_2 & = & \omega_3, \label{w123}
\end{eqnarray}
where ${\bf k}$'s are wavevectors and $\omega$'s are wave frequencies. The first
condition is a statement of wave momentum conservation and the second is a
statement of energy conservation. Alfv\'en waves satisfy the dispersion
relation: $\omega = V_A |k_{\|}|$, where $k_{\|}$ is the component of wavevector
parallel to the background magnetic field. Since only opposite-traveling wave
packets interact, ${\bf k}_1$ and ${\bf k}_2$ must have opposite signs, which
formally means that the cascade is possible only in the perpendicular direction.

In fact, the energy relation is subject to the wave uncertainty principle,
which means that the ambiguity of the order $\delta \omega \sim 1/t_{cas}$ is
acceptable. When $\zeta_l$ is small, $\delta \omega\ll \omega$ and the
energy transfer is happening mostly perpendicular to the {\it local} direction
of magnetic field. As a result of such a cascade the parallel scale $l_{\|}$
is preserved, while the perpendicular scale $l_{\bot}$ decreases. This is the
case of {\it weak Alfv\'enic turbulence}. In incompressible turbulence the energy
flux is
\begin{equation}
\epsilon=u_l^2/t_{cas}\approx\frac{ u_l^4} {V_A^2 \Delta t (l_{\bot}/l_{\|})^2}=const
\label{cascading}
\end{equation}
where Eqs. (\ref{tcas}) and (\ref{zeta}) are used. Taking into account that
$l_{\|}$ is constant, it is easy to see that Eq. (\ref{cascading}) provides
$u_l\sim l_{\perp}^{1/2}$ which in terms of energy spectrum of weak turbulence
provides the relation
\begin{equation}
E_{k, weak}\sim k_{\bot}^{-2},
\label{weak}
\end{equation}
where the relation $k E(k) \sim u_k^2$ is used. Eq. (\ref{weak}) was obtained
on the basis of similar arguments in \citet[][henceforth
LV99]{LazarianVishniac:1999} and later on the basis of a rigorous treatment of
weak turbulence in \cite{Galtier_etal:2000}. Note that $k_{\|}$ stays constant
in the weak cascade.

Note, that the weak turbulence regime should have a limited inertial range.
Indeed, as $k_{\bot}\sim l_{\bot}^{-1}$ increases, the energy change per
collision $\zeta_l$ increases, the cascading time $t_{cas}$ decreases. This
makes the uncertainty in the wave frequency $\delta \omega\sim 1/t_{cas}$
comparable to wave frequency $\omega$ when $\zeta_l$ approaches unity.
Naturally, one expects the nature of the cascade to change. Indeed, the
cascading cannot happen in less than one wave period and therefore the cascading
rate cannot increase further. Similarly with $\delta \omega\sim \omega$ the
constraints given by Eq. (\ref{w123}) cannot prevent the decrease of the parallel
length of wave packets $l_{\|}$. This signifies the advent of a regime of {\it
strong Alfv\'enic turbulence}. The corresponding theory was formulated for the
turbulent injection velocity $u_L=V_A$ by \citet[][henceforth
GS95]{GoldreichSridhar:1995} and was generalized for subAlfv\'enic and
superAlfv\'enic injection velocities in LV99 and \cite{Lazarian:2006},
respectively. Below we follow LV99 in order to obtain the relations for strong
MHD turbulence with subAlfv\'enic energy injection. This type of turbulence is the
most important in the context of turbulent reconnection.

As we explained above the change of the turbulence regime happens when
$\zeta_l\sim 1$, which in terms of the parameters of the interacting
wave packets means that
\begin{equation}
 u_l /l_{\bot}\sim { V_A/ l_{\|}},
\end{equation}
which manifests the famous GS95 {\it critical balance}. This expression was
originally formulated using not scales of the eddies, but wavevectors
$k_{\bot}$ and $k_{\|}$ as the GS95 discussion did not include the fundamental
concept of {\it local} magnetic field direction. Indeed, the weak turbulence
theory can be formulated in terms of mean magnetic field, as the distortions
introduced by turbulence in terms of direction are marginal due to the marginal
change of $l_{\|}$. In the strong turbulence the distinction between the mean
direction of magnetic field and the local direction of the field that a wave
packet is moving along may be significant. This is especially obvious in the
case of transAlfv\'enic and superAlfv\'enic turbulence when the local direction of
magnetic field may poorly correlate with the direction of the mean magnetic
field in the volume. As a result no universal relations exist in the frame of
the mean magnetic field and therefore in the global frame given by wavevectors
$k$. The understanding of the importance of the local magnetic frame for the
GS95 theory was introduced and elaborated in the later publications
\cite[LV99;][]{ChoVishniac:2000, MaronGoldreich:2001, ChoLazarianVishniac:2002}.

The turbulence is injected isotropically at scale $L_i$ with the the velocity
$u_L<V_A$ and the cascading of energy follows the weak turbulence cascade
$u_L^2/t_{cas}$, which for the weak cascading rate gives $u_L^4/(L_i V_A)$.
Starting with the scale corresponding to $\zeta_l=1$, i.e. at the perpendicular
scale
\begin{equation}
l_{trans}\sim L_i(u_L/V_A)^2\equiv L_i M_A^2, \quad M_A<1,
\label{trans}
\end{equation}
where $M_A=u_L/V_A<1$ is the Alfv\'enic Mach number\footnote{ Thus, weak
turbulence has a limited,  i.e. $[L_i, L_i M_A^2]$ inertial range and at small
scales it transits into the regime of strong turbulence. We should stress that
weak and strong are not the characteristics of the amplitude of turbulent
perturbations, but the strength of non-linear interactions (see more discussion
in \cite{Cho_etal:2003}) and small scale Alfv\'enic perturbations can
correspond to a strong Alfv\'enic cascade.}. The turbulence becomes strong and
cascades over one wave period, namely,  $l_\parallel/V_A$.  The cascading of
turbulent energy is $u_l^3/l_{\bot}$, which is similar to Kolmogorov cascade in
the direction perpendicular to the local direction of magnetic field. Due to the
conservation of energy in the cascade the weak and strong turbulence energy
flows should be the same which gives the scaling relations in LV99
\begin{equation}
\ell_{\|}\approx L_i \left(\frac{\ell_{\bot}}{L_i}\right)^{2/3} M_A^{-4/3},
\label{Lambda}
\end{equation}
\begin{equation}
\delta u_{\ell}\approx u_{L} \left(\frac{\ell_{\bot}}{L_i}\right)^{1/3} M_A^{1/3}.
\label{vl}
\end{equation}
Those relations give the GS95 scaling for $M_A\equiv 1$. These are equations
that we will use further to derive the magnetic reconnection rate.

When the measurements are done in the global system of reference, the turbulence
scaling is dominated by perpendicular fluctuations containing most of energy,
and therefore using Eq. (\ref{vl}) with $kE(k)\sim u_k^2$ one can get $E(k)\sim
k^{-5/3}$, which coincides with the Kolmogorov scaling. One can intuitively
understand this result assuming that eddies freely evolve in the direction
perpendicular to magnetic field.

Finally, we want to point out that the isotropic driving of MHD turbulence is
somewhat idealized. For instance, when turbulence is driven by magnetic
reconnection, magnetic field lines are not straight on the injection scale and
therefore the weak cascade ideas are not applicable. This is an important point
for understanding Solar flares and similar reconnection phenomena.

\subsection{Controversy related to GS95 model}

\begin{figure}[t]
\centering
\includegraphics[width=0.46\textwidth]{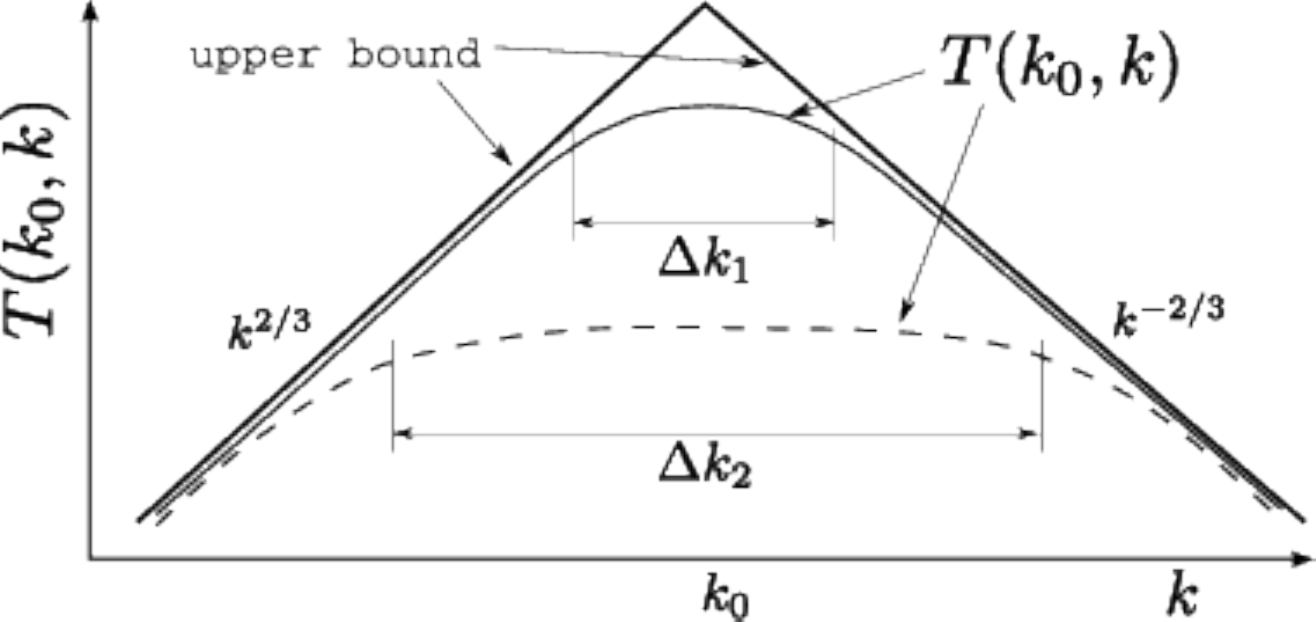}
\includegraphics[width=0.45\textwidth]{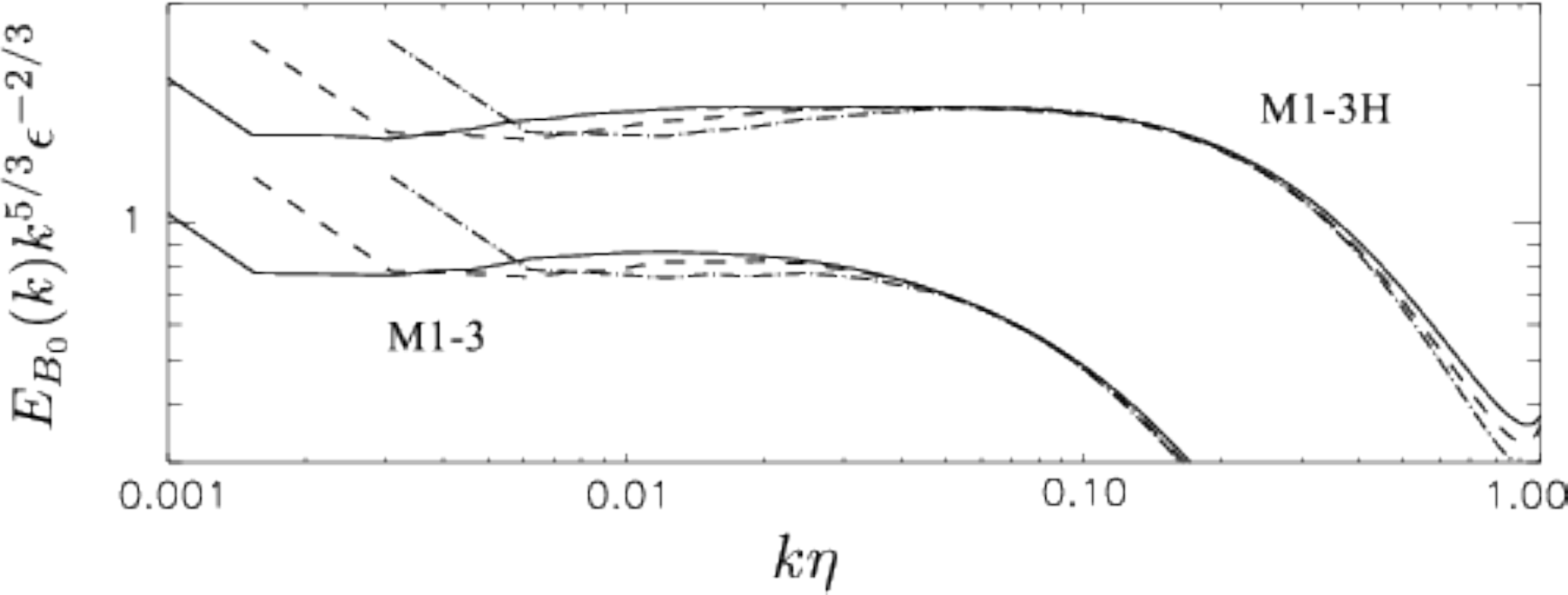}
\caption{ {\it Left Panel} Illustration of ``diffuse locality'' of MHD
turbulence.  The upper bounds for the energy transfer window could be
consistent with both rather ``local'' transfer (upper solid curve) or
``non-local'' and ``diffuse-local'' transfer (lower dashed curve). Reproduced
from \cite{BeresnyakLazarian:2010} by permission of the AAS. {\it Right Panel}.
Spectrum of MHD turbulence in high resolution simulations is consistent with
GS95 predictions. Reproduced from \cite{Beresnyak:2015} by permission of the
AAS. \label{f2}}
\end{figure}

Testing of GS95 turbulence numerically presented a challenging task. The
measurements of the  spectral slope in MHD simulations
\cite[see][]{MaronGoldreich:2001} were better fitted by the spectrum $\sim
k^{-3/2}$ rather than GS95 prediction of $k^{-5/3}$. This resulted in
theoretical attempts to explain the measured slope by
\cite{Boldyrev:2005,Boldyrev:2006}. Another explanation of the slope difference
was suggested in \cite{BeresnyakLazarian:2010}. It was based on the conjecture
that the MHD turbulence is less local compared to hydrodynamic turbulence and
therefore low resolution numerical simulations were not measuring the actual
slope of the turbulence, but the slope distorted by the bottleneck effect. The
latter is generally accepted to be a genuine feature of turbulence and is
attributed to the partial suppression of non-linear turbulent interactions under
the influence of viscous dissipation. Studied extensively for hydrodynamics the
bottleneck effect reveals itself as a pile-up of kinetic energy near the wave
number of maximum dissipation \cite[see][]{Sytine_etal:2000, Dobler_etal:2003}.
With the limited inertial range of existing numerical simulations, the
bottleneck effect may strongly interfere with the attempts to measure the actual
turbulence spectrum.  The locality of turbulence determines whether the
bottleneck produces a localized or more extended bump of the turbulence
spectrum. In the latter case the low resolution numerical simulations may be
affected by the bottleneck effect for the whole range of wave numbers in the
simulation and the distorted spectrum can be mistaken for an inertial range. A
smooth extended bottleneck is expected for MHD turbulence being less local
compared to its hydrodynamic counterpart. This feature of MHD turbulence was
termed by Beresnyak \& Lazarian "diffuse locality"
\cite[see][]{BeresnyakLazarian:2010}. This effect is illustrated by Figure
\ref{f2}.

All in all, the bottleneck is a physical effect and the absence of it in the low
resolution MHD simulations is suggestive that the measured spectral slope is not
the actual slope of the turbulent energy. In fact, the bottleneck effect has
tricked researchers earlier. For instance, the initial compressible simulations
suggested the spectral index of high Mach number hydrodynamic turbulence to be
$-5/3$, which prompted theoretical attempts to explain this
\cite[e.g.][]{Boldyrev:2002}. However, further high resolution research
\citep{Kritsuk_etal:2007} revealed that the flattering of the spectrum observed
was the result of a bottleneck effect, which is more extended in compressible
than in incompressible fluids. Similarly, we believe that the simulations that
reported the spectral slope of $-3/2$ for the MHD turbulence
\citep{MaronGoldreich:2001, MullerGrappin:2005, Mason_etal:2006,
Mason_etal:2008, Perez_etal:2012} are affected by the bottleneck effect.  This
conclusion is supported by the study of scaling properties of turbulence with high
numerical resolution in \cite{Beresnyak:2014}.  This study shows that the
Reynolds number's dependence of the dissipation scale is not fulfilled with the
$-3/2$ spectral slope.

We believe that the new higher resolution simulations
\cite[see][]{Beresnyak:2013, Beresnyak:2014} resolve the controversy and,
indeed, the putative $k^{-3/2}$ spectrum is the result of the bottleneck.
However, whether the slope is $-3/2$ or $-5/3$ has only marginal impact on
the theory of turbulent reconnection. A discussion of turbulent reconnection for
an arbitrary spectral index and arbitrary anisotropy can be found in LV99.

\subsection{Compressible MHD turbulence}

As we discuss later, the Alfv\'enic mode is the most important mode
for turbulent reconnection. Therefore we do not dwell upon compressible MHD
turbulence. In fact, it is important for us to be able to consider Alfv\'enic
perturbations in compressible turbulence. In a sense, in view of our further
discussion we are only interested in (1) whether the treatment of magnetic
reconnection in terms of Alfv\'enic turbulence is adequate in the presence of
fluid compressibility and (2) what portion of energy of driving is going into
the Alfv\'enic component of turbulence.

Original ideas about how Alfv\'enic modes can interact with other fundamental
modes, i.e. slow and fast modes, can be traced back to GS95. They were
elaborated further in
\cite{LithwickGoldreich:2001}. A numerical and theoretical study of the modes
was then performed in \cite{ChoLazarian:2002, ChoLazarian:2003} and
\cite{KowalLazarian:2010}.
These studies answer positively the question (1), i.e. they
show that turbulent Alfv\'enic mode preserves its identity and forms an independent Alfv\'enic
cascade even in compressible fluid \cite[see more in][]{ChoLazarian:2002, ChoLazarian:2003}.
They also address the question (2), i.e. they quantify how much energy is transferred to
compressible motions in MHD turbulence.
The effects of compressibility have been extensively studied in
\cite{Kowal_etal:2007} and scaling relations has been tested in \cite{KowalLazarian:2007}.
A detailed discussion of the
effects of compressibility on MHD turbulence can be found in a review by
\cite{BeresnyakLazarian:2015}.

\section{Relativistic MHD Turbulence}

Some astrophysical fluids involve relativistic motions. In recent years,
interest on MHD turbulence in relativistic fluids has been growing. Can the
ideas of GS95 turbulence be transferred to relativistic fluids? This is the
issue that has been addressed by recent research. Due to advances in numerical
techniques, it is now possible to numerically investigate fully relativistic MHD
turbulence \cite[e.g.][]{ZrakeMacFadyen:2012}.

\subsection{Relativistic force-free MHD turbulence}

Due to its numerical and theoretical simplicity, MHD turbulence in relativistic
force-free regime has been studied first. Relativistic force-free formalism can
be used for a system, such as the magnetosphere of a pulsar or a black hole, in
which the magnetic energy density is much larger than that of matter. In this
case, the Alfv\'en speed approaches the speed of light, and we need relativity
to describe the physics of the system. If we take the flat geometry, the
relativistic MHD equations
\begin{eqnarray}
\partial_\mu (\rho u^\mu ) =0, \\
\partial_\mu T^{ \mu \nu}=0, \\
\partial_t {\bf B} = \nabla \times ({\bf v}  \times {\bf B} ), \\
\nabla \cdot {\bf B}=0,
\end{eqnarray}
where $u^{\mu}$ is the fluid four velocity and $T^{\mu \nu}$ is the stress-energy
tensor of the fluid and the electromagnetic field, reduce to
\begin{equation}
   \frac{ \partial {\bf Q} }{ \partial t }
 + \frac{ \partial {\bf F} }{ \partial x^1 }
 =0,
\end{equation}
where
\begin{eqnarray}
{\bf Q}=(S_1,S_2,S_3,B_2,B_3), \\
 {\bf F}=(T_{11},T_{12},T_{13},-E_3,E_2), \\
 T_{ij}=-(E_iE_j + B_iB_j)+\frac{ \delta_{ij} }{2} (E^2+B^2), \\
 {\bf S}={\bf E}\times {\bf B}, \\
 {\bf E}=-\frac{1}{B^2} {\bf S}\times {\bf B}.
\end{eqnarray}
Here, ${\bf E}$ is the electric field, ${\bf S}$ is the Poynting flux vector,
and we use units such that the speed of light and $\pi$ do not appear in the
equations \cite[see][]{Komissarov:2002}. After solving equations along $x^1$
direction, we repeat similar procedures for $x^2$ and $x^3$ directions with
appropriate rotation of indexes.

\begin{figure}[t]
\includegraphics[width=\textwidth]{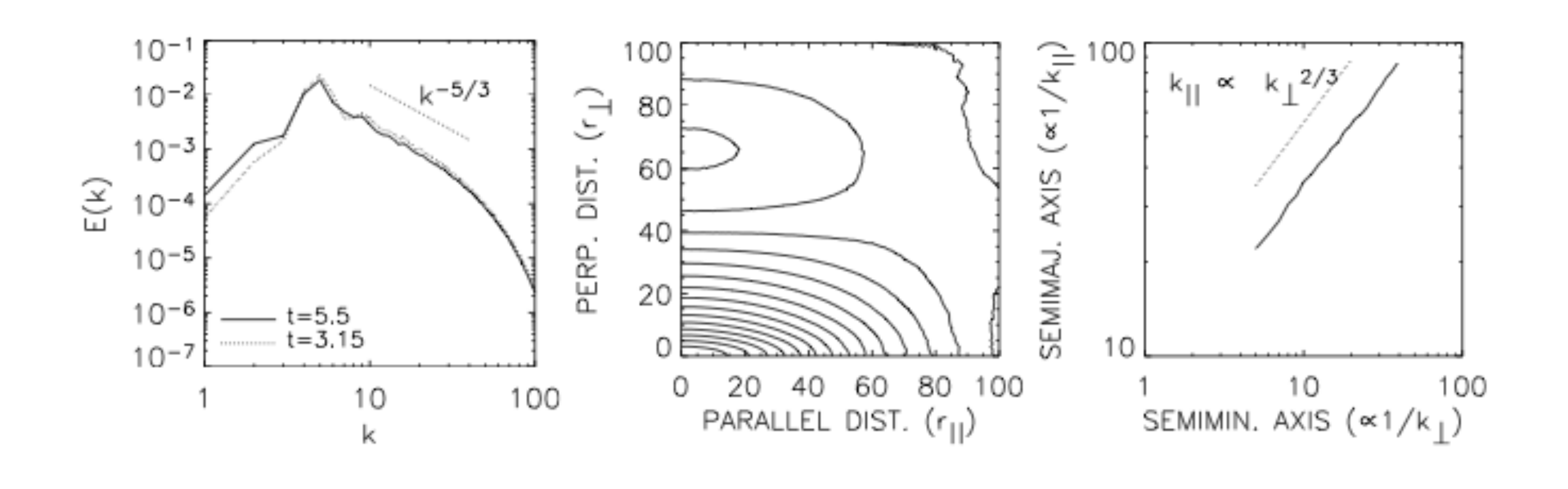}
\caption{Simulation of decaying relativistic force-free MHD turbulence.
   (Left) Energy spectrum is compatible with a Kolmogorov one.
   (Middle) Eddy shapes, represented by contours, show scale-dependent
   anisotropy: smaller eddies are more elongated.
   (Right) The anisotropy of eddy shape follows a
   Goldreich-Sridhar type anisotropy. Reproduced from \cite{Cho:2005} by permission of the AAS.}
\label{fig1_rel}
\end{figure}

Scaling relations for relativistic Alfv\'enic MHD turbulence were first
derived by \cite{ThompsonBlaes:1998} and were numerically tested by
\cite{Cho:2005}. \cite{Cho:2005} performed a numerical simulation of a decaying
relativistic force-free\footnote{ One can obtain the force-free condition from
Maxwell's equations and the energy-momentum equation: $\partial_{\mu}
T^{\nu\mu}_{(f)}=-F_{\nu\mu}J^{\mu}=0$. Here, $F^{\nu\mu}$ is the
electromagnetic field tensor. } MHD turbulence with numerical resolution of
$512^3$ and calculated energy spectrum and anisotropy of eddy structures. At the
beginning of the simulation, only Alfv\'en modes are present and the condition
for critical balance, $ \chi \equiv (b k_{\perp} )/( B_0k_{\|} ) \sim 1, $ is
satisfied (see \cite{Cho:2005, Cho:2014} for heuristic discussions on the
critical balance in relativistic force-free MHD turbulence). The left panel of
Figure \ref{fig1_rel} shows energy spectrum of magnetic field at two different
times. Although only large-scale (i.e. small $k$) Fourier modes are excited at
t=0 (not shown), cascade of energy produces small-scale (i.e. large $k$) modes
at later times. After $t \sim 3$, the energy spectrum decreases without changing
its slope. The spectrum at this stage is very close to a Kolmogorov spectrum:
\begin{equation}
  E(k) \propto k^{-5/3}.
\end{equation}
Contours in the middle panel of Figure \ref{fig1_rel} shows shapes of eddies
revealed by the second-order structure function for magnetic field. Note that
the shape of eddies is measured in a local frame, which is aligned with the
local mean magnetic field \cite[see][for details]{Cho_etal:2002,
ChoVishniac:2000, KowalLazarian:2010}. The contour plot clearly shows existence
of scale-dependent anisotropy: smaller eddies are more elongated.
The relation between the semi-major axis ($\sim l_{\|} \sim 1/k_{\|}$) and the
semi-minor axis ($\sim l_{\perp} \sim 1/k_{\perp}$) of the contours fits very
well the Goldreich-Sridhar type anisotropy:
\begin{equation}
  k_{\|} \propto k_{\perp}^{2/3}
\end{equation}
(see the right panel of Figure \ref{fig1_rel}).
All these results are consistent with the theoretical predictions in
\cite{ThompsonBlaes:1998}. A driven turbulence simulation in \cite{Cho:2014}
also confirms the scaling relations.

Although the similarity between relativistic and non-relativistic Alfv\'enic
turbulences may not be so surprising because the conditions for critical balance
are identical, it has many astrophysical implications. So far, we do not fully
understand turbulent processes in extremely relativistic environments, such as
black hole/pulsar magnetospheres or gamma-ray bursts. The close similarities
between extremely relativistic and non-relativistic Alfv\'enic turbulences
enable us to understand the physical processes, e.g.,~reconnection, particle
acceleration, etc., in such media better.

\begin{figure}[t]
\includegraphics[width=\textwidth]{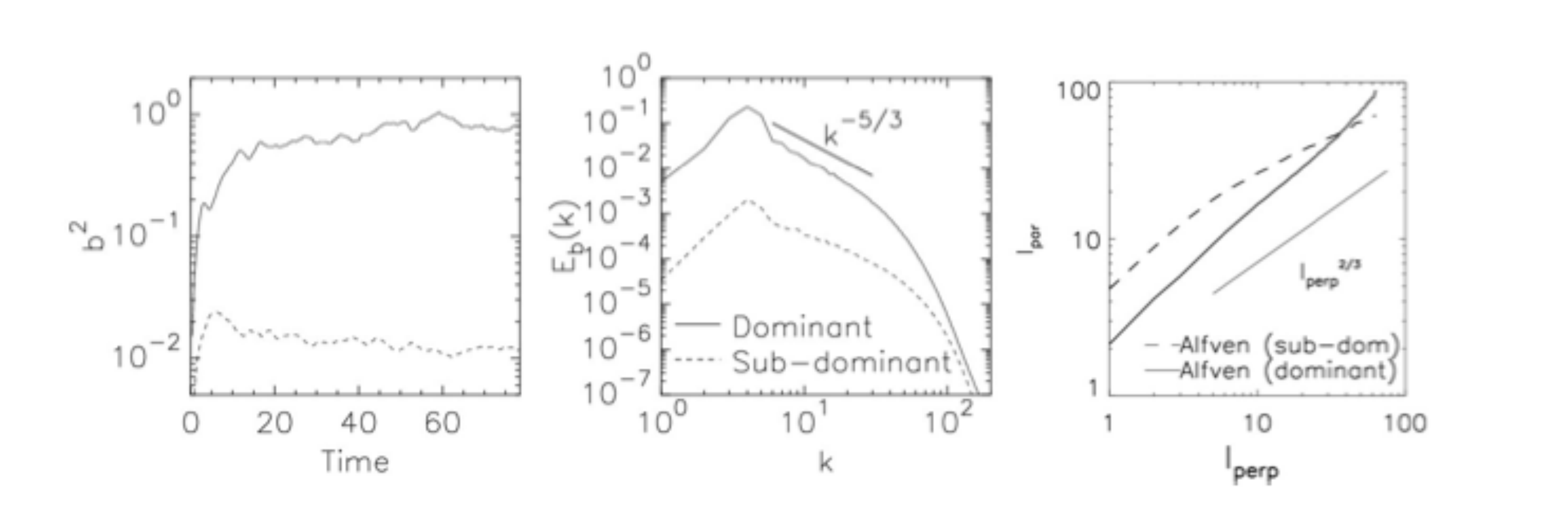}
\caption{Simulation of imbalanced relativistic force-free MHD turbulence.
   (Left) About a factor of 4 difference in energy injection rates results in a huge
   imbalance in energy densities.
   (Middle) The spectra for the dominant and the sub-dominant waves have
   different slopes: the dominant waves have a steeper spectrum.
   (Right) The degrees of anisotropy for the dominant and the sub-dominant waves are
      different: the dominant waves have a weaker anisotropy.
      Reproduced from \cite{ChoLazarian:2014} by permission of the AAS.}
\label{fig2_rel}
\end{figure}

Due to the similarity, it is also possible that we can test non-relativistic
theories using relativistic turbulence simulations. For example,
\cite{ChoLazarian:2014} performed numerical simulations of imbalanced
relativistic force-free MHD turbulence. The results of a simulation for $512^3$
resolution is presented in Figure \ref{fig2_rel}, in which the energy injection
rate for Alfv\'en waves moving in one direction (dominant waves) is 4 times
larger than that for waves moving in the other direction (sub-dominant waves).
The left panel of Figure \ref{fig2_rel} shows that, even though the ratio of the
energy injection rates is about $\sim$4, the ratio of the energy densities is
about $\sim$100. The middle panel of the figure shows that the spectrum for the
dominant waves is a bit steeper than a Kolmogorov spectrum, while that for the
sub-dominant waves is a bit shallower. The right panel of the figure shows that
the anisotropy of the dominant waves is a bit weaker than the Goldreich-Sridhar
type anisotropy, while that of the sub-dominant waves is a bit stronger. All
these results are consistent with the model of \cite{BeresnyakLazarian:2008} for
non-relativistic Alfv\'enic MHD turbulence.

\subsection{Fully relativistic MHD Turbulence}

Fully relativistic MHD turbulence has been studied since 2009
(\cite{Zhang_etal:2009, Inoue_etal:2011, BeckwithStone:2011,
ZrakeMacFadyen:2012, ZrakeMacFadyen:2013, GarrisonNguyen:2015}, see
also \cite{RadiceRezzolla:2013} for non-magnetized turbulence). The results in
\cite{ZrakeMacFadyen:2012, ZrakeMacFadyen:2013} for the mean lab-frame Lorentz
factor of $\sim$1.67 and numerical resolutions of up to $2048^3$ confirm that
there exists an inertial sub-range of relativistic velocity fluctuations with a
-5/3 spectral index. They also found that intermittency based on the scaling
exponents of the longitudinal velocity structure functions follows the
\cite{SheLeveque:1994} model fairly well. On the other hand, simulations for
unmagnetized relativistic turbulence with average Lorentz factors up to
$\sim$1.7 revealed that relativistic effects enhance intermittency, so that the
scaling exponents for high-order structure functions deviate from the prediction
of the She-Leveque model significantly \citep{RadiceRezzolla:2013}.

We note that the decomposition of the relativistic MHD cascade into fundamental
MHD modes has not been performed yet. The corresponding study in
\cite{ChoLazarian:2002, ChoLazarian:2003} and \cite{KowalLazarian:2010} provided
the framework for considering the separate Alfv\'enic, slow and fast mode
cascades. We, however, expect that in analogy with what we already learned about
the MHD turbulence, the results for relativistic and non-relativistic cases will
not be much different.

\section{Turbulent MHD reconnection}

\subsection{Sweet-Parker model and its generalization to turbulent media}

The model of turbulent reconnection in LV99 generalizes the classical
Sweet-Parker model \citep{Parker:1957, Sweet:1958}\footnote{The basic idea of
the model was first discussed by Sweet and the corresponding paper by Parker
refers to the model as ``Sweet model''.}.  In the latter model two regions with
uniform {\it laminar} magnetic fields are separated by a thin current sheet.  The
speed of reconnection is given roughly by the resistivity divided by the sheet
thickness, i.e.
\begin{equation}
V_{rec1}\approx \eta/\Delta.
\label{eq.1}
\end{equation}
For {\it steady state reconnection} the plasma in the current sheet must be
ejected from the edge of the current sheet at the Alfv\'en speed, $V_A$.  Thus
the reconnection speed is
\begin{equation}
V_{rec2}\approx V_A \Delta/L_x,
\label{eq.2}
\end{equation}
where $L_x$ is the length of the current sheet, which requires $\Delta$ to be
large for a large reconnection speed.  As a result, the overall reconnection
speed is reduced from the Alfv\'en speed by the square root of the Lundquist
number, $S\equiv L_xV_A/\eta$, i.e.
\begin{equation}
V_{rec, SP}=V_A S^{-1/2}.
\label{SP}
\end{equation}
The corresponding Sweet-Parker reconnection speed is negligible in astrophysical
conditions as $S$ may be $10^{16}$ or larger.

\begin{figure}[t]
\centering
\includegraphics[width=0.65\textwidth]{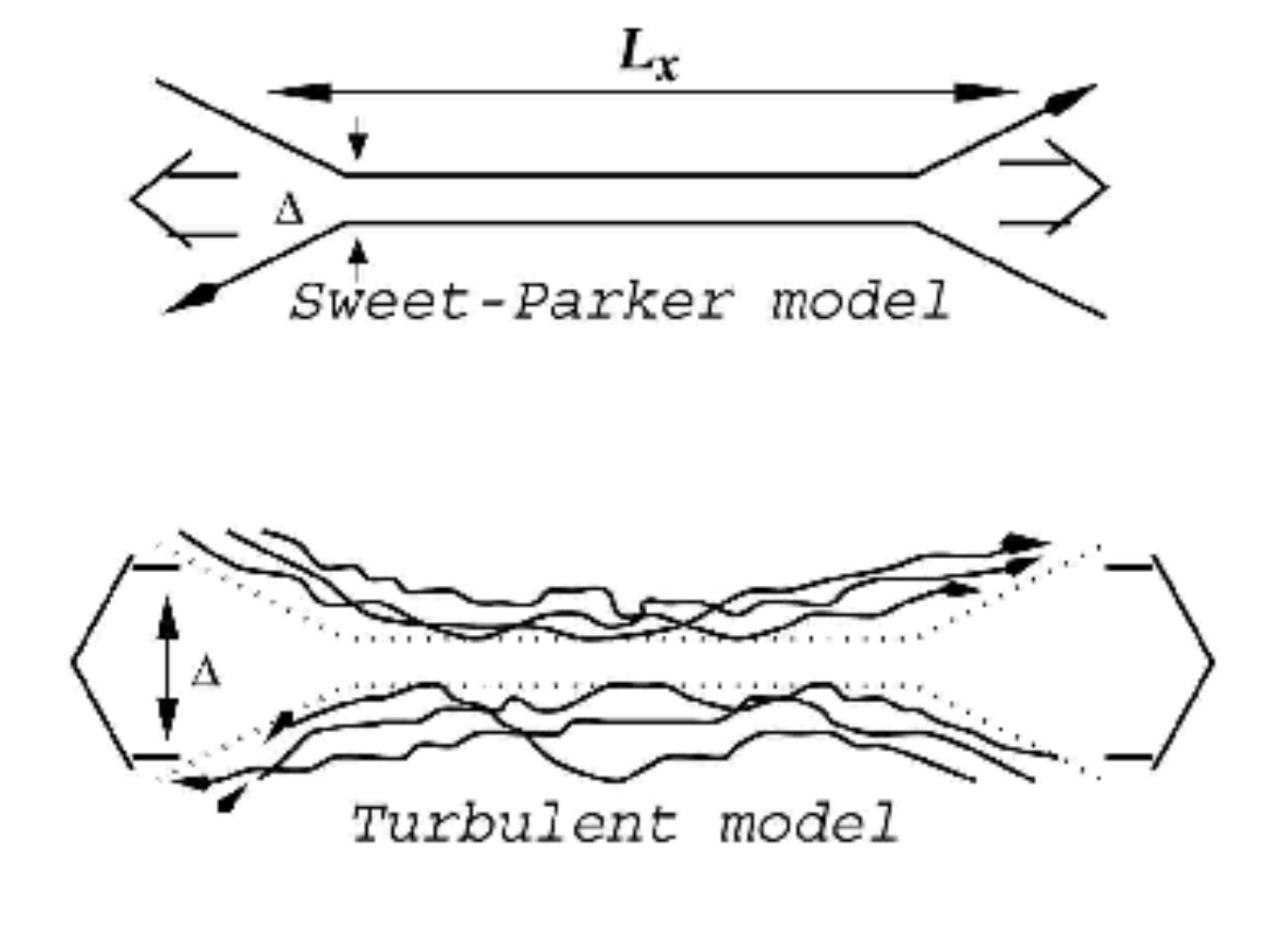}
\caption{{\it Upper plot}: Sweet-Parker model of reconnection.  The outflow is
limited to a thin width $\delta$, which is determined by Ohmic diffusivity.  The
other scale is an astrophysical scale $L_x \gg \delta$.  Magnetic field lines
are laminar. Modified from \cite{Lazarian_etal:2004}. Reproduced by permission
of the AAS.}
\label{recon}
\end{figure}

It is evident that the Sweet-Parker reconnection should be modified in the
presence of turbulence. Figure~\ref{recon} illustrates the modification that
takes place. It is evident that the outflow in the turbulent flow is not limited
by the microscopic region determined by resistivity, but is determined by
magnetic field wandering. Therefore there is no disparity between $L_x$ and
$\Delta$, e.g. for transAlfv\'enic turbulence they can be comparable. Actually,
Figure~\ref{recon} provides the concise illustration of the LV99 model of
reconnection.

Adopting that the field wandering is the cause of the reconnection zone opening
up, it is easy to calculate $\Delta$ in the regime when the turbulence injection
scale $L_i$ is less than $L_x$. Substituting $l_{\|}=L_i$ in Eq. (\ref{Lambda})
one finds that the perpendicular extend of the eddy at the injection scale is
$L_i M_A^2$. The transverse contributions from different eddies at the injection
scale are not correlated and therefore $\Delta$ is a result of random walk with
a step of $L_i M_A^2$. The number of the steps along $L_x$ is $L_x/L_i$ and thus
\begin{equation}
\Delta\approx \left(\frac{L_x}{L_i}\right)^{1/2} L_i M_A^2, ~~~~L_i<L_x,
\label{Delta1}
\end{equation}
and therefore
\begin{equation}
v_{rec, LV99}\approx V_A \left(\frac{L_i}{L_x}\right)^{1/2} M_A^2, ~~~~ L_i<L_x,
\label{LV99_1}
\end{equation}
which coincides with the LV99 result in this limit.

The result for $L_i>L_x$ can be obtained using the concept of Richardson
dispersion following the approach in \cite{Eyink_etal:2011}. Richardson
diffusion/dispersion can be illustrated with a simple hydrodynamic model.
Consider the growth of the separation between two particles $dl(t)/dt\sim v(l),$
which for Kolmogorov turbulence is $\sim \alpha_t l^{1/3}$, where $\alpha_t$ is
proportional to the energy cascading rate, i.e. $\alpha_t\approx V_L^3/L$ for
turbulence injected with superAlv\'{e}nic velocity $V_L$ at the scale $L$.  The
solution of this equation is
\begin{equation}
l(t)=[l_0^{2/3}+\alpha_t (t-t_0)]^{3/2},
\label{sol}
\end{equation}
which at late times leads to Richardson dispersion or $l^2\sim t^3$ compared
with $l^2\sim t$ for ordinary diffusion. This superdiffusive and even
superballistic behavior, i.e. $l^2$ increases faster than $t^2$, can be easily understood
if one takes into account that for points separated by the distance less than
turbulence injection scale, the larger the separation of the points the larger
the eddies that induce the point separation.

Both terms ``diffusion and dispersion'' can be used interchangeably, but keeping
in mind that the Richardson process results in superdiffusion \cite[see][and
references therein]{LazarianYan:2014} we feel that it is advantageous to use the
term ``dispersion''.

We again start with the Sweet-Parker reconnection.  There magnetic field lines
are subject to Ohmic diffusion.  The latter induces the mean-square distance
across the reconnection layer that a magnetic field-line can diffuse by
resistivity in a time $t$ given by
\begin{equation}
\langle y^2(t)\rangle \sim \lambda t.
\label{diff-dist}
\end{equation}
where $\lambda=c^2/(4\pi\sigma)$ is the magnetic diffusivity.  The field lines are
advected out of the sides of the reconnection layer of length $L_x$ at a
velocity of order $V_A$.  Therefore, the time that the lines can spend in the
resistive layer is the Alfv\'en crossing time $t_A=L_x/V_A$. Thus, field lines
that can reconnect are separated by a distance
\begin{equation}
\Delta = \sqrt{\langle y^2(t_A)\rangle} \sim \sqrt{\lambda t_A} = L_x/\sqrt{S},
\label{Delta}
\end{equation}
where $S$ is Lundquist number.  Combining Eqs. (\ref{eq.2}) and (\ref{Delta})
one gets again the well-known Sweet-Parker result, $v_{rec}=V_A/\sqrt{S}$.

The difference with the turbulent case is that instead of Ohmic diffusion one
should use the Richardson one \citep{Eyink_etal:2011}.  In this case the mean
squared separation of particles is $\langle |x_1(t)-x_2(t)|^2 \rangle\approx
\epsilon t^3$, where $t$ is time, $\epsilon$ is the energy cascading rate and
$\langle...\rangle$ denote an ensemble averaging \cite[see][]{Kupiainen:2003}.
For subAlfv\'enic turbulence $\epsilon\approx u_L^4/(V_A L_i)$ (see LV99) and
therefore analogously to Eq. (\ref{Delta}) one can write
\begin{equation}
\Delta\approx \sqrt{\epsilon t_A^3}\approx L_x(L_x/L_i)^{1/2}M_A^2,
\label{D2}
\end{equation}
where it is assumed that $L_x<L_i$.  Combining Eqs. (\ref{eq.2}) and (\ref{D2})
one obtains
\begin{equation}
v_{rec, LV99}\approx V_A (L/L_i)^{1/2}M_A^2,~~~L_i>L_x,
\label{LV99}
\end{equation}
that together with Eq. (\ref{LV99_1}) provides the description of the
reconnection for turbulent reconnection in the presence of sub-Alfv\'enic
turbulence. Naturally, LV99 can be easily generalized for the case of
superAlfv\'enic turbulence.

\subsection{Temporal and spatial Richardson diffusion}

We would like to stress that two formally different ways of obtaining LV99
reconnection rates have clear physical connection. In both cases we are dealing
with magnetic field lines stochasticity, but the case of Richardson dispersion
considers the evolution of magnetic fields lines in turbulent fluids, while
magnetic field wandering presents the spatial distribution of magnetic field
lines for a given moment of time. In a sense the dispersion of magnetic
field lines that was quantified in LV99\footnote{The magnetic field wandering
was discussed for an extended period to explain the diffusion of cosmic rays
perpendicular to the mean magnetic field, but, as was shown in
\cite{LazarianYan:2014}, those attempts employed scalings that were erroneous
even for the hypothetical Kolmogorov turbulence of magnetic fields, for which
they were developed.} presents the Richardson dispersion in space.

While we employed the Alfv\'enic incompressible motions to describe the physics of
Richardson dispersion, the process also takes place in compressible MHD
turbulence. This is due to the fact, that Alfv\'enic cascade is a part and parcel
of compressible MHD turbulence \citep{ChoLazarian:2003}. We can, however, note
parenthetically that even for turbulence of shocks, i.e. Burgers turbulence,
the phenomenon of Richardson diffusion is present \citep{Eyink_etal:2013}.

\subsection{Turbulent reconnection and violation of magnetic flux freezing}

Magnetic flux freezing is a concept that is widely used in astrophysics. It is
based on the Alfv\'en theorem, the proof of which is rather trivial for perfectly
conductive laminar fluids. For laminar fluids of finite conductivity, the
violation of Alfv\'en theorem becomes negligible as fluid conductivity increases.
This, however, is not true for turbulent fluids. Turbulent reconnection as we
discussed above induces reconnection diffusion. Mathematically the failure of
the flux freezing is discussed in \citet{Lazarian_etal:2015a}. The numerical
proof based on demonstrating of Richardson dispersion of magnetic field lines is
in \cite{Eyink_etal:2013}.

\subsection{Turbulent reconnection in compressible media}
\label{ssec:tricm}

Two new effects become important in compressible media as compared to its
incompressible counterpart that we discussed above. First of all, the density of
plasmas changes in the reconnection region and therefore the mass conservation
takes the form
\begin{equation}
\rho_{\rm i} v_{\rm rec,comp} L_x = \rho_{\rm s} V_A \Delta,
\label{dens}
\end{equation}
where $\rho_{\rm i}$ is the density of the incoming plasma far from the reconnection
layer and $\rho_{\rm s}$ is the density of plasma in the reconnection layer.

In addition, the derivation of the magnetic field wandering rate that we
discussed above was performed appealing to the Alfv\'enic component of MHD
turbulence. Numerical simulations in \cite{ChoLazarian:2002, ChoLazarian:2003}
demonstrated that the Alfv\'enic component develops independently from the
compressible MHD components in agreement with theoretical considerations in
GS95. Therefore one can estimate the amplitude in incompressible Alfv\'enic
perturbations by subtracting the contribution of the slow and fast modes from
the total energy of the turbulent motions
\begin{equation}
u_L^2\approx V_{total}^2-V_{comp}^2.
\label{Alfenic}
\end{equation}

Using both Eq. (\ref{dens}) and Eqs. (\ref{Delta1}) and (\ref{D2}) one can
generalize the expression for the reconnection rate (compare to Eqs.
(\ref{LV99}), (\ref{LV99_1})):
\begin{equation}
v_{rec, comp} \approx V_A \frac{\rho_i}{\rho_s} \min \left[ \left( \frac{L_i}{L_x} \right)^{1/2}, \left( \frac{L_x}{L_i} \right)^{1/2} \right] \frac{V_{total}^2-V_{comp}^2}{V_A^2}.
\label{rec_comp}
\end{equation}

If our turbulence driving is incompressible, another form of presenting the
reconnection rate is useful if one takes into account the relation between the
Alfv\'enic modes and the generated compressible modes obtained in
\cite{ChoLazarian:2002}
\begin{equation}\label{eq: comalfrat}
\frac{V_{comp}^2}{V_{Alf}^2}\approx C_1 \frac{v_{inj}}{V_{Alf}},
\end{equation}
where $C_1$ is a coefficient which depends on the media equation of state.
Taking into account the relation between the injection velocity and the
resulting velocity in weak turbulence given by Eq. (\ref{eq: comalfrat}), one can
rewrite Eq. (\ref{rec_comp}) as
\begin{equation}
v_{rec, comp} \sim V_A \frac{\rho_i}{\rho_s} \min \left[ \left( \frac{L_i}{L_x} \right)^{1/2}, \left( \frac{L_x}{L_i} \right)^{1/2} \right] \frac{v_{inj}^2 (1-C_1(v_{inj}/V_A))}{V_A^2}.
\label{recon_relat}
\end{equation}

\subsection{Turbulent reconnection in partially ionized gas}

\begin{figure}[t]
\centering
\raisebox{-0.5\height}{\includegraphics[width=0.48\textwidth]{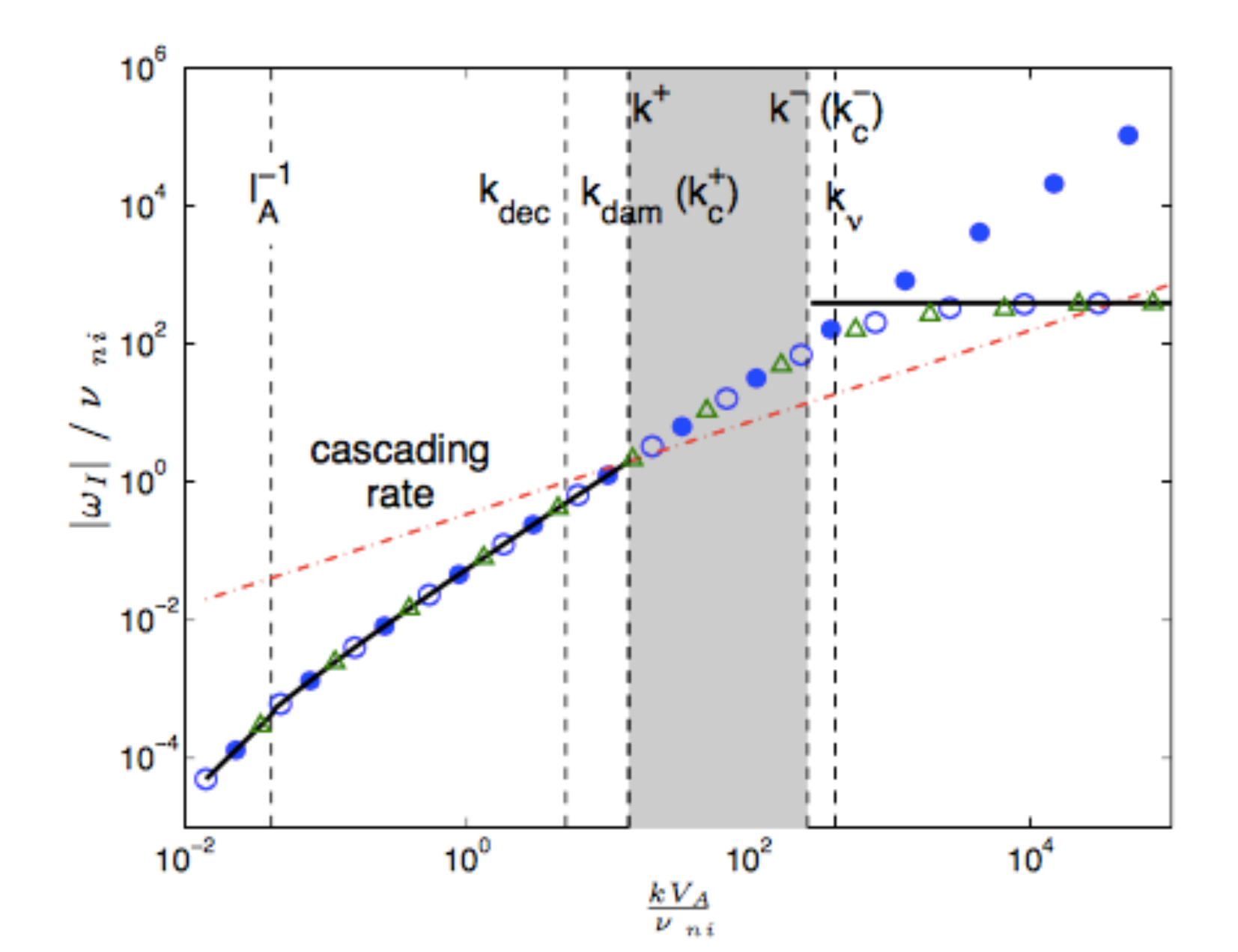}}
\caption{Damping of Alfv\'enic turbulence in low beta partially
ionized gas. The damping of turbulence happens when the rate of damping (solid line)
intersects the dashed line corresponding to the cascading rate. From \cite{Xu_etal:2015}.}
\label{damp1}
\end{figure}

Partially ionized gas represents a complex medium where the ions co-exist with
neutrals. Complex processes of ionization and recombination are taking place in
turbulent partially ionized gas. However, in view of turbulent reconnection, the
major feature of the partially ionized gas is that the turbulent motions in
partially ionized gas are subject to damping which arises from both neutral-ion
collisions and the viscosity associated with neutrals \cite[see][for a detailed
discussion of the latter process]{Lazarian_etal:2004, Xu_etal:2015}. Figure
\ref{damp1} illustrates the damping of Alfv\'en modes in a typical environment
of molecular cloud. The corresponding damping scales are substantially larger
than those in the fully ionized gas, which poses the question of how turbulent
reconnection is modified.

The reconnection in partially ionized gas was discussed on the basis of the
Richardson dispersion in \cite{Lazarian_etal:2015b}. The essence of the approach
is that on the scales at which Richardson dispersion is applicable, the
magnetic fields are not frozen in and therefore magnetic reconnection is fast.
Therefore the issue at which scale the reconnection is fast boils down to what
is the scale of the onset of the Richardson dispersion description for the
magnetic field lines.

It is natural to identify the magnetic field lines as subject to the Richardson
dispersion as soon as the separation of the lines exceeds the size of the
smallest turbulence eddy, i.e. the size of the critically damped eddy. In
partially ionized gas the ion-neutral damping or viscosity determines this size.
As the eddies are anisotropic, we would associate the damping scale with the
parallel scale of the eddies $l_{\|, crit}$. Due to the shear induced by
perpendicular motions associated with these eddies the magnetic field lines
which are initially separated by $r_{init}$ are spreading further and
further from each other. The rate of line separating $dr/dl$ is proportional to
the $r/l_{\bot, crit}$ and this provides an exponential rate of separation. It
is easy to show that separation becomes equal to $l_{\bot, crit}$ after the field lines
are traced over a distance of
\begin{equation}
L_{RR}\approx l_{\|, crit} \ln (l_{\bot, crit}/r_{init}),
\label{RR}
\end{equation}
which was introduced by \cite{RechesterRosenbluth:1978} in the framework of
"turbulence" with a single scale of driving. We follow
\cite{NarayanMedvedev:2003} and \cite{Lazarian:2006} associating this scale with
the smallest turbulent eddies \cite[cf.][]{Chandran_etal:2000}, as the smallest
scales induce the largest shear. For $r_{init}$ it is natural to associate this
length with the separation of the field lines arising from the action
of Ohmic resistivity on the scale of the critically damped eddies
\begin{equation}
r_{init}^2=\eta l_{\|, crit}/V_A,
\label{int}
\end{equation}
where $\eta$ is the Ohmic resistivity coefficient. Taking into account Eq.
(\ref{int}) and that
\begin{equation}
l_{\bot, crit}^2=\nu l_{\|, crit}/V_A,
\end{equation}
where $\nu$ is the viscosity coefficient, one can rewrite Eq. (\ref{RR}) as:
\begin{equation}
L_{RR}\approx l_{\|, crit}\ln Pt,
\label{RR2}
\end{equation}
where $Pt=\nu/\eta$ is the Prandtl number. This means that when the current
sheets are much longer than $L_{RR}$, magnetic field lines undergo
Richardson dispersion and according to \cite{Eyink_etal:2011} the reconnection
follows the laws established in LV99.  At the same time on scales less than
$L_{RR}$ magnetic reconnection may be slow\footnote{Incidentally, this can
explain the formation of density fluctuations on scales of thousands of AU, that
are observed in the ISM.}.

Additional effects, e.g. diffusion of neutrals perpendicular to magnetic field
might potentially influence the reconnection rate \cite[see][]{VishniacLazarian:1999}. Indeed,
ions can recombine in the reconnection zone and this can allow the matter to
outflow as a flow of neutrals. This outflow is not directly constrained by
magnetic field and therefore \cite{VishniacLazarian:1999} obtained large
reconnection rates even for laminar magnetic fields provided that magnetic
fields are perfectly anti-parallel and astrophysical medium is pure ionized
hydrogen (see also a numerical study by \cite{HeitschZweibel:2003}). The
reconnection rates plummet in the presence of the guide field and heavy ions
(``metals'') which are subject to ionization by the ambient field. Therefore the
effect of ``ambipolar reconnection'' is of marginal importance for most of the
settings involving realistically turbulent media\cite[see][]{Lazarian_etal:2004}.

\section{Testing turbulent reconnection}

\begin{figure*}[t]
\centering
\raisebox{-0.5\height}{\includegraphics[trim = 20mm -10mm 20mm 0mm, clip, width=0.48\textwidth]{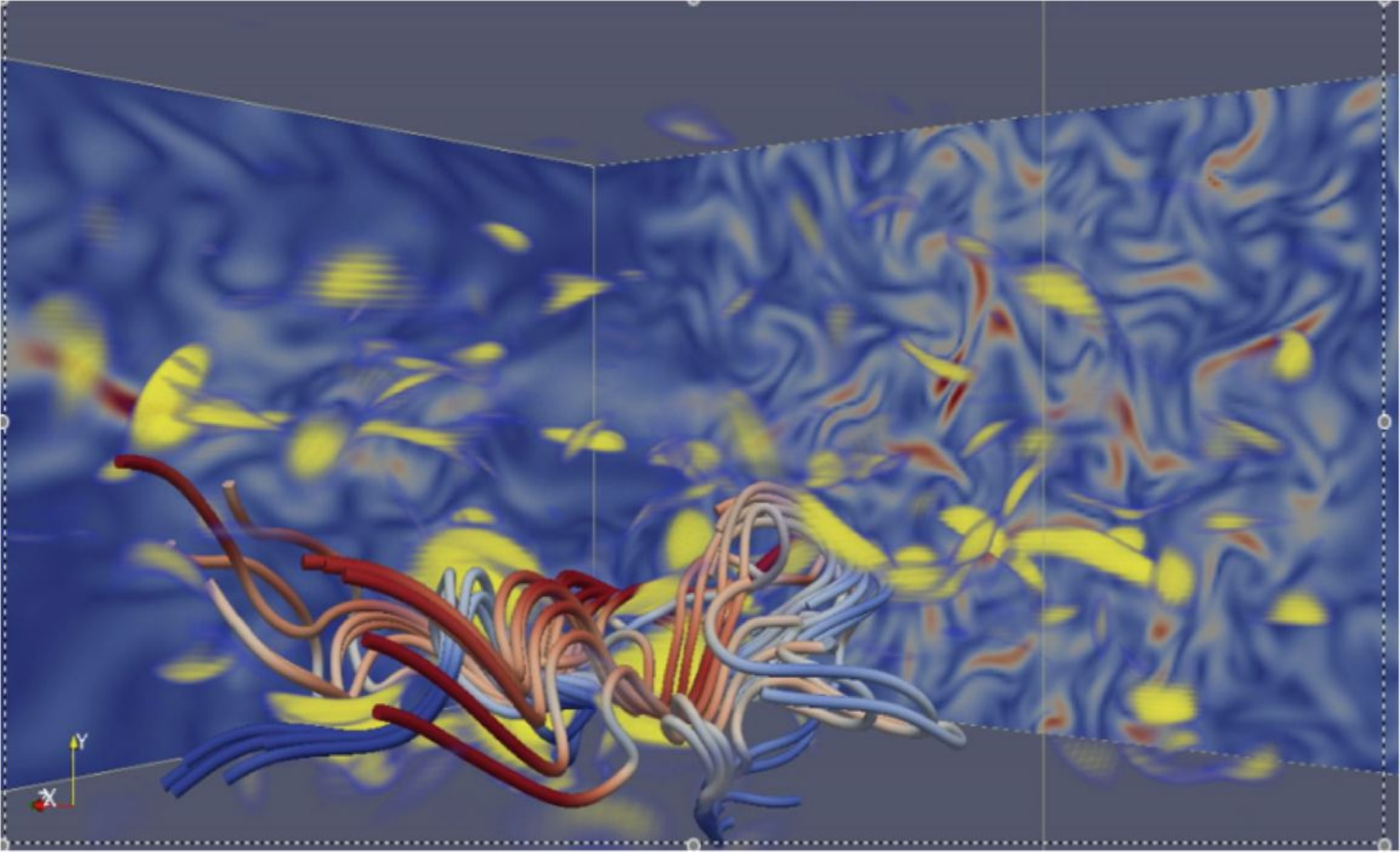}}
\raisebox{-0.5\height}{\includegraphics[width=0.25\textwidth]{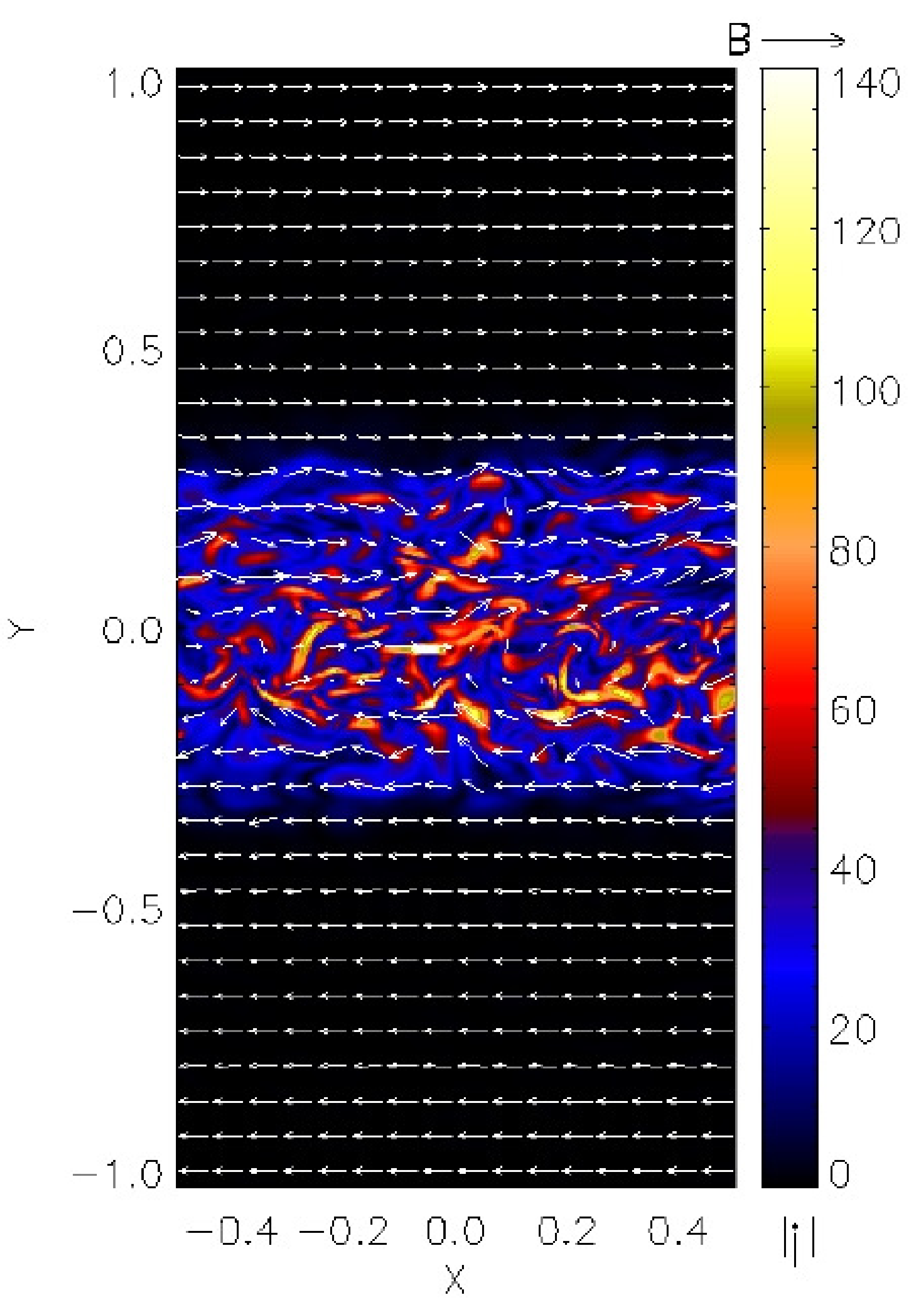}}
\raisebox{-0.5\height}{\includegraphics[width=0.25\textwidth]{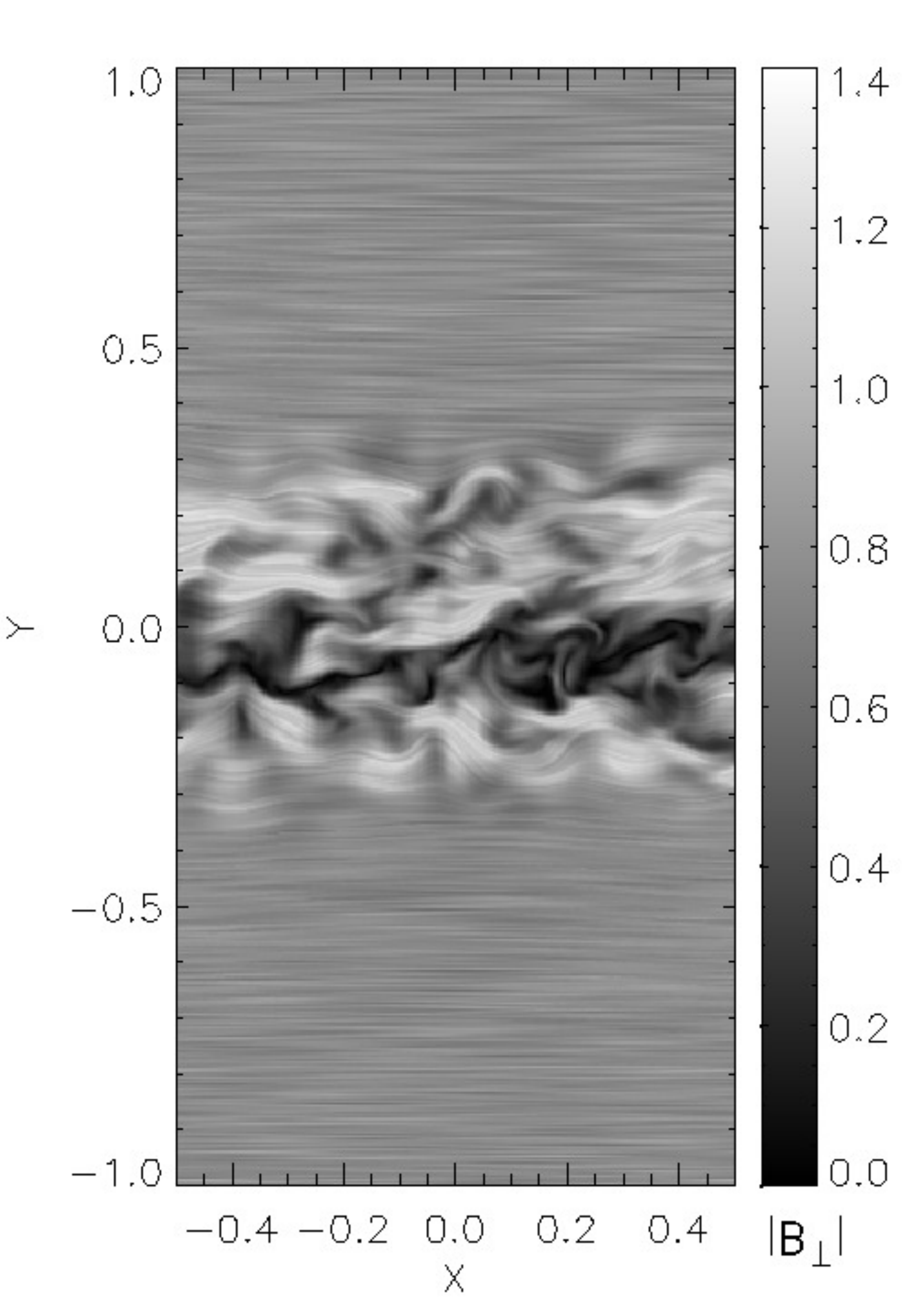}}
\caption{ Visualization of reconnection simulations in \cite{Kowal_etal:2009, Kowal_etal:2012}.
{\it Left panel}: Magnetic field in the reconnection region.
{\it Central panel}: Current intensity and magnetic field configuration during
stochastic reconnection.  The guide field is perpendicular to the page. The
intensity and direction of the magnetic field is represented by the length and
direction of the arrows.  The color bar gives the intensity of the current.
{\it Right panel}: Representation of the magnetic field in the reconnection zone
with textures.
Reproduced from \cite{Kowal_etal:2009} by permission of the AAS.
\label{visual}}
\end{figure*}

Figure~\ref{visual} illustrates results of numerical simulations of turbulent
reconnection with turbulence driven using wavelets in \cite{Kowal_etal:2009} and
in real space in \cite{Kowal_etal:2012}.

As we show below, simulations in \cite{Kowal_etal:2009, Kowal_etal:2012}
confirmed LV99 prediction that turbulent reconnection is fast,  i.e. it does not
depend on resistivity, and provided a good correspondence with the LV99
prediction on the injection power.

\begin{figure}[t]
\centering
\includegraphics[width=0.48\textwidth]{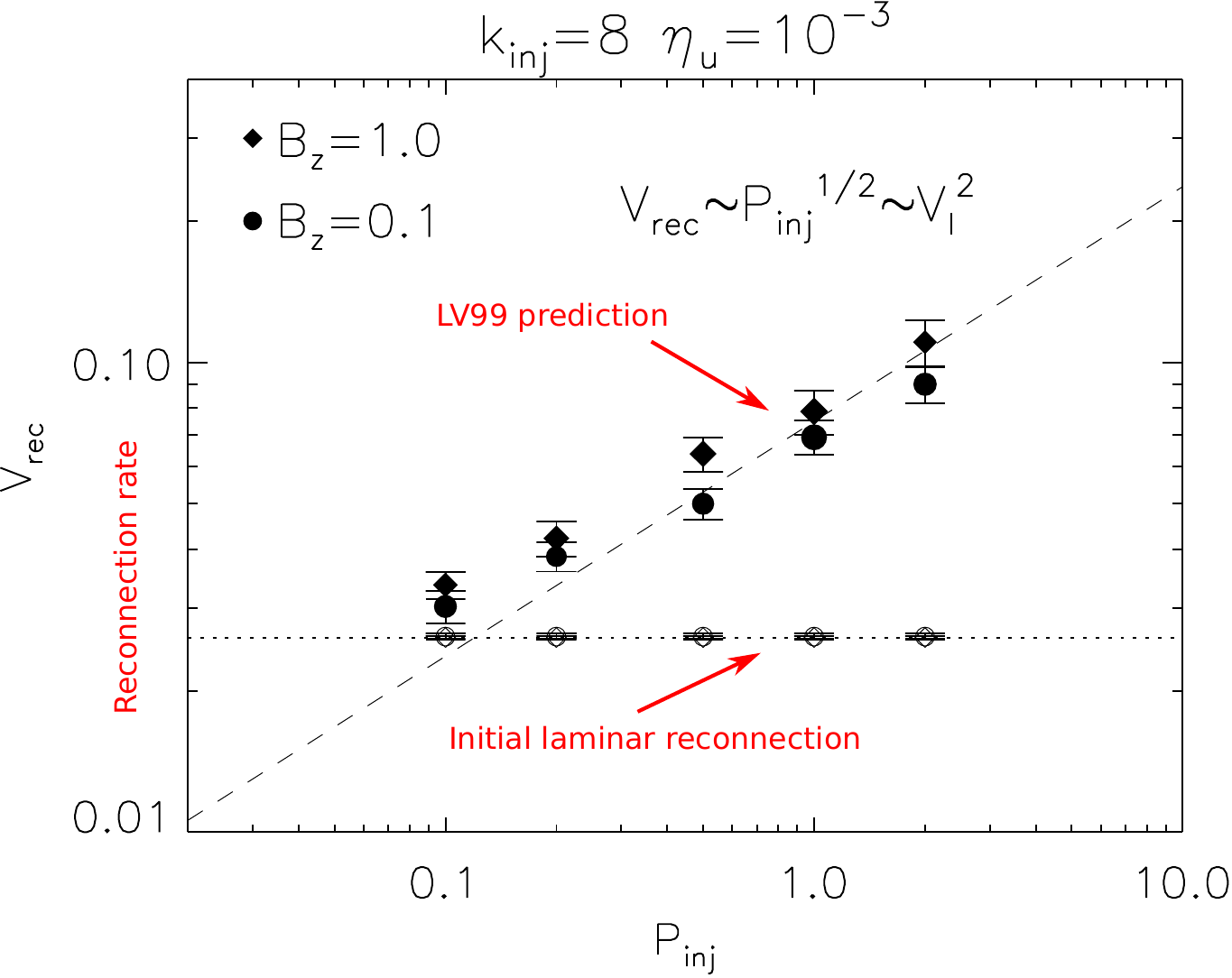}
\includegraphics[width=0.48\textwidth]{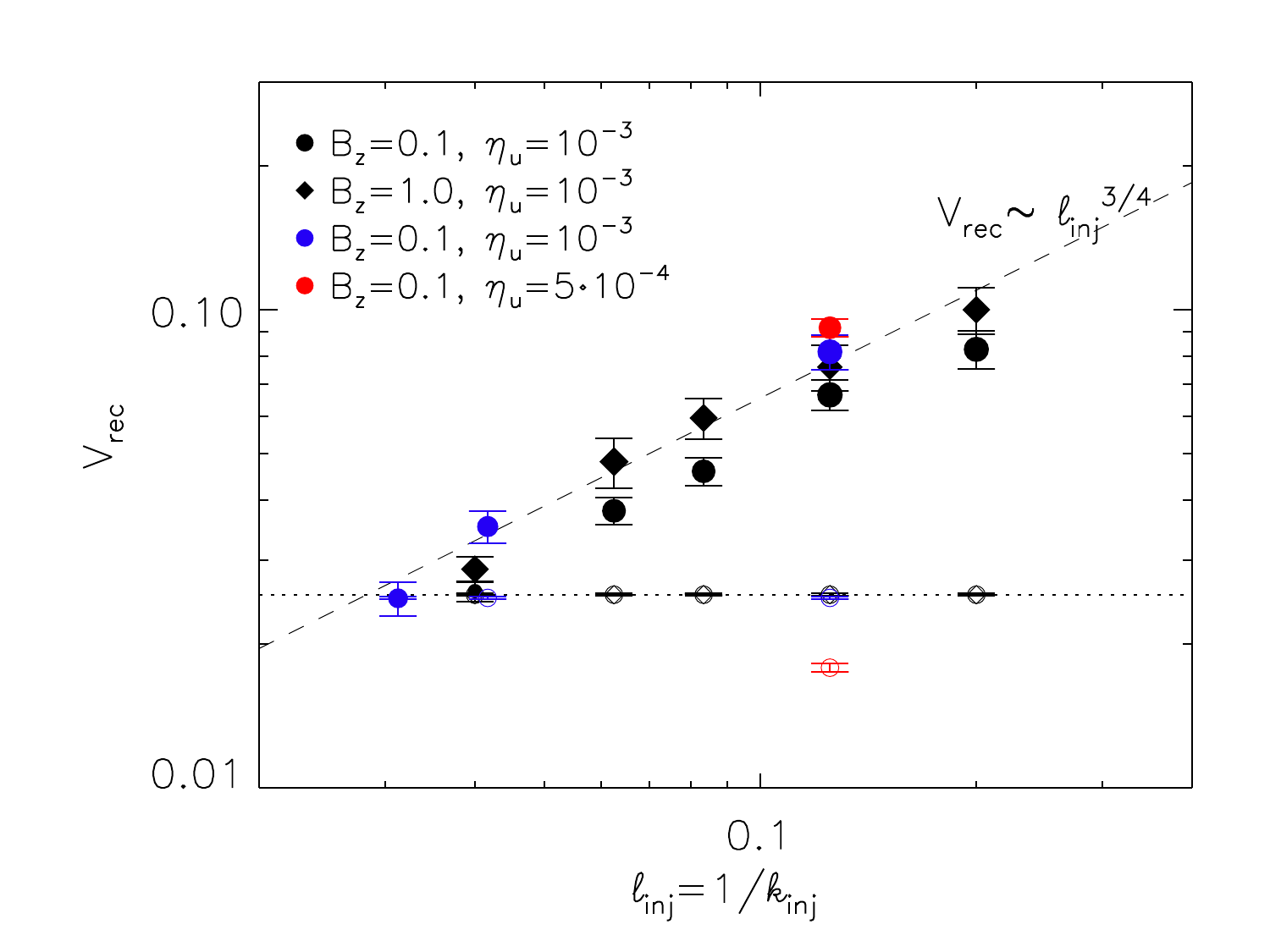}
\caption{
{\it Left Panel} The dependence of the reconnection velocity on the injection
power for different simulations with different drivings. The predicted LV99
dependence is also shown. $P_{inj}$ and $k_{inj}$ are the injection power and
scale, respectively, $B_z$ is the guide field strength, and $\eta_u$ is the value
of uniform resistivity coefficient.
{\it Right Panel} The dependence of the reconnection velocity on the injection
scale. Reproduced from \cite{Kowal_etal:2012}.
\label{figure6}}
\end{figure}

In the simulations subAlfv\'enic turbulence was induced, i.e. with the energy of
kinetic motions less than the energy of magnetic field. Indeed, according to Eq.
(\ref{LV99_1}) $v_{rec, LV99}\sim u_l^2$. At the same time for the weak
turbulence the injected power
\begin{equation}
P_{inj}\sim v_{inj}^2/\Delta t_{inj}
\label{energy}
\end{equation}
is equal to the cascading power given by Eq. (\ref{cascading}). This provides a relation
\begin{equation}
  v_{rec, LV99} \sim u_l^2\sim v_{inj}\sim P_{inj}^{1/2}.
\label{relation}
\end{equation}
The corresponding dependence is shown in Figure~\ref{figure6}, left panel.

We also see some differences from the idealized theoretical predictions. For
instance, the injection of energy in LV99 is assumed to happen at a given scale
and the inverse cascade is not considered in the theory. Therefore, it is not
unexpected that the measured dependence on the turbulence scale differs from the
predictions. In fact, it is a bit more shallow compared to the LV99 predictions
(see Figure~\ref{figure6}, right panel).

\begin{figure}[t]
\centering
\includegraphics[width=0.48\textwidth]{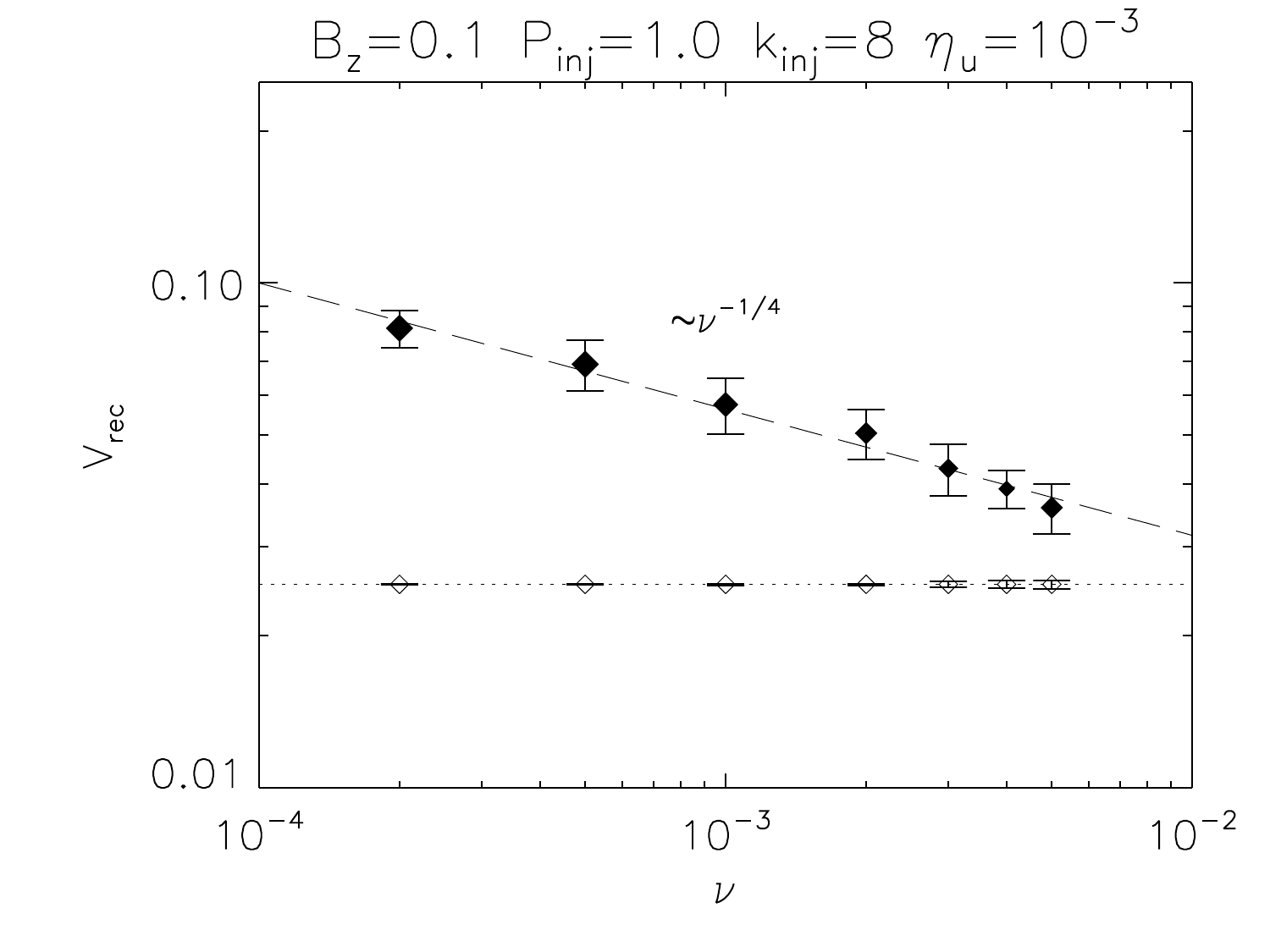}
\includegraphics[width=0.48\textwidth]{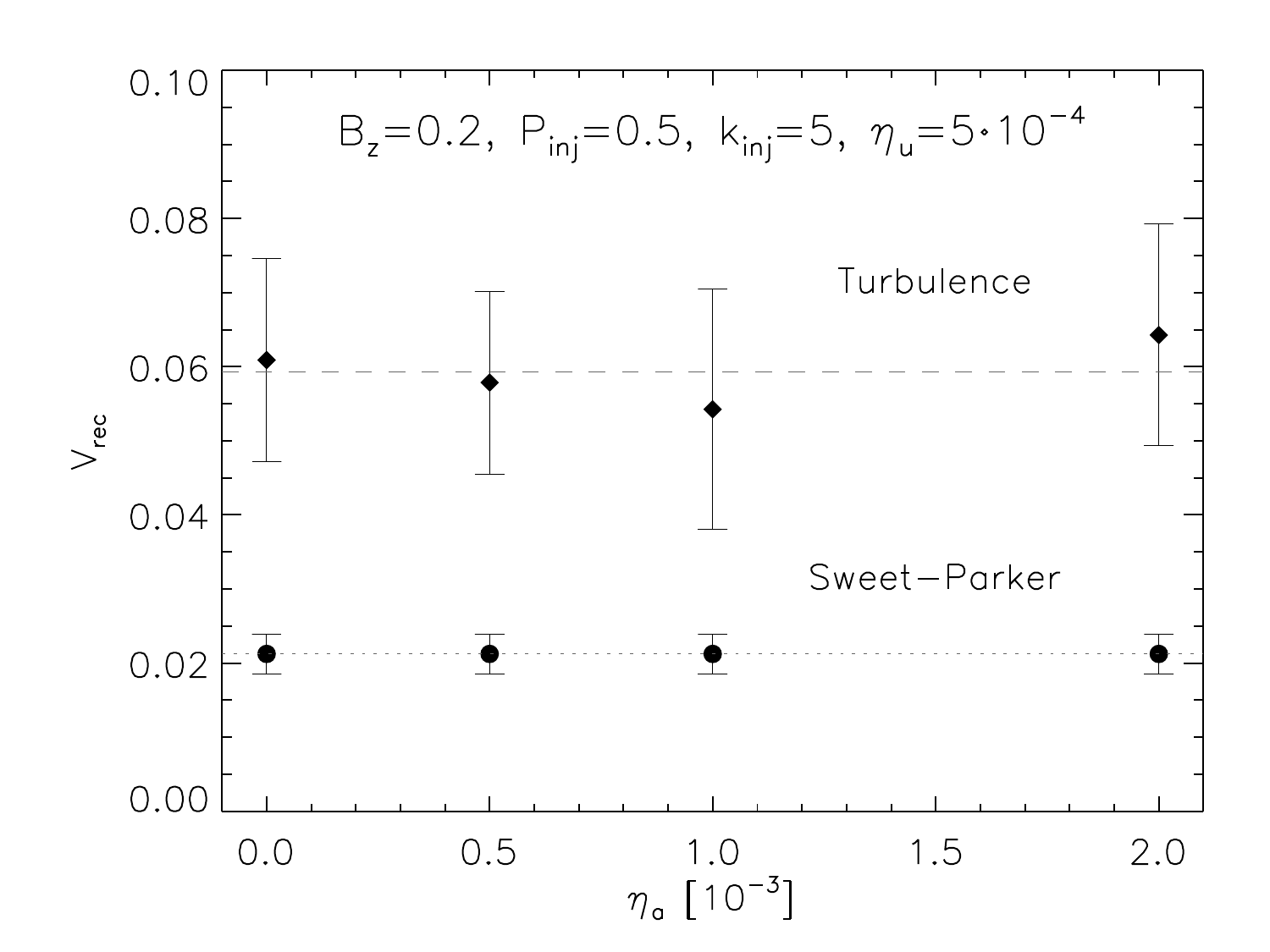}
\caption{{\it Left panel}. The dependence of the reconnection velocity on
uniform viscosity in the 3D isothermal models of Sweet-Parker reconnection (open
symbols) and reconnection enhanced by the presence of turbulence (closed
symbols) from \cite{Kowal_etal:2012}.  }  {\it Right
panel.} The reconnection rate in models with anomalous resistivity for
Sweet-Parker case (filled circles) and in the presence of turbulence (filled
diamonds). We observe no dependence of the reconnection rate on the strength of
anomalous effects. Reproduced from \cite{Kowal_etal:2009}.
\label{fig:viscosity}
\end{figure}

The left panel of Figure~\ref{fig:viscosity} shows the dependence of the
reconnection rate on explicit uniform viscosity obtained from the isothermal
simulations of the magnetic reconnection in the presence of turbulence
\citep{Kowal_etal:2012}. The open symbols show the reconnection rate for the
laminar case when there was no turbulence driving, while closed symbols
correspond to the mean values of reconnection rate in the presence of saturated
turbulence.  All parameters in those models were kept the same, except the
uniform viscosity which varied from $10^{-4}$ to $10^{-2}$ in the code units.
We believe that the dependence can be explained as the effect of the finite
inertial range of turbulence than the effect of energy balance affected by
viscosity or boundary conditions.  For an extended range of motions, LV99 does
not predict any viscosity dependence, if the dissipation scale lies much below
the scale of current sheet.  However, for numerical simulations the range of
turbulent motions is very limited and any additional viscosity decreases the
resulting velocity dispersion and the field wandering thus affecting
the reconnection rate.

LV99 predicted that in the presence of sufficiently strong turbulence, plasma
effects should not play a role.  The accepted way to simulate plasma effects
within MHD code is to use anomalous resistivity.  The results of the
corresponding simulations are shown in the right panel of
Figure~\ref{fig:viscosity} and they confirm that the change of the anomalous
resistivity does not change the reconnection rate.

As we discussed in section 3, the LV99 expressions can be obtained by applying
the concept of Richardson dispersion to a magnetized layer. Thus by testing the
Richardson diffusion of magnetic field, one also provides tests for the theory
of turbulent reconnection. A successful direct testing of the temporal
Richardson dispersion of magnetic field lines was performed in
\cite{Eyink_etal:2013}. The study confirmed that magnetic fields are not frozen
in highly conducting fluids, as this follows from the LV99 theory.

Within the derivation adopted in LV99 current sheet is broad with individual
currents distributed widely within a three dimensional volume  and the
turbulence within the reconnection region is similar to the turbulence within a
statistically homogeneous volume. Numerically, the structure of the reconnection
region was analyzed by \cite{Vishniac_etal:2012} based on the numerical work by
\cite{Kowal_etal:2009}. The results support LV99 assumptions about reconnection
region being broad, the magnetic shear is more o less coincident with the
outflow zone, and the turbulence within it is broadly similar to turbulence in a
homogeneous system.

Another prediction that follows from LV99 theory is that the turbulence required
for the process of turbulent reconnection can be generated by the process of
turbulent reconnection. In particular, a theory of reconnection flares in low
$\beta$ (highly magnetized) plasmas was discussed in
\cite{LazarianVishniac:2009}, while the expressions presenting reconnection
rates in high $\beta$ plasmas are presented in \cite{Lazarian_etal:2015b}.

\section{Towards theory of turbulent relativistic reconnection}

Recently, it has been recognized that the relativistically magnetized plasma,
so-called Poynting-dominated plasma, plays an important role for many high
energy astrophysical phenomena with relativistic outflows, such as pulsar wind
nebulae, relativistic jets, and gamma-ray bursts. Those phenomena are believed
to have a strongly magnetized compact object with rapid spin which naturally
explains collimated jets or magnetized winds. The energy stored in the magnetic
field initially needs to be converted into kinetic and radiation energy to
explain the observations. However, the usual collisional magnetic field
dissipation fails to explain the necessary dissipation rate. Relativistic
turbulent magnetic reconnection is considered to be one of the most probable
mechanism for the magnetic dissipation, and we review our recent development of
the relativistic version of turbulent reconnection theory reported in
\cite{Takamoto_etal:2015}.

We have seen that the relativistic balanced MHD turbulence in terms of theory is
a clone of the GS95 model. We noticed that the imbalanced Alfv\'enic turbulence
simulations provide very similar results in relativistic and non-relativistic
cases \footnote{Thus we can expect that the theory of imbalanced relativistic MHD
can be also very similar to \cite{BeresnyakLazarian:2008} model.}. As properties
of Alfv\'enic turbulence dominate the LV99 reconnection, one can expect that the
LV99 theory can be reformulated in terms of relativistic physics. However,
instead of reformulating LV99 in terms of relativistic variables, for the time
being, we shall use the theory in its non-relativistic formulation and seek its
correspondence with the simulations of the relativistic turbulent reconnection.

For instance, it is obvious that effects of compressibility are likely to be more
important in relativistic reconnection compared to its non-relativistic
counterpart. This is because in Poynting-flux-dominated plasmas the magnetic
field can induce a relativistic velocity in current sheets but the Alfv\'en
velocity is limited by the light velocity, which allows the induced turbulence
to be trans-Alfv\'enic one. Therefore, we use our results from section \ref{ssec:tricm}, in
particular Eq. (\ref{rec_comp}).

The comparison between the theoretical expectations and numerical simulations was
performed in \cite{Takamoto_etal:2015}. The simulation is calculated using the
relativistic resistive MHD code developed in \cite{TakamotoInoue:2011}. The
initial current sheet is assumed to be the relativistic Harris current sheet
with uniform temperature $k_B T / m c^2 = 1$ where $k_B, T, m, c$ are the
Boltzmann constant, temperature, rest mass, and light velocity, respectively.
The relativistic ideal gas is assumed, $h = 1 + (p/\rho c^2) (\Gamma/(\Gamma - 1))$
with $\Gamma = 4/3$ where $h, p, \rho$ are the specific enthalpy, gas pressure,
rest mass density. Following \cite{Kowal_etal:2009}, we used the open boundary
in the direction perpendicular to the current sheet and parallel to the magnetic
field, which corresponds to x and z direction. The periodic boundary is used in y
direction. The guide field is basically omitted other than the runs used for
obtaining Figure \ref{general}. Turbulence is driven by injecting a randomly determined
turbulent flow every fixed time step. The turbulent flow has a flat kinetic
energy spectrum and a characteristic wavelength is distributed around sheet
width scale. More detailed setup is provided in \cite{Takamoto_etal:2015}. To
quantify the reconnection rate the approach based on measurements of the change
of the absolute value of magnetic flux in \cite{Kowal_etal:2009} was used. In
the calculations, we investigated turbulent reconnection in plasmas with the
magnetization parameter from $0.04$ (matter dominated) to $5$ (Poynting
dominated). Note that we assume a relativistic temperature, so that the plasma
is always relativistic.

\begin{figure}[t]
\centering
\includegraphics[width=0.5\textwidth]{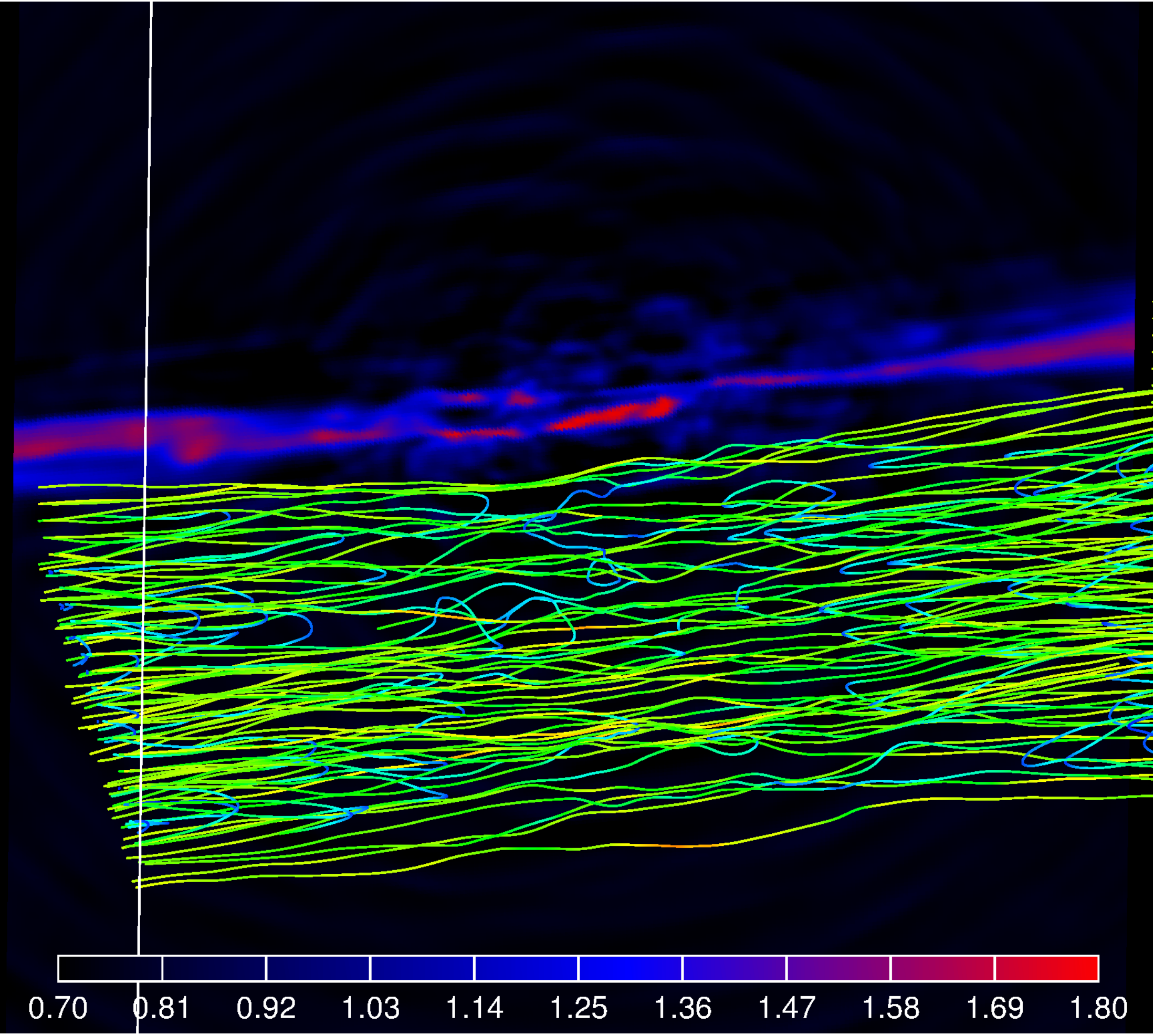}
\caption{
Visualization of relativistic reconnection simulations in the case of the
magnetization parameter $\sigma=5$ from \cite{Takamoto_etal:2015}. The lines
describe the magnetic field lines relating magnetic reconnection. The background
plane shows the gas pressure profile in the unit of the upstream magnetic
pressure. Similarly to the non-relativistic case, the magnetic field lines are
wandering due to the injected turbulence, even in a Poynting-dominated plasma,
which results in a wider reconnection exhaust region and large reconnection
rate.
\label{fig:RMRsheet}}
\end{figure}

Figure \ref{fig:RMRsheet} illustrates the magnetic field structure and gas pressure
profile obtained by the simulations in \cite{Takamoto_etal:2015}. The
magnetization parameter is $\sigma=5$ and there is no guide field. The lines
describe the magnetic field, and the background plane shows the gas pressure
profile in units of the upstream magnetic pressure. It indicates the
turbulence induces reconnecting points around the central sheet region. It also
shows the magnetic field is wandering similarly to the non-relativistic case,
which is responsible for determining the size of exhaust region and reconnection
rate in LV99 theory. Note that the injected turbulence is sub-Alfv\'enic
velocity but it can cause the stochastic magnetic field lines even in the case
of Poynting-dominated plasma whose Alfv\'en velocity is relativistic.

\begin{figure}[t]
\centering
\includegraphics[width=0.45\textwidth]{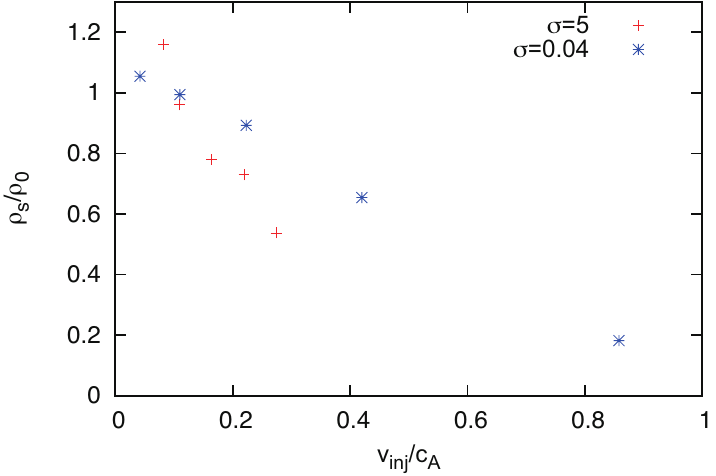}
\includegraphics[width=0.45\textwidth]{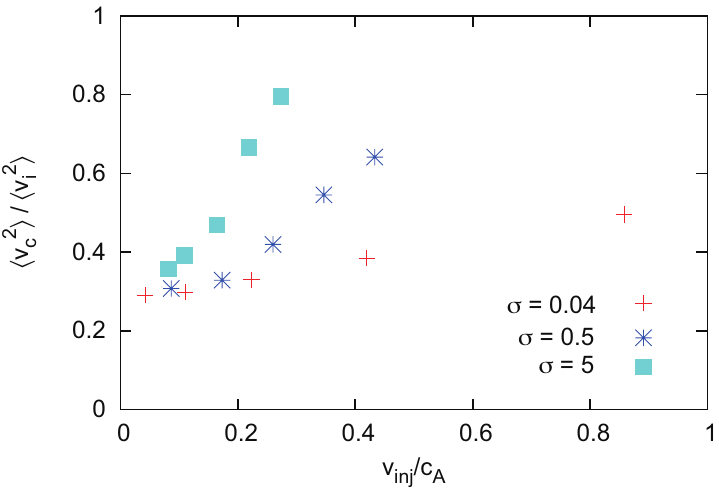}
\caption{{\it Left Panel} Variations of plasma density in relativistic reconnection.
{\it Right Panel} Generation of compressible modes in relativistic reconnection.
From \cite{Takamoto_etal:2015}.
\label{density_velocity}}
\end{figure}

Figure \ref{density_velocity} (left panel) illustrates that in the process of
relativistic magnetic reconnection the density inside the sheet changes
substantially as the injected turbulent energy increases comparing with the
energy flux of the reconnection outflow. This is expected as can be seen from
the simple arguments in \cite{Takamoto_etal:2015}. Indeed, the conservation of
energy flux can be written as:
\begin{equation}
  \rho_{\rm i} h_{\rm i} c^2 \gamma_{\rm i}^2 \left(1 + \sigma_{\rm i} \right) v_{\rm i} L \
= \rho_{\rm s} h_{\rm s} c^2 \gamma_{\rm s}^2 \left(1 + \sigma_{\rm s} \right) v_{\rm s} \delta
  ,
  \label{eq:R1}
\end{equation}
where $\rho, h, \gamma$ are the mass density, specific enthalpy, and Lorentz
factor, respectively. $c$ is the light velocity, and $\sigma \equiv B^2/4 \pi
\rho h c^2 \gamma^2$ is the magnetization parameter. The subscript $i, s$ means
the variables defined in the inflow and outflow region, respectively. If we
inject turbulence externally, this can be written as an input of turbulent
energy into sheet, so that Equation (\ref{eq:R1}) becomes
\begin{eqnarray}
  \rho_{\rm i} h_{\rm i} c^2 \gamma_{\rm i}^2 \left(1 + \sigma_{\rm i} \right) v_{\rm i} L + (\rho_{\rm in} + B_0^2) \epsilon_{\rm inj} l_x l_z
= \rho_{\rm s} h_{\rm s} c^2 \gamma_{\rm s}^2 \left(1 + \sigma_{\rm s} \right) v_{\rm s} \delta
\nonumber
\\
= \rho_{\rm s} h_{\rm s} c^2 \gamma_{\rm s}^2 \left(1 + \sigma_{\rm s} \right) v_{\rm s} \sqrt{\epsilon t_A^3}
  ,
  \label{eq:R2}
\end{eqnarray}
where $\epsilon$ is the energy injection rate of the turbulence, and we used
${\vec E}_{\rm turb}^2 = ({\vec v}_{\rm turb} \times {\vec B}_0)^2 \sim B_0^2
v_{\rm turb}^2 / 2$. $l_x, l_z$ are the injection size along x and z axis. and
Equation \ref{D2} is used in the second line. Note that the outflow energy flux is
measured at the boundary of the reconnection outflow, and we assumed all the
injected energy into the sheet is ejected as the outflow flux along the sheet,
that is, the escaping components as compressible modes is assumed at least less
than the Alfv\'enic component. Equation (\ref{eq:R2}) shows that when the
injected turbulence, $\epsilon$, is small, the 2nd term in the left-hand side of
the equation can be neglected, and the inflow velocity $v_{\rm i}$ increases as
$\delta \propto \sqrt{\epsilon}$. However, if the injected turbulence is
sufficiently strong, the neglected term increases as $\epsilon$, and becomes
comparable to the outflow flux which increases more slowly as $\epsilon^{1/2}$.
In this case, combining with the conservation of mass,
\begin{equation}
  \label{eq:R3}
  \rho_{\rm i} \gamma_{\rm i} v_{\rm i} L = \rho_{\rm s} \gamma_{\rm s} v_{\rm s} \delta,
\end{equation}
equation (\ref{eq:R2}) gives
\begin{equation}
  \label{eq:R4}
  \frac{\rho_{\rm s}}{\rho_{\rm i}} = \frac{1}{(1 + \sigma - \gamma_{\rm s}) \gamma_{\rm s}} \left[2 \sigma \gamma_{\rm s}^2
   - (1 + \sigma) \frac{\epsilon_{\rm inj}}{\sqrt{\epsilon t_A^3}}
    \frac{l_x l_z}{v_{\rm s} c^2} \right].
\end{equation}
This shows that the density ratio decreases as $\epsilon^{1/2} \propto v_{\rm inj}$,
indicated as Figure \ref{density_velocity} (left panel). The change of the
matter density is an important factor in expression for the turbulent
reconnection rate given by Eq. (\ref{rec_comp}).

The other factor that we have to account is the decrease of the energy in
Alfv\'enic turbulence as more energy is getting transferred to compressible
modes for highly magnetized plasmas as illustrated by Figure
\ref{density_velocity} (right panel). Note, that the compressible component is
obtained through the Helmholtz decomposition into solenoidal and compressible
part rather than through the mode decomposition as in \cite{ChoLazarian:2002,
ChoLazarian:2003} or \cite{KowalLazarian:2010}. The latter procedure has not
been adopted for the relativistic turbulence so far. Interestingly, the
compressible component increases with the increase of the
$\sigma$-parameter\footnote{This may indicates a relation similar to one
predicted by \cite{GaltierBanerjee:2011}, i.e. that the compressible component
is proportional to $B_0^2$, exists even in relativistic MHD turbulence.}.
All in all, we conjecture that the compressible generalization of LV99 theory,
see Eq. (\ref{recon_relat}), can provide the description of relativistic reconnection.

\begin{figure}
\centering
\raisebox{-0.5\height}{\includegraphics[width=\textwidth]{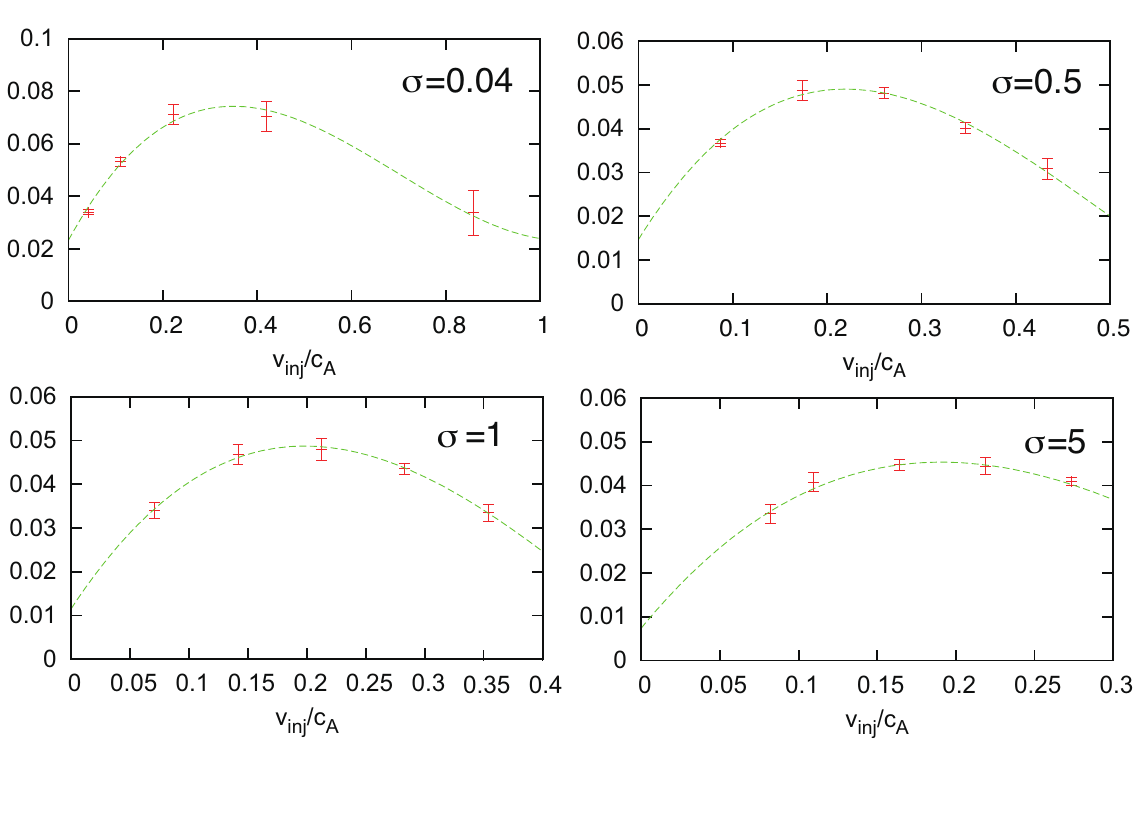}}
\caption{Reconnection rate in terms of various magnetization parameters: $\sigma
= 0.04, 0.5, 1, 5$. The green dashed curves are the modified turbulent
reconnection law taking into account the effect of density decrease and
compressible turbulence effects. \label{relativ_result}}
\end{figure}

\begin{figure}[t]
\centering
\includegraphics[width=0.45\textwidth]{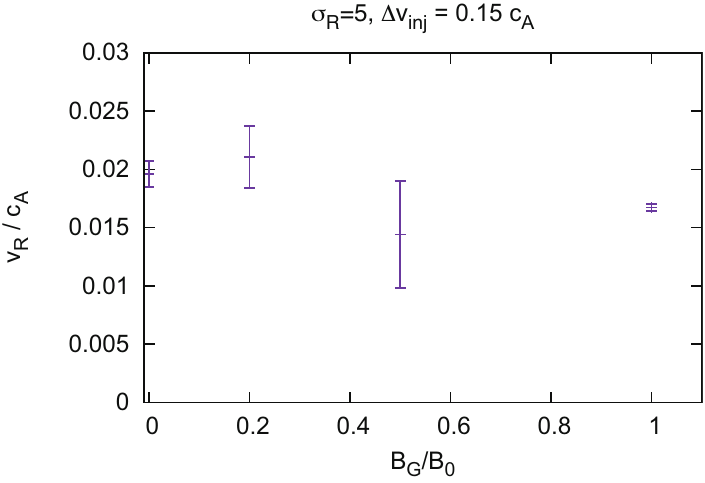}
\includegraphics[width=0.45\textwidth]{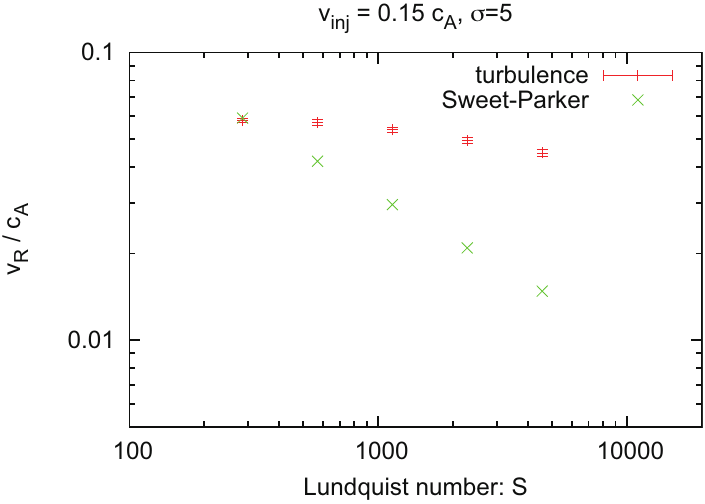}
\caption{{\it Left Panel}. Dependence of the reconnection rate on the guide
field. {\it Right Panel}. Dependence of the reconnection rate on resistivity.
From \cite{Takamoto_etal:2015}.
\label{general}}
\end{figure}

Accounting for both effects \cite{Takamoto_etal:2015} obtained a good
correspondence between the theoretical predictions and numerical results. Figure
\ref{relativ_result} illustrates the dependence of reconnection rate on the
strength of the injected turbulence with different magnetization cases. It
shows that the maximal reconnection rate increases with the driving intensity (cf.
Figure \ref{figure6} (left panel)) in the sub-Alfv\'enic Mach number region.
This can be basically explained by the law of turbulent reconnection given by
LV99. However, as the injected Alfv\'en Mach number approaches to
trans-Alfv\'enic region, the reconnection rate reaches a maximum value and even
decreases with injected power. This is because the injected turbulence becomes
compressible and the effect of compressibility should be accounted for (see \S
4.4).  The detailed discussion including the compressiblity effects for
relativistic reconnection is given in \cite{Takamoto_etal:2015}.

The guiding field effect is plotted in the left panel of Figure \ref{general}.
Following \cite{Kowal_etal:2009, Kowal_etal:2012}, we increased the guiding
field while fixing the strength of reconnecting magnetic field component, that
is, the total $\sigma$-parameter increases as increasing the guide field. The
figure shows the reconnection rate that marginally depends of the guide field,
which is very similar to the non-relativistic results obtained in
\cite{Kowal_etal:2009, Kowal_etal:2012} presented in Figure \ref{figure6}. Thus
we conclude that turbulent reconnection in relativistic and non-relativist cases
is similar and a compressible generalization of the LV99 theory does reflect the
main features of relativistic reconnection.

The right panel of Figure \ref{general} shows that the reconnection rate does
not show dependence on the resistivity. This supports the idea that the
turbulent relativistic reconnection is fast.

The obtained results indicate that the reconnection rate can approach $0.3c$
if we assume a sufficient injection scale $l$, and this is enough to explain
most cases of relativistic reconnection \cite[see][for
review]{LyutikovLazarian:2013}. Note that that this result that satisfies
the observational requirements is obtained with
pure MHD rather than appealing to complicated collisionless physics, which
manifests that fast relativistic reconnection is a robust process that takes place
in various environments irrespectively on the the plasma collisionality.

Naturally, there are many important issues that must be studied in relation to
turbulent relativistic reconnection and related processes. For instance, it is
interesting and important to relate this reconnection with the relativistic
analog of Richardson dispersion.

\section{Reconnection with self-generated turbulence}

Turbulence that drives turbulent reconnection is not necessarily pre-existent
but can be generated as a result of reconnection.  This was discussed in various
publications starting with LV99 \cite[see also][]{LazarianVishniac:2009}, but it
is only with high resolution simulations that it became possible to observe this
effect.
Simulations by \cite{Lapenta:2008} that showed a transfer to fast
reconnection in the MHD regime can be interpreted as spontaneous turbulent
reconnection.
The turbulence generation is seen in PIC simulations
\cite[see][]{Karimabadi_etal:2014}, incompressible simulations
\citep{Beresnyak:2013} and compressible simulations \citep{Oishi_etal:2015}. The
latter two papers identified the process of reconnection with fast turbulent
reconnection.
The shortcoming of these papers, however, is the use of
periodic boundary conditions that do not allow studying steady state magnetic
reconnection.

Below we present results from our calculations in
\cite{Kowal_etal:2015} where the open boundary conditions are adopted and the
calculations are performed over several crossing times.

\begin{figure}[t]
\centering
\includegraphics[width=0.95\textwidth]{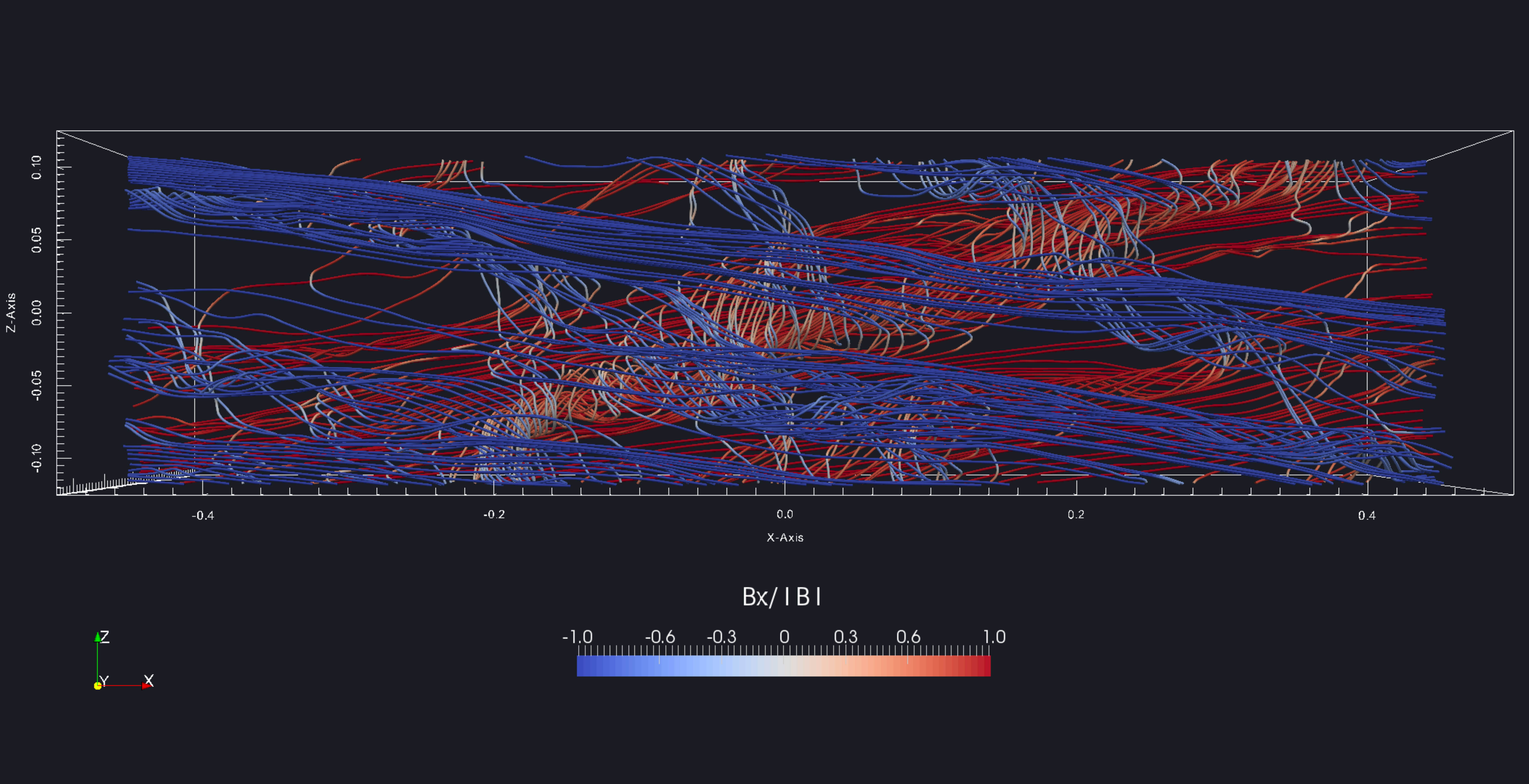}
\caption{Visualization of the magnetic field lines in the reconnection with
self-generated turbulence as seen from the below of the current sheet plane. The
colors correspond to the line orientation with respect to the X direction with
red and blue being parallel and antiparallel to the X axis, respectively. We can
recognize the organized field above and below the reconnection region and
strongly turbulent flux tubes within the reconnection region. From
\cite{Kowal_etal:2015}. \label{blines}}
\end{figure}

\begin{figure}[t]
\centering
\includegraphics[width=0.7\textwidth]{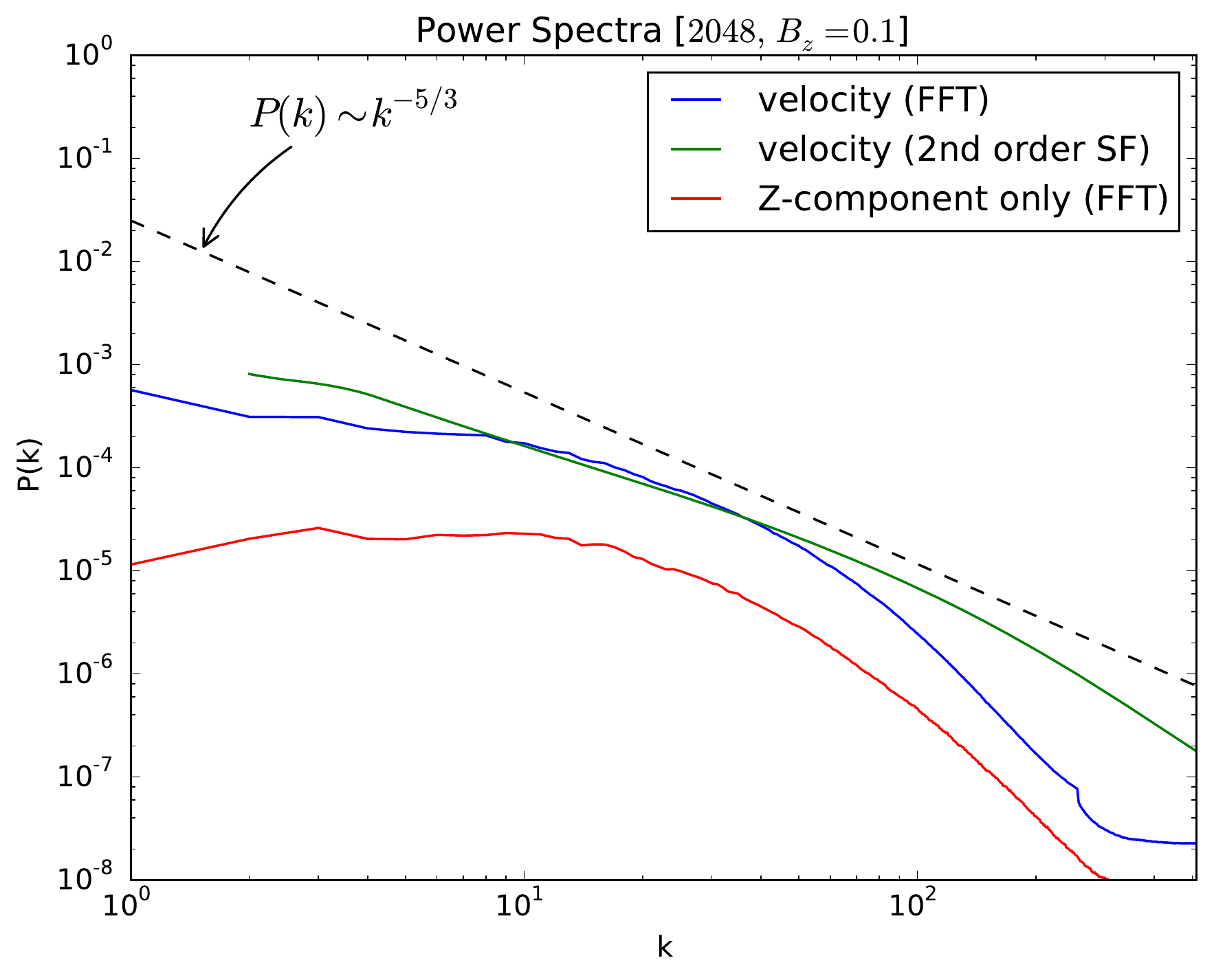}
\caption{Velocity power spectra obtained in a few different ways corresponding
to the simulation snapshot shown in the previous figure. We show the power
spectrum of the velocity obtained using the fast Fourier transform and the
second order structure function (blue and green lines, respectively). The
spectrum from the structure function approaches the Kolmogorov slope (dashed
line) better, most probably because it is not sensitive to the type of boundary
conditions. For comparison we show the power spectrum of Z-component (red line).
From \cite{Kowal_etal:2015}. \label{vspectra}}
\end{figure}

In the reconnection with a laminar configuration, the presence of initial noise
in the velocity or magnetic fields results in the development of instabilities
of the current sheet layer inducing its deformation and fragmentation. In such
situations, we would expect that any deformation of the current sheet layer
would grow, fed by the continuous plasma ejection from the local reconnection
events ejecting more kinetic energy into the surrounding medium. In such a picture
the injection scale would be related to the spatial separation of the randomly
oriented small ``jets'' of the outflows from the local reconnection events.
Those local outflows are estimated to have speeds comparable to the local
Alfv\'enic speeds, i.e., capable of deforming local field. The corresponding
bending of magnetic field lines is presented in Figure~\ref{blines} in one of the
models presented in \cite{Kowal_etal:2015}. The view from the bottom is shown,
with the current sheet being perpendicular to the line of sight. In the initial
configuration the magnetic field lines in the upper and bottom half of the domain are
antiparallel with a small inclination due to the presence of the guide field.
After some time, a turbulent region is developed around the midplane of the box
due to the stochastic reconnection taking place there. This turbulent region is
characterized by the magnetic line topology change. The lines are bent and
twisted in this region, as seen in Figure~\ref{blines}. The color corresponds to
the degree of line alignment with -1 (blue) being perfectly antiparallel and 1
(red) being perfectly parallel to the X direction.

In the next figure, Figure~\ref{vspectra}, we show the velocity power spectra
calculated in two different ways for the snapshot shown in Figure~\ref{blines}.
The blue line shows the classical power spectrum using the Fourier transform.
However, since our domain is not periodic (periodicity is enforced only along
the Z direction, otherwise the boundaries are open), the Fourier transform may
not be a proper way to obtain the power spectra. Therefore, in the same figure we
plot the velocity power spectrum obtained using the second-order structure
function (SF) which is calculated in the real space and is insensitive to the
type of the boundaries. Figure~\ref{vspectra} shows that the power spectrum
obtained from the structure function is more regular and approaches the
Kolmogorov (dashed lines) slope better. This is a clear indication of the
turbulence developed in such simulation.

For comparison, we also show the Fourier power spectrum of the Z-component of
the velocity (red line in Fig.~\ref{vspectra}) for which should be less
sensitive to the open boundaries since along this component we impose the
periodicity. The power spectrum of this component is significantly weaker in
amplitudes, especially in the large scale regime (small wave numbers $k$). This
component is perpendicular both to the Y component, along which the new magnetic
flux is brought, and to the X component, along which the reconnected flux is
removed. In fast reconnection, both these components are comparable to the
Alfv\'en speed. The weak amplitudes of the Z-component of velocity may indicate
strong anisotropies of the velocity eddies in the generated turbulence.

Some other features of the self-generated turbulence like the growth of the
turbulence region were presented in \cite{Lazarian_etal:2015}. For more detailed
description of these models and analysis of the Kelvin-Helmholtz instability as
the suspected mechanism of the injection, we refer to \cite{Kowal_etal:2015}.

\section{Observational testing}

The criterion for the application of LV99 theory is that the outflow region is
much larger than the ion Larmor radius $\Delta \gg \rho_i$. This is definitely
applicable for solar atmosphere, solar wind, but not for the magnetosphere. In
the latter case the corresponding scales are comparable and plasma effects are
important for reconnection.

\subsection{Solar Reconnection}

Solar reconnection was studied by \cite{CiaravellaRaymond:2008} in order to test
LV99 prediction of thick outflows. As we discussed earlier, the driving by
magnetic reconnection is not isotropic and therefore the turbulence is strong
from the injection scale. In this case
\begin{equation}
V_{rec}\approx U_{obs, turb} (L_{inj}/L_x)^{1/2},
\label{obs}
\end{equation}
where $U_{obs, turb}$ is the spectroscopically measured turbulent velocity
dispersion. Similarly, the thickness of the reconnection layer should be defined
as
\begin{equation}
\Delta\approx L_x (U_{obs, turb}/V_A) (L_{inj}/L_x)^{1/2}.
\label{delta_obs}
\end{equation}
The expressions given by Eqs.~(\ref{obs}) and (\ref{delta_obs}) can be compared
with observations in \cite{CiaravellaRaymond:2008}.  There, the widths of the
reconnection regions were reported in the range from 0.08$L_x$ up to 0.16$L_x$
while the the observed Doppler velocities in the units of $V_A$ were of the
order of 0.1.  It is easy to see that these values are in a good agreement with
the predictions given by Eq.~(\ref{delta_obs}). The agreement obtained in the
original comparison by \cite{CiaravellaRaymond:2008} was based on the original
expressions in LV99 that assume isotropic driving and weak turbulent cascading.
Therefore the correspondence that the authors got was not so impressive and the
authors concluded that both LV99 and Petschek X-point reconnection are
potentially acceptable solutions.

At the same time, triggering of magnetic reconnection by turbulence generated in
adjacent sites is a unique prediction of LV99 theory. This prediction was
successfully tested in \cite{Sych_etal:2009}, where the authors explained
quasi-periodic pulsations in observed flaring energy releases at an active
region above the sunspots as being triggered by the wave packets arising from
the sunspots.

\subsection{Solar Wind, Parker Spiral, Heliospheric Current Sheet}

\begin{figure}[t]
\centering
\begin{tabular}{c}
\includegraphics[width=0.8\textwidth]{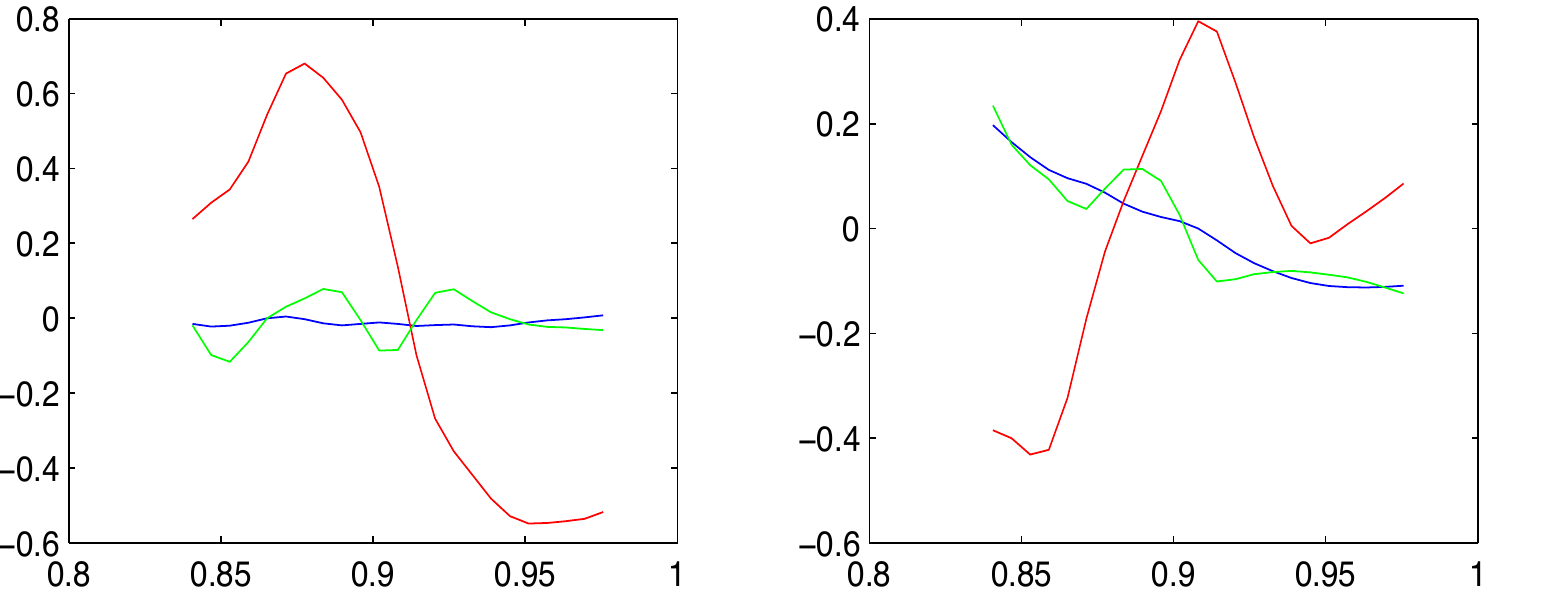}\\
\includegraphics[width=0.8\textwidth]{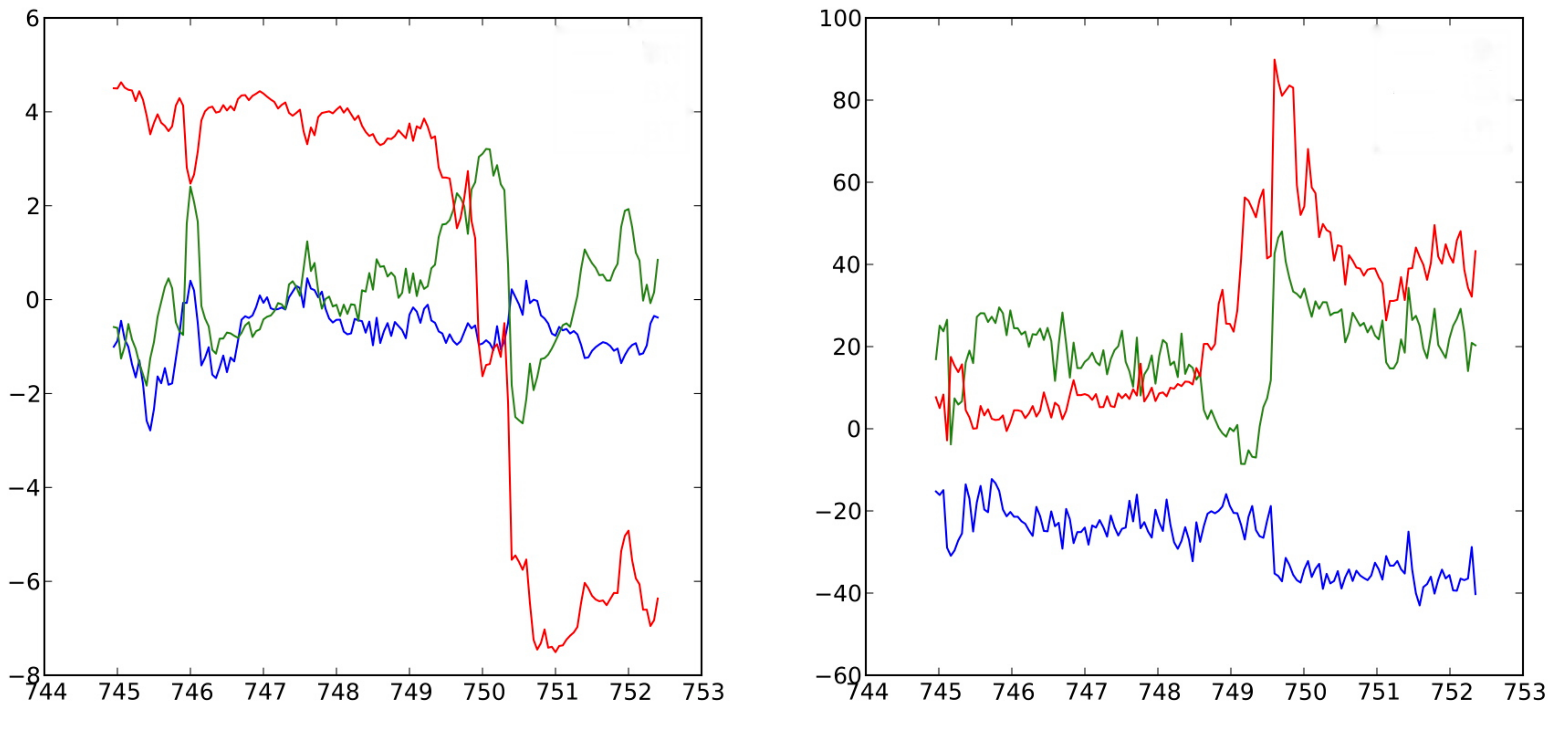}
\end{tabular}
\caption{Candidate Events for Turbulent Reconnection. {\it MHD Turbulence
Simulation (Top Panels) and High-Speed Solar Wind (Bottom Panels)}. The left
panels show magnetic field components and the right panels velocity components,
both rotated into a local minimum-variance frame of the magnetic field. The
component of maximum variance in \textcolor{red}{{\bf red}} is the apparent
reconnecting component, the component of medium variance in
\textcolor{green}{green} is the nominal guide-field direction, and the
minimum-variance direction in \textcolor{blue}{blue} is perpendicular to the
reconnection layer. Reprinted figure with permission from
\cite{Lalescu_etal:2015}. Copyright (2015) by the American Physical Society.
\label{Gosling-criterion}}
\end{figure}

Solar wind reconnection was considered in \cite{KarimabadiLazarian:2013} review
from the point of view of tearing plasma reconnection. The possibility of
turbulent MHD reconnection was not considered in spite of the fact that  $\Delta
\gg \rho_i$, the deficiency of this review that was compensated in more recent
review by \cite{Lazarian_etal:2015b}. There on the basis of studies of Solar
wind in \cite{Lalescu_etal:2015} it was concluded that the Solar wind
reconnection is well compatible with LV99 theory (see Figure
\ref{Gosling-criterion}). The general features of the turbulent reconnection in
MHD simulations correspond to the features of solar wind reconnection searched
to identify reconnection events in the Solar wind \citep{Gosling:2007}.

Similarly \cite{Eyink:2014} discussed some implications of turbulent
reconnection for heliospheric reconnection, in particular for deviations from
the Parker spiral model of interplanetary magnetic field. The latter model
assumed frozen-in condition for magnetic field, which according to turbulent
reconnection should be violated. Indeed, \cite{Burlaga_etal:1982} studied the
magnetic geometry and found ``notable deviations'' from the spiral model using
Voyager 1 and 2 data at solar distances R=1-5 AU . The deviations from the
theoretical expectations based on the frozen-in condition were substantiated by
\cite{KhabarovaObridko:2012}, who presented evidence on the breakdown of the
Parker spiral model for time- and space-averaged values of the magnetic field
from several spacecraft (Helios 2, Pioneer Venus Orbiter, IMP8, Voyager 1) in
the inner heliosphere at solar distances 0.3-5 AU. The latter authors interpret
their observations as due to ``a quasi-continuous magnetic reconnection,
occurring both at the heliospheric current sheet and at local current sheets''.
\cite{Eyink:2014} estimated the magnetic field slippage velocities and related
the deviation from Parker original predictions to LV99 reconnection. In
addition, \cite{Eyink:2014} analyzed the data relevant to the region associated
with the broadened heliospheric current sheet (HCS), noticed its turbulent
nature and provided arguments on the applicability of LV99 magnetic reconnection
model to HCS.

\subsection{Indirect observational testing}

Magnetic reconnection is extremely difficult to observe directly in generic
astrophysical situations. Observations of the Sun, direct measurements of the
Solar wind are notable exceptions. However, turbulent reconnection is
happening everywhere in astrophysical turbulent magnetized environments and one
can test the properties of reconnection by comparing the predictions that
follow from the turbulent reconnection theory for particular astrophysical
phenomena. This is an indirect way of testing turbulent reconnection and testing
of different applications of the turbulent reconnection theory, including those
that we cover in our review, also put turbulent reconnection at test.

We believe that the spectrum of turbulent fluctuations observed in astrophysical
settings, e.g. in molecular clouds, galactic atomic hydrogen \cite[see][for
review]{Lazarian:2009} testifies in favor of turbulent reconnection. Indeed, the
measurements are consistent with numerical simulations
\cite[see][]{KowalLazarian:2010}, which are performed in situations when
turbulence induces fast reconnection.  As it is discussed in
\cite{Lazarian_etal:2015b} it is the turbulent reconnection that makes the GS95
theory of strong turbulence self-consistent.

Similarly, we can say that testing of the processes of rapid diffusion of
magnetic field in turbulent fluids that are mediated by turbulent reconnection,
i.e. processes of reconnection diffusion that we discuss in \S 9.1, is
also a testing of the underlying turbulent reconnection predictions. The same
can be said about testing of the theories of gamma ray bursts, accretion disks,
black hole sources that are based on the theory of turbulent reconnection (see
examples of the comparison of theoretical predictions and observations in \S 9.3). We are sure that further detailed modeling of these phenomena based on
the predictions of turbulent reconnection theory is an exciting avenue of
research.

\section{Implications of the theory}

\subsection{Reconnection diffusion: star formation and accretion disk evolution}

Within the textbook theory of star formation, magnetic fields can influence and
even control star formation at different stages, from the formation of the
molecular cloud to the evolution of an accretion disk around a newly formed star
\cite[see][]{Shu:1983}. The basic pillar of the corresponding
theoretical constructions is that magnetic field is well frozen in highly
conducting ionized component of the media so that the characteristic
displacement of the magnetic field lines arising from Ohmic effects $\sqrt{\eta
t}$ is much less than the scale of the system for any characteristic times of
the system existence. This, however, assumes slow reconnection. But, because
the media is typically only partially ionized, the segregation of matter and
magnetic field can still happen at higher rate which is controlled by the
differential drift of ions and neutrals, i.e. by the process that is termed
{\it ambipolar diffusion}.

On the basis of LV99 theory, \cite{Lazarian:2005} suggested that the diffusion
of magnetic fields in turbulent systems will be fast and independent of
resistivity \cite[see also][]{LazarianVishniac:2009}. This resulted in the
concept that was termed ``reconnection diffusion" in analogy to the earlier
concept of ambipolar diffusion. A detailed discussion of the reconnection
diffusion concept and its relation to star formation is presented in \cite{Lazarian:2014}.

A formal theoretical proof that magnetic fields are not frozen in turbulent
fluids is presented in \cite{Lazarian_etal:2015b}, while the numerical proof of
the violation of flux freezing in turbulent media is provided through confirming
the Richardson dispersion in  \cite{Eyink_etal:2013}. This also follows
directly from LV99 and this is what motivated the reconnection diffusion
concept. The corresponding review dealing with the reconnection diffusion is in
\cite{Lazarian:2014} and we refer the interested reader to this extended work.
In what follows, we just briefly discuss the
reconnection diffusion concept as well as its implications.

\begin{figure}[t]
\centering
\includegraphics[width=0.9\textwidth]{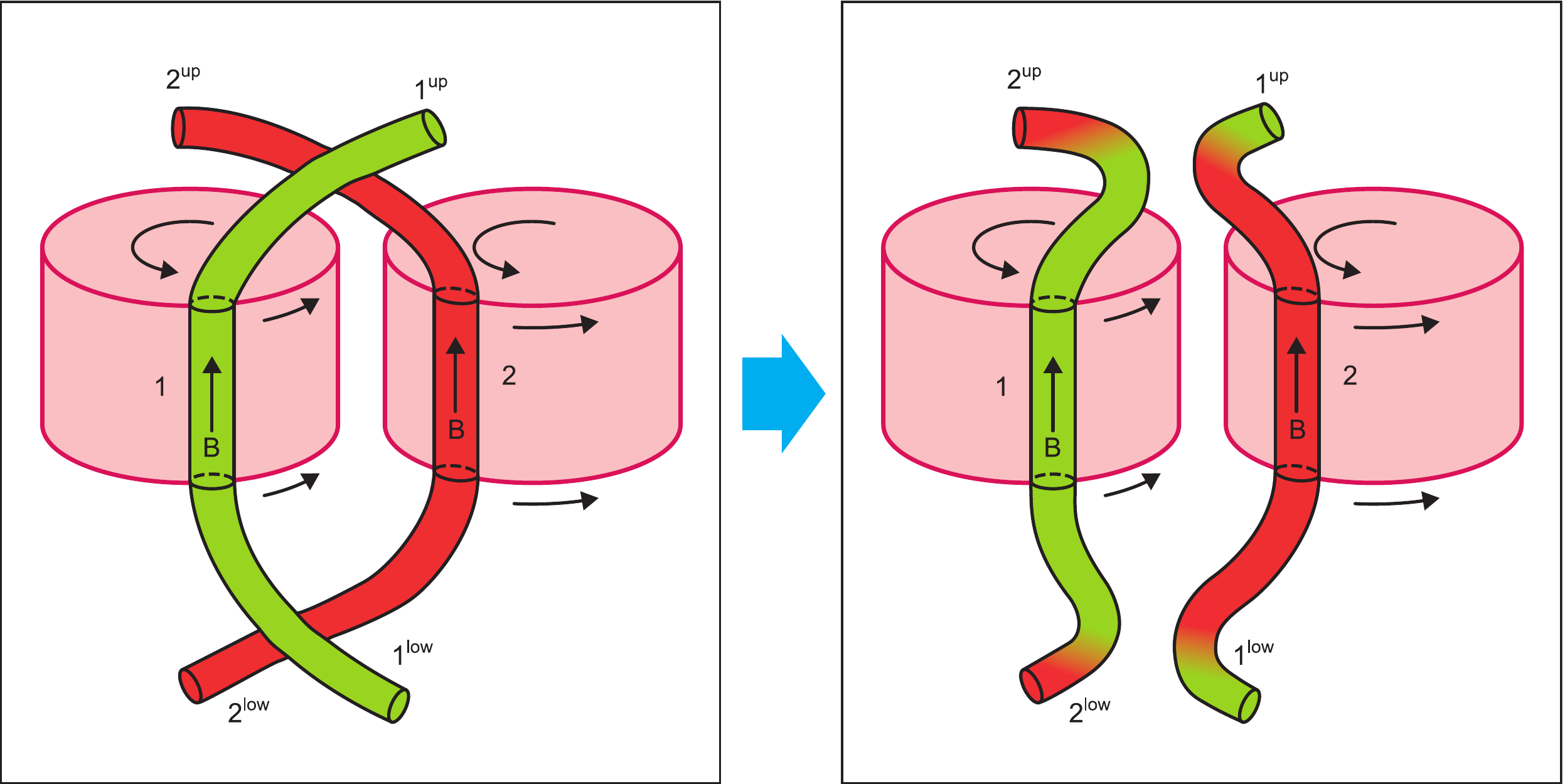}
\includegraphics[width=0.8\textwidth]{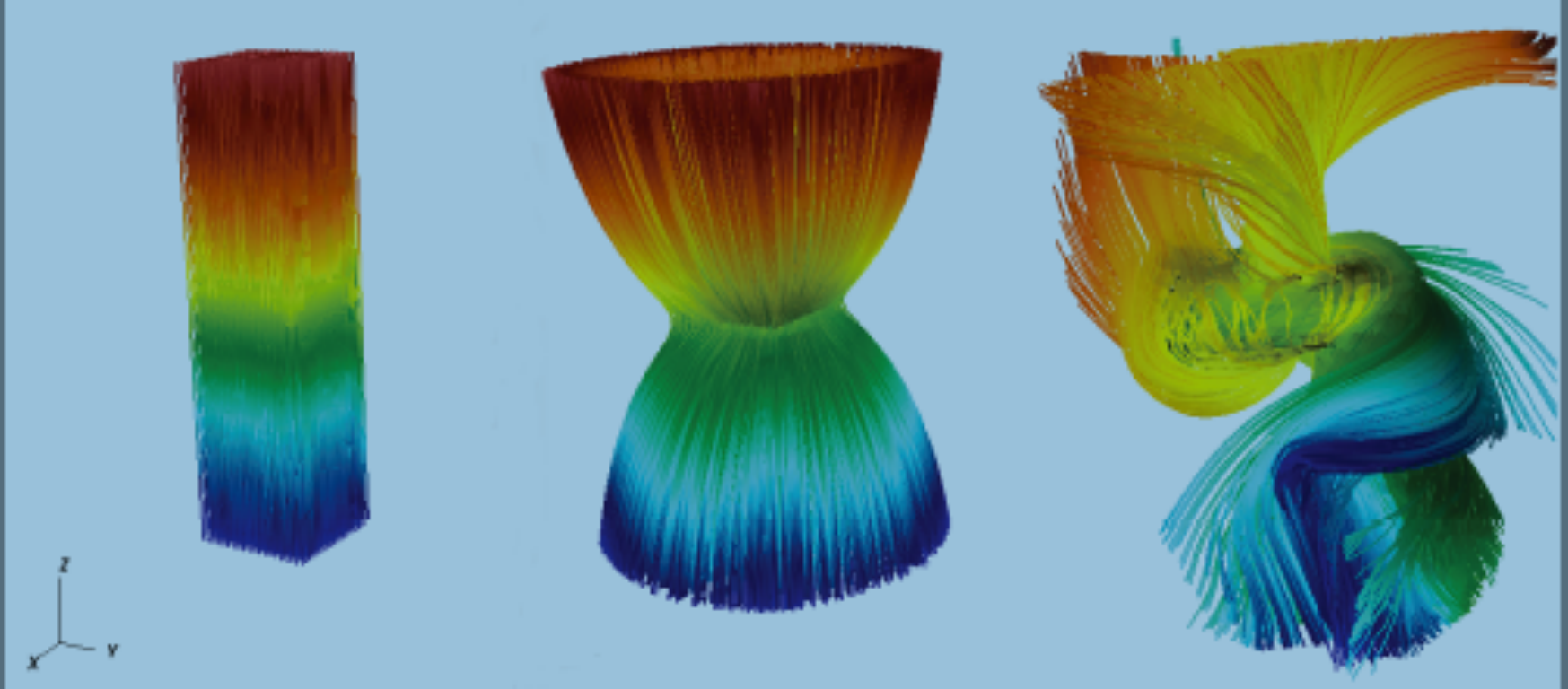}
\caption{Upper panel: Reconnection diffusion: exchange of flux with
entrained matter. Illustration of the mixing of matter and magnetic fields due
to reconnection as two flux tubes of different eddies interact. Only one scale
of turbulent motions is shown. In real turbulent cascade such interactions
proceed at every scale of turbulent motions. From \cite{Lazarian:2011}. Lower
panel: Visualization of magnetic field lines in a turbulent accretion
disk. Illustration of the process using smoothed lines. From
\cite{Casanova_etal:2015}.
\label{mix}}
\end{figure}

The process of reconnection diffusion can be illustrated by Figure (\ref{mix},
upper), where the reconnection of flux tubes belonging to two adjacent eddies
is shown. It is evident that, as a result of magnetic reconnection, the matter is
being exchanged between the flux tubes and that in the presence of the cascade
of turbulent motion at different scales the concept of the flux tube has a
transient character, as the flux tubes evolve constantly being reformed by the
motion of eddies at different scales.

Naturally, the concept of reconnection diffusion is applicable beyond the star
formation range of problems. Quantitatively it boils down to understanding that
on the scales larger than the turbulent injection scale the transfer of matter
and magnetic field in superAlfv\'enic turbulence is happening through the
turbulent advection by the eddies at the injection scale and the corresponding
diffusion coefficient coincides with that in hydrodynamics, i.e. $k_{rec. diff,
super}\approx 1/3 u_L L$.
At the same time, for subAlfv\'enic turbulence the
transfer is enabled by the strong turbulence eddies of the $l_{trans}$ size and
therefore the diffusion is reduced by the third power of Alfv\'en Mach number,
i.e. $k_{rec.diff, sub}\approx 1/3 u_L L (u_L/V_A)^3$ \citep{Lazarian:2006}.
Interestingly enough, the same law, i.e. the reconnection diffusion coefficient
proportional to $M_A^3$, follows from the theory of weak turbulence induced
by the motions at the scales larger than $t_{trans}$ (Lazarian 2006, see also
\citealt[][henceforth ELV11]{Eyink_etal:2011}).
On the scales smaller than the injection scale the transport of magnetic field and
matter follows the Richardson dispersion and accelerates as the scale under
study increases. A discussion of the reconnection diffusion from the point of view
of plasma slippage \cite[see][]{Eyink:2014} is presented in
\citet{Lazarian_etal:2015b}.

We would like to clarify that when we are talking about the suppression of
reconnection diffusion in subAlfv\'enic turbulence, this is the suppression at the
{\it large scales}, comparable with and larger than the injection scale. The
local mixing of magnetic field lines at the scales $l$ less than the scales at
which turbulence is transferred to the strong regime, i.e. at scales smaller than
$l_{trans}=LM_A^2$ (see Eq. (\ref{trans})), is still given by the product $l v_l$
and the corresponding small scale reconnection diffusion is governed by the
Richardson dispersion and exhibits superdiffusive behavior. It relative inefficiency of
reconnection diffusion at scales larger than $l_{trans}$ that that impedes reconnection diffusion for large scales $\gg
l_{trans}$.

The first numerical work that explored the consequences of reconnection
diffusion for star formation was performed by \cite{SantosLima_etal:2010}, where
the reconnection diffusion was applied to idealized setting motivated by
magnetized diffuse interstellar medium and molecular clouds. A later paper by
\cite{Leao_etal:2013} provided a numerical treatment of the reconnection
diffusion for 3D collapse of self-gravitating clouds. In addition, the problem
of the transport of angular momentum in protostellar accretion disks was
considered in \cite{SantosLima_etal:2012, SantosLima_etal:2013}, where it was
numerically demonstrated that the long-standing problem of magnetic breaking in
the formation of protostellar accretion disks can be solved if disks are formed
from turbulent media and therefore, reconnection diffusion takes place. A more
recent study by \cite{Casanova_etal:2015} (see Figure (\ref{mix}, lower)
provides more evidence on the dominant role that reconnection diffusion plays in
the disks.

\begin{figure}[t]
\centering
\includegraphics[width=0.4\textwidth]{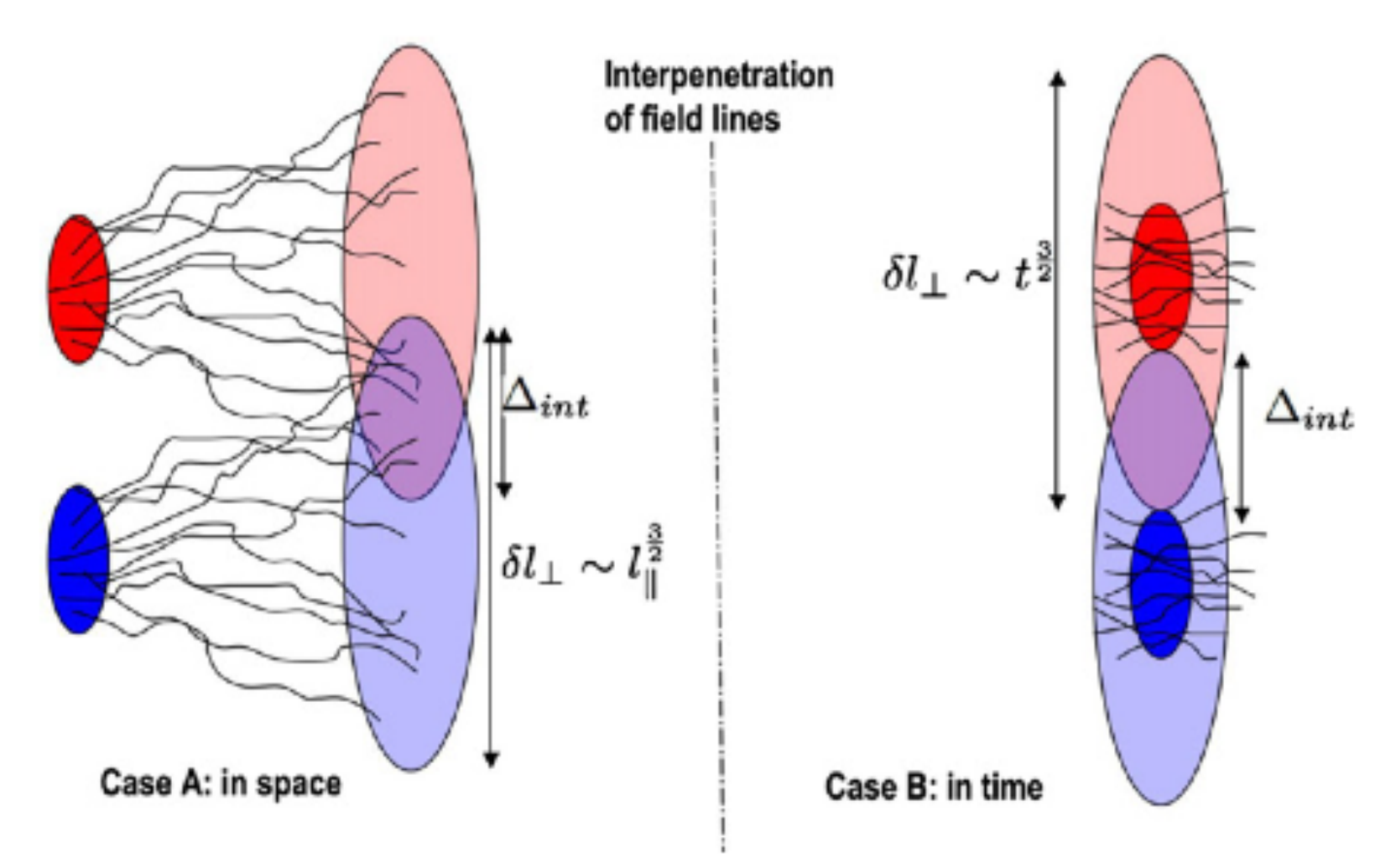}
\includegraphics[width=0.4\textwidth]{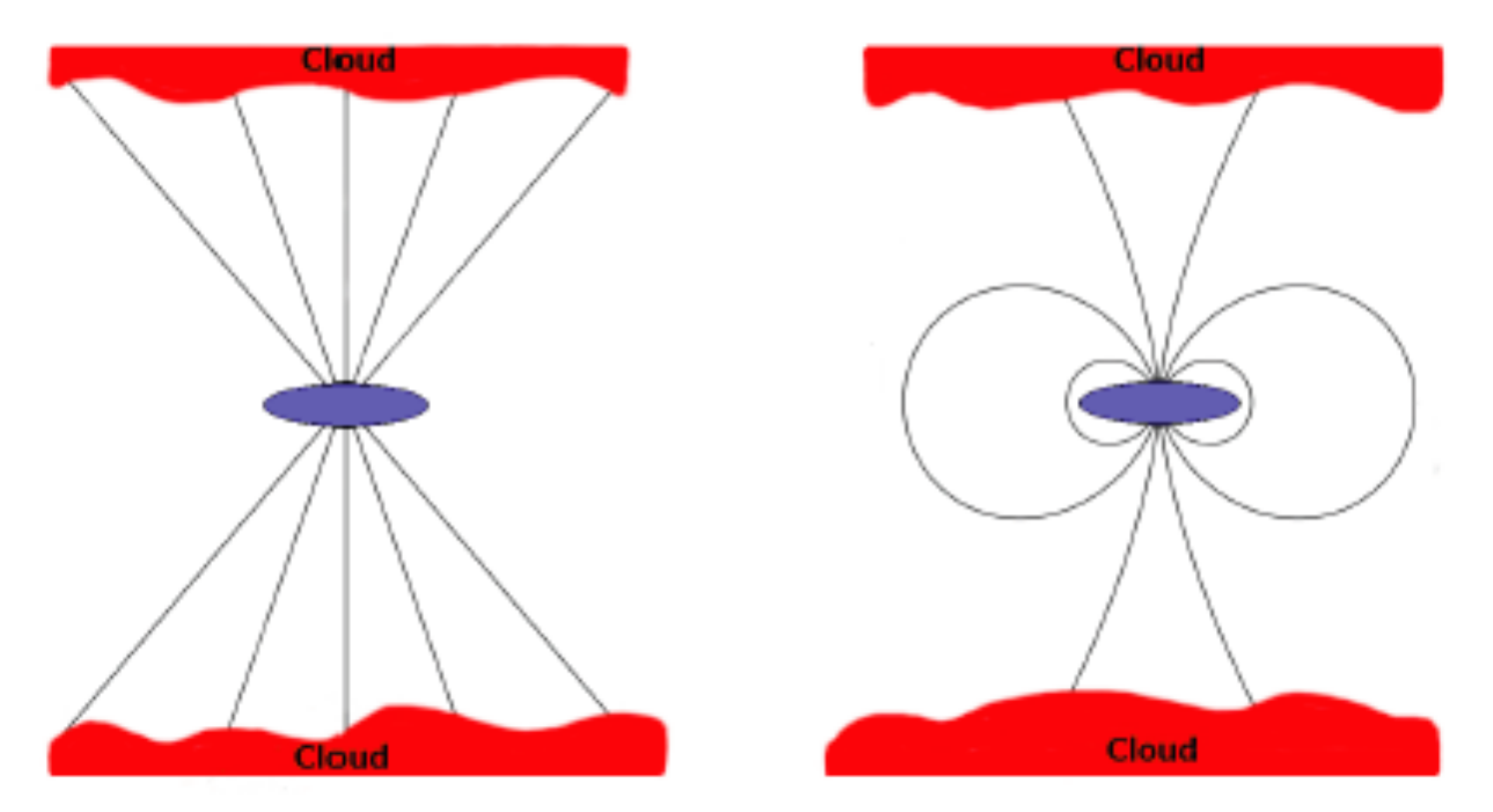}
\caption{{\it Left panel}. Case A. Microscopic physical picture of reconnection
diffusion. Magnetized plasma from two regions is spread by turbulence and mixed
up over the region $\Delta$.
Case B: description of the magnetic field line spreading with time.
{\it Right Panel}. Change of the magnetic field configuration
from the split monopole on the left to the dipole configuration on the right
decreases the degree of coupling of the disk with the surrounding ISM without
removing magnetic field from the disk. From \cite{Casanova_etal:2015}.
\label{rec_diff}}
\end{figure}

It is important to understand that reconnection diffusion does not require
magnetic fields changing their direction in space to the opposite one. In fact, in subAlfv\'enic turbulence, reconnection diffusion proceeds with
magnetic field lines roughly directed along the mean magnetic field as shown in
Figure \ref{rec_diff}, (left panel). When reconnection diffusion happens on the
scales smaller than the turbulence injection scale, the spread of magnetic field
lines obeys superdiffusive/superballistic Richardson dispersion law. In
addition, the process of reconnection diffusion that does not change the
topology of magnetic flux in the statistical sense and the process of
reconnection that radically changes the magnetic field topology can happen
simultaneously, as it is illustrated in Figure \ref{rec_diff} (right panel). There
an additional process that facilitates the accretion disk formation is shown,
namely, the change of magnetic field topology from the so-called ``split
monopole'' to the dipole configuration. This change in turbulent interstellar
plasmas is induced by turbulent reconnection and, similar to the reducing of the
flux through the accretion disk that the reconnection diffusion entails, this
topology change decreases the coupling of the accretion disk to the surrounding
gas. Thus in reality both incarnations of turbulent reconnection process work
together to solve the so-called ``magnetic breaking catastrophe'' problem.

A number of important implications of reconnection diffusion is discussed in
\cite{Lazarian_etal:2012}. Those include the independence of the star formation
rate on the metallicity  in galaxies, star formation in galaxies with high
ionization of matter, e.g. star formation in ultra-luminous infrared galaxies or
ULIRGs \citep{Papadopoulos_etal:2011}, the absence of correlation between the
magnetic field strength and gaseous density, etc.

These observational facts contradicting to the paradigm based on ambipolar
diffusion naturally follow from the reconnection diffusion theory.
Potential implications of reconnection diffusion in dynamos are also
addressed  in \cite{deGouveiaDalPino_etal:2012}.

Finally, we would like to compare the concept of reconnection diffusion with
that of "turbulent ambipolar diffusion" \citep{FatuzzoAdams:2002, Zweibel:2002}.
The latter concept is based on the idea that turbulence can create gradients of
neutrals and those can accelerate the overall pace of ambipolar diffusion. The
questions that naturally arise are (1) whether this process can proceed without
magnetic reconnection and (2) what is the role of ambipolar diffusion in the
process. \cite{Heitsch_etal:2004} performed numerical simulations with 2D
turbulent mixing of a layer with magnetic field perpendicular to the layer and
reported fast diffusion that was of the order of turbulent diffusivity number
$V_L L$, {\it independent of ambipolar diffusion coefficient}.
However, this sort of mixing can happen without reconnection only
in a {\it degenerate case of 2D mixing} with exactly parallel magnetic field
lines. In any realistic 3D case turbulence will bend magnetic field lines
and the mixing process does inevitably involve reconnection. Therefore
the 3D turbulent mixing in magnetized fluid must be treated from
the point of view of reconnection theory. If reconnection is slow, the
process in \cite{Heitsch_etal:2004} cannot proceed due to the inability of
magnetic field lines to freely cross each other (as opposed to the 2D case!). This should arrest the mixing and makes the conclusions obtained in the degenerate 2D case inapplicable to
the 3D diffusion.
If, however, reconnection is fast as predicted in LV99, then the mixing and turbulent diffusion
will take place. However, such a diffusion is not expected to depend on the ambipolar diffusion processes, which is, incidentally, in agreement with results in \cite{Heitsch_etal:2004}, and
will proceed at the the same rate in partially ionized gas as in fully ionize gas. The answers
to the questions above are that turbulent diffusion in partially ionized gas is (1) impossible without fast turbulent reconnection and (2) independent of ambipolar diffusion physics.
In this situation we believe that it may be misleading to talk about "turbulent ambipolar diffusion" in any astrophysical 3D setting.In fact, the actual diffusion in turbulent media is controlled by magnetic reconnection and is independent of ambipolar diffusion process!

\subsection{Acceleration of energetic particles}

Magnetic reconnection results in shrinking of magnetic loops and the charged
particles entrained over magnetic loops get accelerated.  This process was
proposed by \citet[][henceforth GL05]{deGouveiaDalPinoLazarian:2005} in
the setting of LV99 reconnection.

\begin{figure}[t]
\centering
\includegraphics[width=0.49\textwidth]{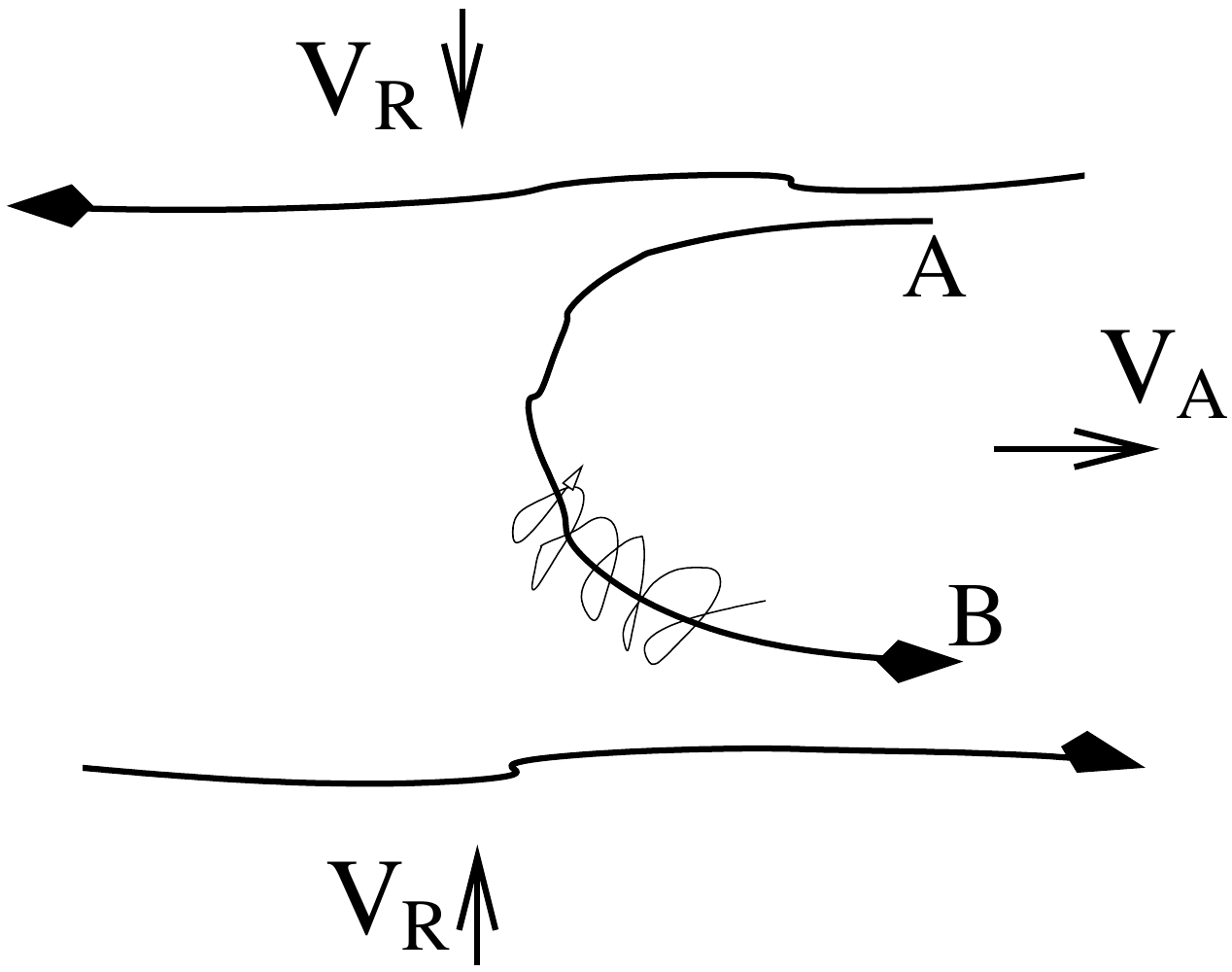}
\includegraphics[width=0.49\textwidth]{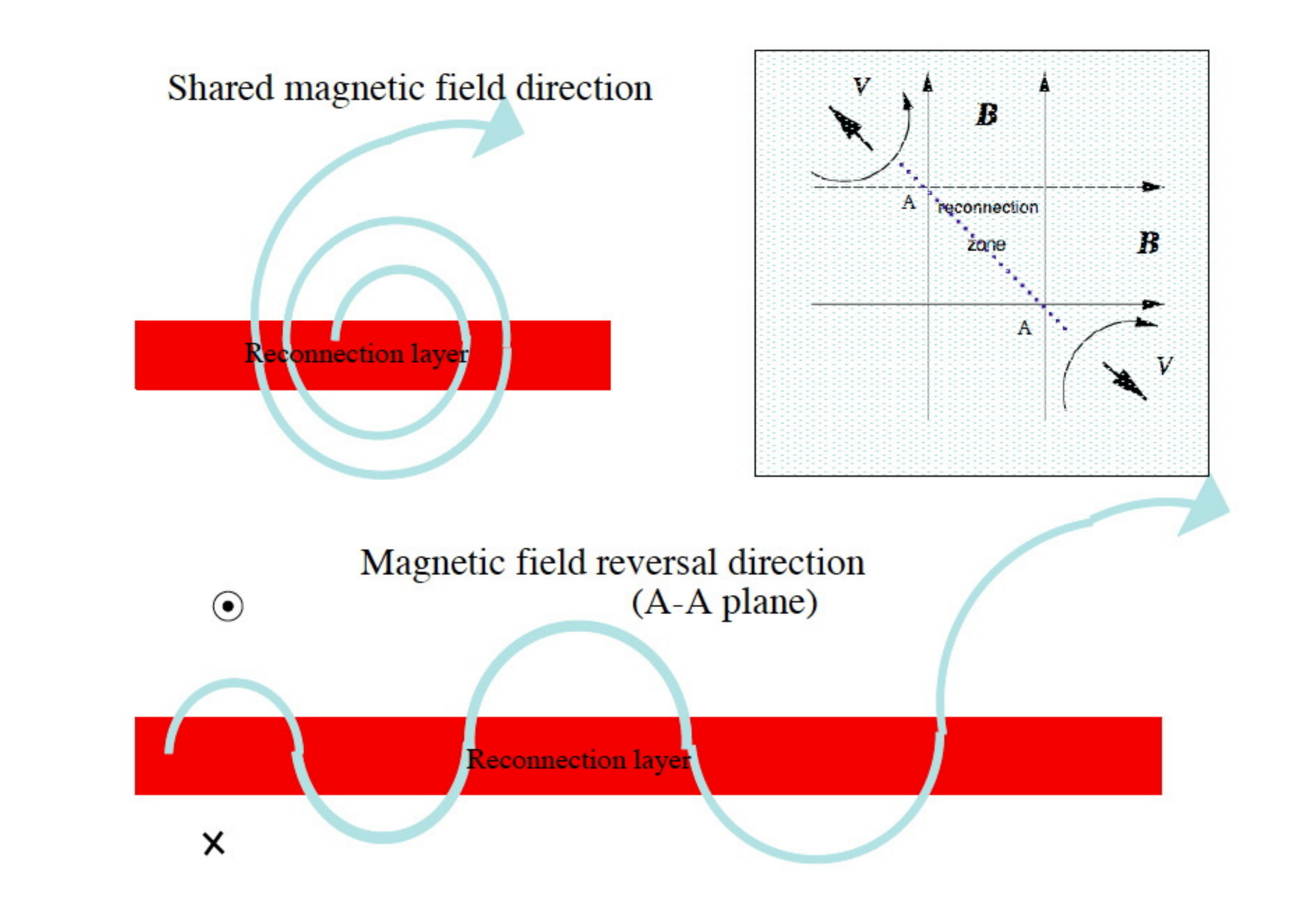}
\caption{ {\it Left Panel:} Cosmic rays spiral about a reconnected magnetic
field line and bounce back and forth at points A and B. The reconnected regions
move towards each other with the reconnection velocity $V_R$.  Reprinted figure
with permission from \cite{Lazarian:2005}. Copyright (2005), AIP Publishing LLC.
{\it Right Panel:} Particles with a large Larmor radius gyrate about the
guide/shared magnetic field. As the particle interacts with converging
magnetized flow the particle gets energy gain during every gyration. Reprinted
from \cite{Lazarian_etal:2012}. \label{acceler}}
\end{figure}

The acceleration process is illustrated by Figure \ref{acceler}. Particles
bounce back and forth between converging magnetic fluxes and undergo a
first-order Fermi acceleration. An easy way to understand the process is by
making an analogy with shock acceleration. As in shocks  particles  trapped
within two converging magnetic flux tubes  (moving to each other with the
reconnection velocity $V_R$), will bounce back and forth undergoing head-on
interactions with magnetic fluctuations and their energy after a round trip will
increase by $<\Delta E/E> \sim V_R / c$, which implies a first-order Fermi
process with an exponential energy growth after several round trips.
Disregarding the particles backreaction one can get the spectrum of
accelerated cosmic rays (GL05):
\begin{equation}
N(E)dE=const_1 E^{-5/2}dE.
\label{-5/2}
\end{equation}
This result of GL05 is valid for particle  acceleration in the absence of
compression \cite[see][for a study of the effects of compression
which may result a flatter power-law spectrum]{Drury:2012}.

Before the study in GL05, reconnection was discussed in the context of particle acceleration
\citep[e.g.,][]{Litvinenko:1996, ShibataTanuma:2001, ZenitaniHoshino:2001}, but
the first-order Fermi nature of the acceleration process was not revealed in these
studies. A process similar to that in GL05 was later suggested as a driver of
particle acceleration within  collisionless reconnection in
\cite{Drake_etal:2006}. The physics of the acceleration is the same, although
GL05 appealed to 3D magnetic bundles (see, e.g.,  Figure \ref{acceler}), while
\cite{Drake_etal:2006} considered 2D shrinking islands. The latter is actually
an artifact of the constrained 2D geometry. The difference in dimensions
 affects the acceleration efficiency as demonstrated numerically in
\cite{Kowal_etal:2011}. Indeed, another way to view the first-order Fermi acceleration of
particles entrained on the contracting helical magnetic loops in the embedded
turbulence can be envisioned from the Liouville's theorem.  In the process of
loop contraction a regular increase of the particles energies takes
place. The contraction of helical 3D loops is very different from the contraction of
2D islands. The most striking difference is that the latter stop contracting as the islands
get circular in shape, while 3D loops shrink without experiencing such a constrain.

Several other studies explored particle acceleration in magnetic reconnection
discontinuities considering collisionless plasmas
\cite[e.g.][]{Drake_etal:2010, Jaroschek_etal:2004, ZenitaniHoshino:2008,
Zenitani_etal:2009, Cerutti_etal:2013, Cerutti_etal:2014, SironiSpitkovsky:2014,
Werner_etal:2014, Guo_etal:2015}, where magnetic islands or Petschek-like
X-point configurations of fast reconnection can be driven  by kinetic
instabilities \citep{Shay_etal:2004, Yamada_etal:2010}, or anomalous resistivity
\cite[e.g.,][]{Parker:1979}, see also \cite{HoshinoLyubarsky:2012} for a review.
But as we discussed above, turbulence arises naturally in the process of reconnection
which makes it necessary to consider turbulent reconnection for systems with sufficiently
high Reynolds numbers.

\begin{figure}[t]
\centering
\includegraphics[width=0.9\textwidth]{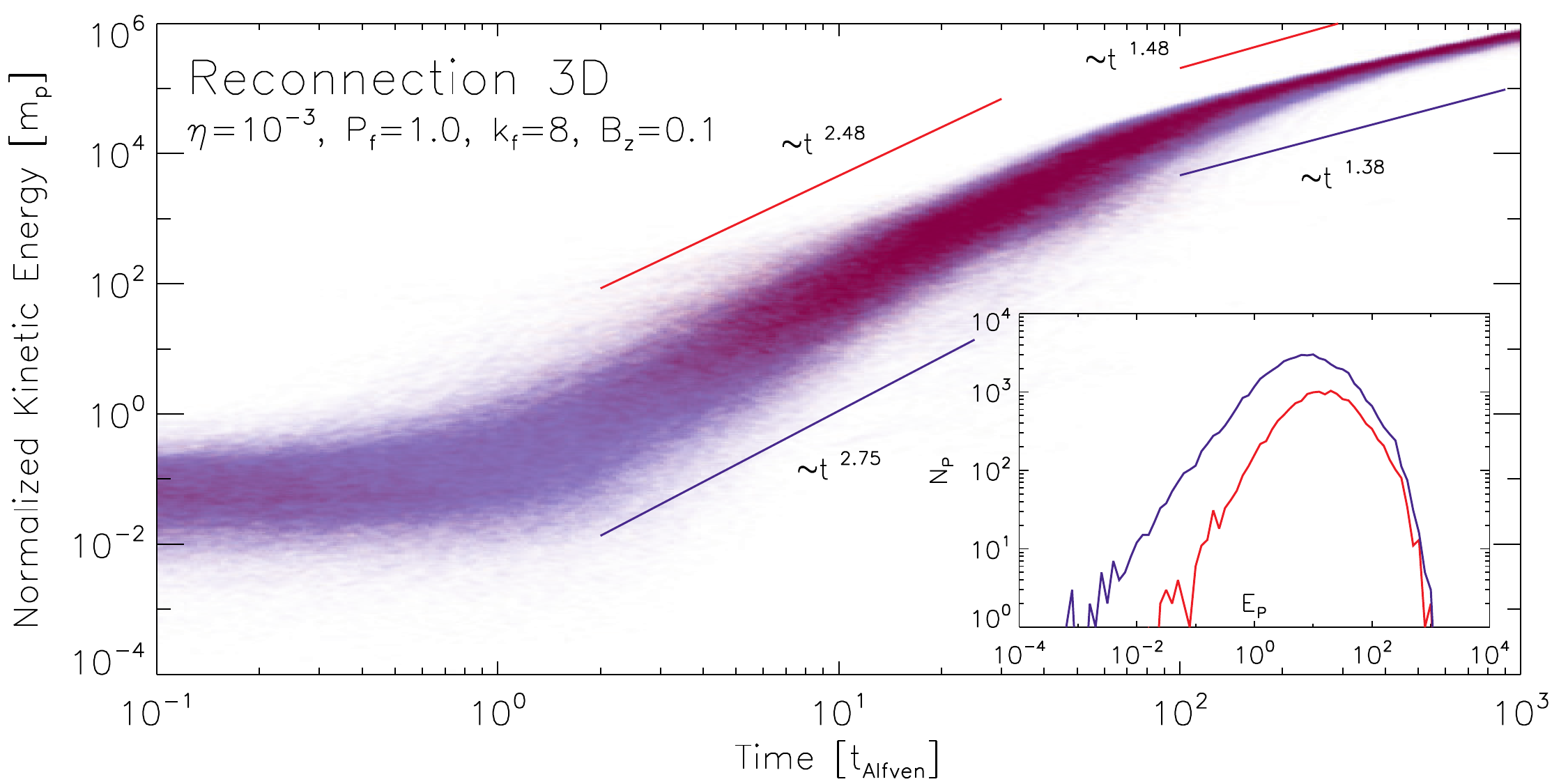}
\caption{Particle kinetic energy distributions for 10,000 protons injected in
the fast magnetic reconnection domain.  The colors indicate which velocity
component is accelerated (red or blue for parallel or perpendicular,
respectively).  The energy is normalized by the rest proton mass. Subplot shows
the particle energy distributions at $t=5.0$.  Models with $B_{0z} = 0.1$,
$\eta=10^{-3}$, and the resolution 256x512x256 is shown. Reprinted figure with
permission from \cite{Kowal_etal:2012}. Copyright (2012) by the American
Physical Society.
\label{fig:accel2}}
\end{figure}

Testing of particle acceleration in a large-scale current sheet with embedded
turbulence to make reconnection fast was performed in \cite{Kowal_etal:2012} and
its results are presented in Figure~\ref{fig:accel2}. The simulations were
performed considering 3D MHD domains of reconnection with the injection of
10,000 test particles. This study showed that the process of acceleration by
large-scale turbulent reconnection can be adequately described by
magnetohydrodynamics.  Figure~\ref{fig:accel2} depicts the evolution of the
kinetic energy of the particles.  After injection, a large fraction of test
particles accelerates and the particle energy grows exponentially (see also the
energy spectrum at $t=5$ in the detail at the bottom right).  This is explained
by a combination of two effects: the presence of a large number of converging
small scale current sheets and the broadening of the acceleration region due to
the turbulence. The acceleration process is clearly a first-order Fermi process
and involves large number of particles since the size of the acceleration zone
and the number of scatterers naturally increases by the presence of turbulence.
Moreover, the reconnection speed, which in this case is independent of
resistivity (LV99, \cite{Kowal_etal:2009}), determines the velocity at which the
current sheets scatter particles that can be a substantial fraction of $V_A$.  During this stage the
acceleration rate is $\propto E^{-0.4}$ \citep{Khiali_etal:2015} and the
particle power law index in the large energy tail is very flat (of the order of
$1-2$).

The process of fast magnetic reconnection acceleration is expected to be
widespread.  In particular, it has been discussed in \cite{LazarianOpher:2009}
as a cause of the anomalous cosmic rays observed by Voyagers and in
\cite{LazarianDesiati:2010} as a source of the observed cosmic ray anisotropies.

Magnetic reconnection was also discussed in the context of acceleration of
energetic particles in relativistic environments, like pulsars \cite[e.g.][see
also Uzdensky this volume]{Cerutti_etal:2013, Cerutti_etal:2014,
SironiSpitkovsky:2014, UzdenskySpitkovsky:2014}  and
relativistic jets of active galactic nuclei (AGNs) \cite[e.g.][]{Giannios:2010}.
The turbulent reconnection, that we argue in this review to be a natural
process in astrophysical environments was considered for the environments
 around of black hole
sources  \cite[GL05, ][]{deGouveiaDalPino_etal:2010a, Kadowaki_etal:2015,
Singh_etal:2015, Khiali_etal:2015, KhialideGouveiaDalPino:2015};
and  gamma ray bursts (GRBs) \cite[e.g.][]{ZhangYan:2011}. The aforementioned
studies are based on non-relativistic turbulent reconnection. However, the
evidence that we provided above points out to a close similarity between the
relativistic and non-relativistic turbulent reconnection. Thus we also expect
that the process of particle acceleration in turbulent relativistic and non-relativistic
reconnection to be similar. Naturally, this is an exciting topic for future research.
 A more detailed discussion of the
acceleration of energetic particles by turbulent reconnection can be found in
\cite{deGouveiaDalPino_etal:2014} and \cite{deGouveiaDalPinoKowal:2015}.

\subsection{Flares of magnetic reconnection and associated processes}

It is obvious that in magnetically dominated media the release of energy must
result in the outflow that induces turbulence in astrophysical high Reynolds number plasmas.
This inevitably increases the reconnection rate and therefore the energy release.
As a result we get a {\it reconnection instability}.
The details of such energy release and the
transfer to turbulence have not been sufficiently studied yet. Nevertheless, on
the basis of LV99 theory a simple quantitative model of flares was presented in
\cite{LazarianVishniac:2009}, where it is assumed that since stochastic
reconnection is expected to proceed unevenly, with large variations in the
thickness of the current sheet, one can expect that some fraction of this energy
will be deposited inhomogeneously, generating waves and adding energy to the
local turbulent cascade. A more detailed discussion of the model is provided in
\cite{Lazarian_etal:2015b}.

\begin{figure}[t]
\centering
\includegraphics[width=0.3\textwidth]{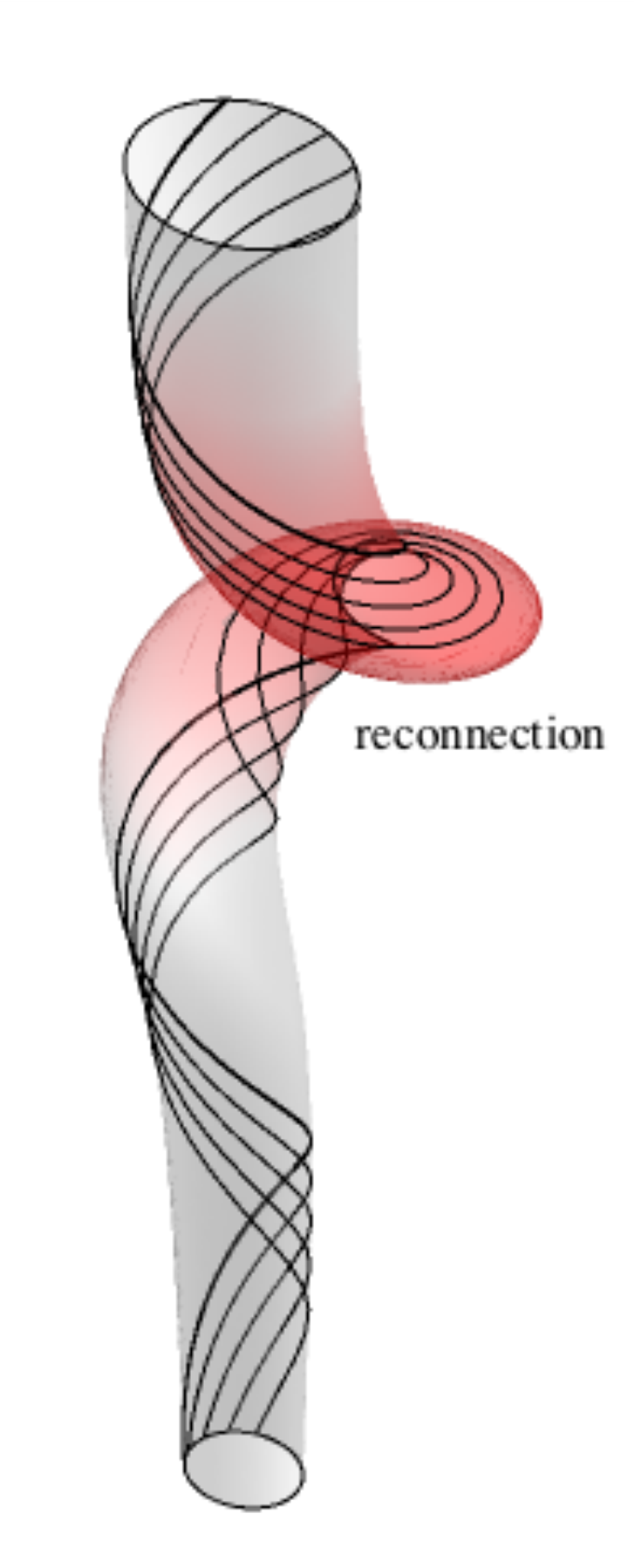}
\includegraphics[width=0.65\textwidth]{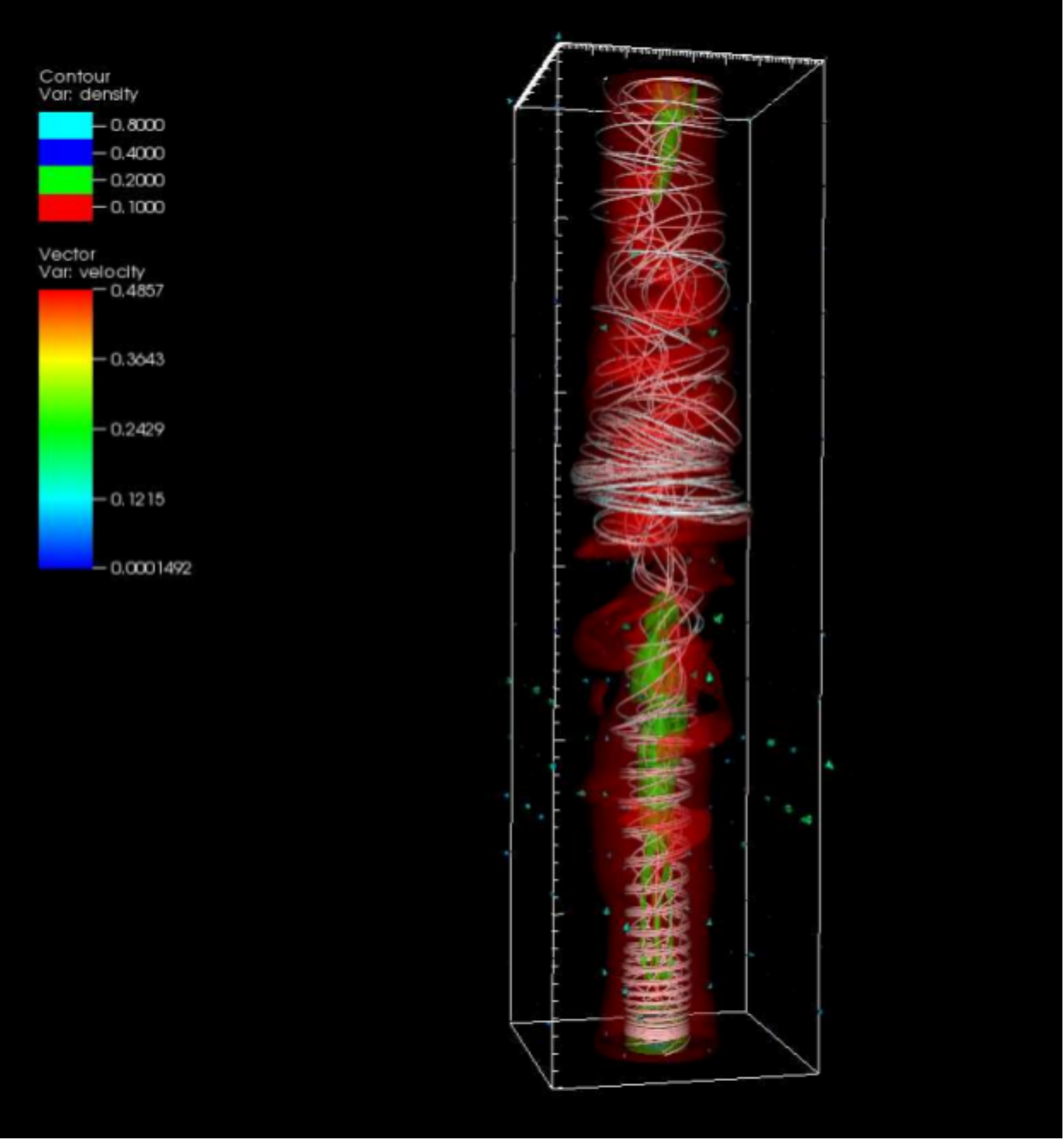}
\caption{{\it Left panel}. In the model by \cite{LazarianMedvedev:2015}
magnetized jet with spiral magnetic field is being ejected. The spiral undergoes
kink instability which results in turbulent reconnection. {\it Right panel}
Numerical simulations of 3D relativistic jet that is subject to the kink
instability and turbulent reconnection. From \cite{Mizuno_etal:2015}. \label{kink}}
\end{figure}

The applications of the theory range from solar flares to gamma ray bursts
(GRBs). In particular, a  model for GRBs based on LV99 reconnection was
suggested in \cite{Lazarian_etal:2003}.  It was elaborated and compared with
observations in \cite{ZhangYan:2011}, where collisions of magnetic turbulent
fluxes were considered.  A different version of gamma ray bursts powered by
turbulent reconnection proposed by \cite{LazarianMedvedev:2015} is based on kink
instability. It is illustrated in Figure~\ref{kink}, left panel. Naturally, the
model appeals to the relativistic turbulent reconnection that we described
above. The difference of this model from other kink-driven models
of  GRBs \cite[e.g.][]{DrenkhahnSpruit:2002, GianniosSpruit:2006, Giannios:2008,
McKinneyUzdensky:2012} is that the kink instability also induces turbulence
\citep{GalsgaardNordlund:1997, GerrardHood:2003} which
drives magnetic fast reconnection as we discuss at length in this review.
Within the GRB models in
\cite{ZhangYan:2011} and \cite{LazarianMedvedev:2015} turbulent reconnection
provides a good fit to the dynamics of GRBs.

In a similar line of research, \cite{Mizuno_etal:2015}, have performed 3D
relativistic MHD simulations of rotating jets subject to the kink instability,
considering several  models with different initial conditions (i.e.,  different
density ratios between the jet and  the environment, different angular
velocities, etc.) in order to explore  fast  magnetic reconnection, magnetic
energy dissipation and a potential transition from  magnetic to a matter
dominated regime as predicted for GRBs and AGN jets \cite[see
also][]{RochadaSilva_etal:2015, McKinneyUzdensky:2012}. The results indicate
that a complex structure develops in the helical magnetic field due to the kink
instability, developing several regions with large current densities, suggestive
of intense turbulent reconnection (see Figure \ref{kink}, right panel).
Naturally,
insufficient resolution limits the development of turbulence with a sufficiently
extended inertial range. Therefore further
numerical investigation of the process is required and, in fact, is in progress.

Turbulent reconnection is, unlike the Sweet Parker one, is a volume-filling reconnection.
The magnetic energy is being released in the volume and is being transferred into the kinetic
energy of fluid and energetic particles. Combined with fast rates of magnetic reconnection
this makes the first-order Fermi process of particle acceleration efficient, which makes it
plausible that a substantial part of the energy in
magnetic field should be transferred to the accelerated particles. Therefore we believe that
magnetic reconnection in the case of AGN jets (see \cite{Giannios:2010}) can be an copious
source of high energy particles.
In fact, numerical
simulations of $in$ $situ$  particle acceleration by magnetic reconnection in
the turbulent regions of relativistic jets (see
\cite{deGouveiaDalPinoKowal:2015}) demonstrate that the process can be
competitive with the acceleration in shocks.

\begin{figure}[t]
\centering
\includegraphics[width=0.8\textwidth]{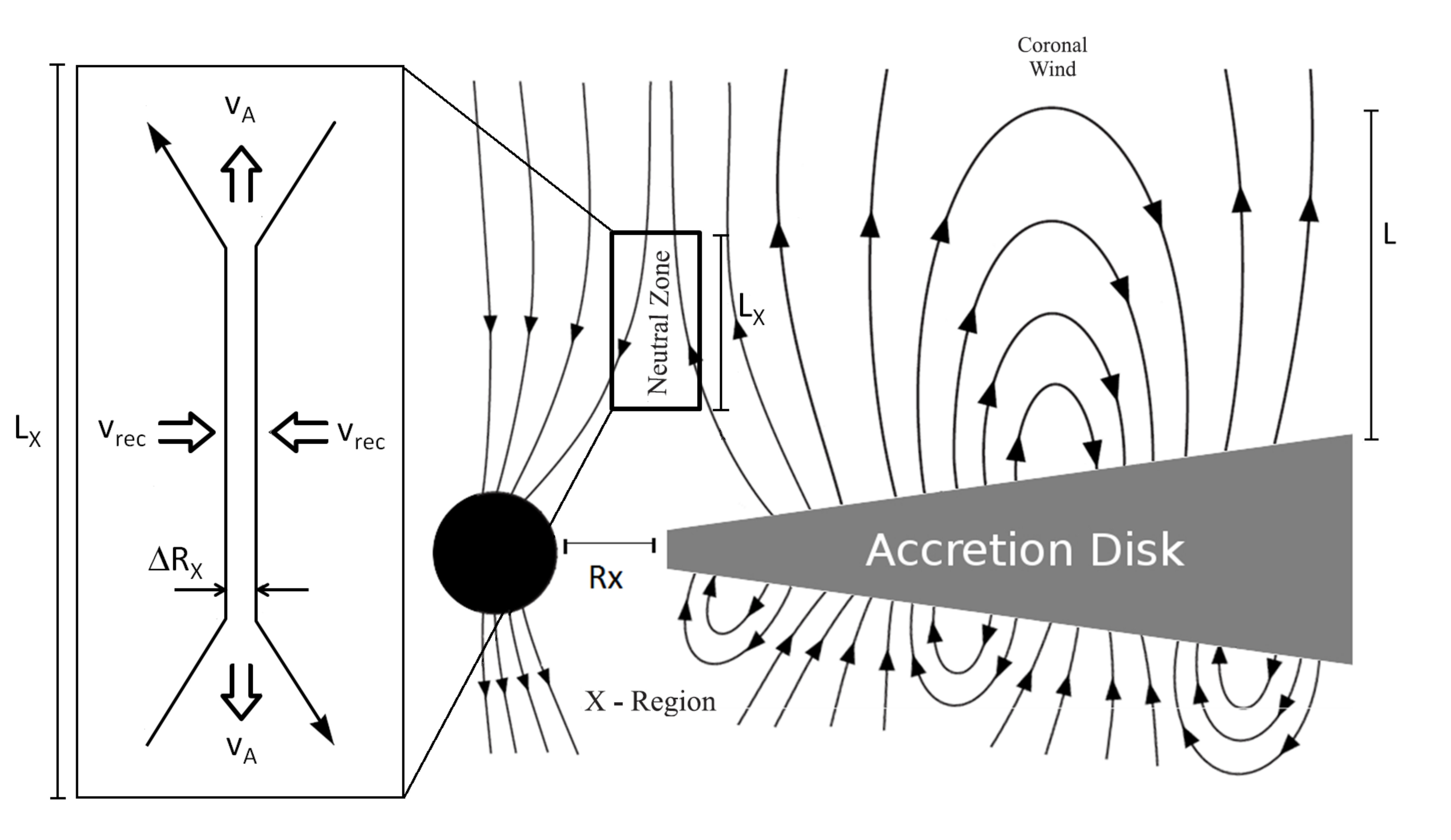}
\caption{Scheme of magnetic reconnection between the lines arising from the
accretion disk and the lines of  the BH magnetosphere. Reconnection is made fast
by the presence of embedded turbulence in the reconnection zone (see text for
more details).  Particle acceleration occurs in the magnetic reconnection zone
by a first-order Fermi process. Reprinted from \cite{Kadowaki_etal:2015} by
permission of the AAS.}
 \label{figaccretionBH}
\end{figure}

Particle acceleration arising from magnetic reconnection in the surrounds of
black hole sources like active galactic nuclei (AGNs) and galactic black hole
binaries (GHBs)  has been also studied. In particular, GL05 (see also
\cite{deGouveiaDalPino_etal:2010a, deGouveiaDalPino_etal:2010b}) proposed that
fast turbulent reconnection events occurring between  the magnetic field lines
arising from the inner accretion disk and the magnetosphere of the BH  (see
Figure~\ref{figaccretionBH}) could be efficient enough to accelerate the
particles and produce the observed  core radio outbursts in GBHs and AGNs.

\begin{figure}[t]
 \centering
  \includegraphics[width=0.8\textwidth] {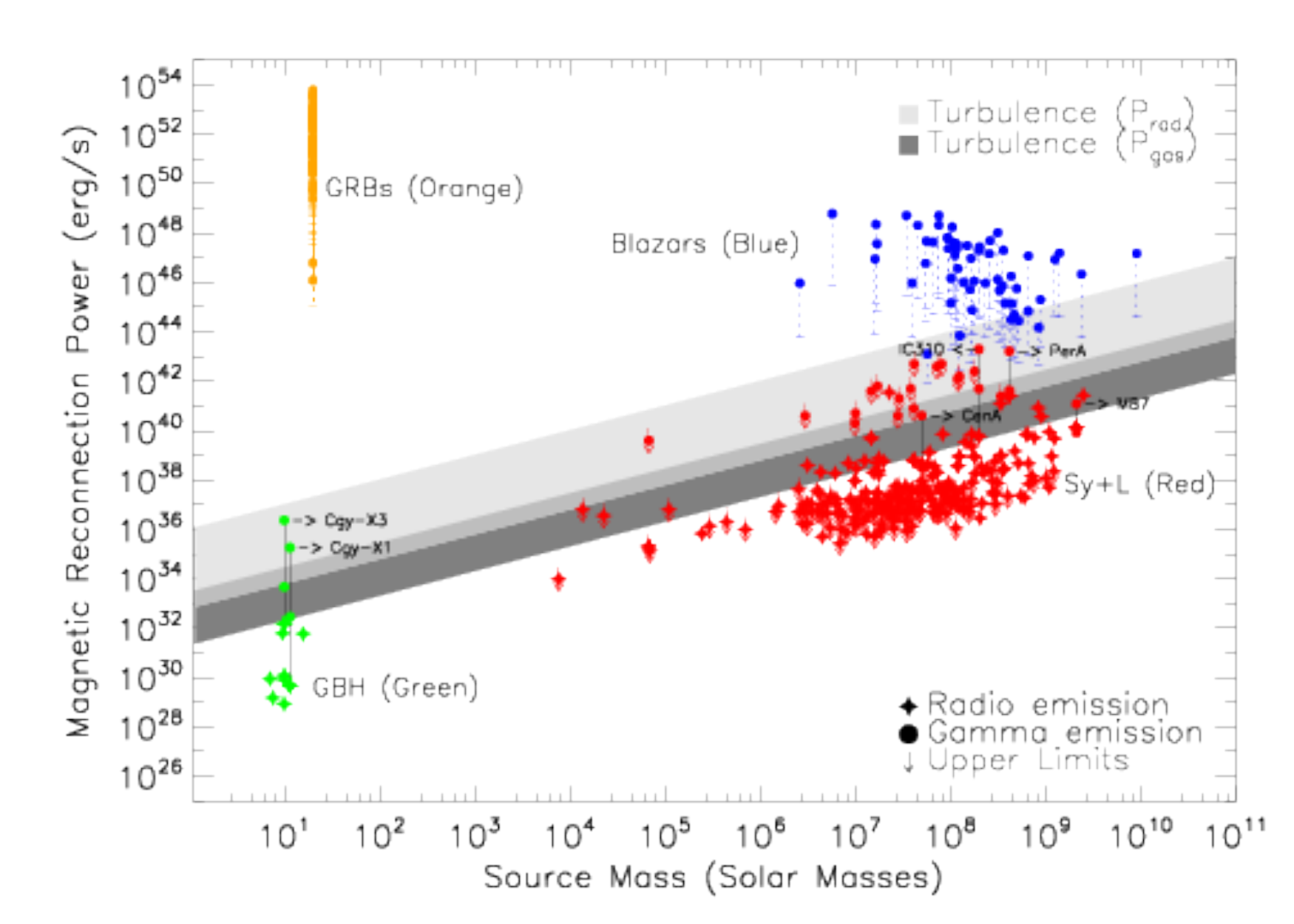}
\caption{ Turbulent driven magnetic reconnection power against BH source mass
(gray region) compared to the observed  emission of low luminous AGNs (LLAGNs:
LINERS and Seyferts), galactic black hole binaries (GBHs), high luminous AGNs
(blazars) and gamma ray burst (GRBs). The core radio emission of the GBHs and
LLAGNs is represented by red and green diamonds, the gamma-ray emission of these
two classes is represented by red and green circles, respectively.  In the few
cases for which there is observed gamma-ray luminosity it is plotted the maximum
and minimum values linking both circles with a vertical black line that extends
down to the radio emission of each  of these sources. The inverted arrows
associated to some sources indicate upper limits in gamma-ray emission.  For
blazars and GRBs only the gamma-ray emission is depicted, represented in blue
and orange circles, respectively. The vertical dashed lines correct the observed
emission by  Doppler boosting effects. The calculated reconnection power
clearly matches the observed radio and gamma-ray emissions from LLAGNs and GBHs,
but not that from blazars and GRBs. This result confirms early expectations that
the emission in blazars and GRBs is produced along the jet and not in the core
of the sources.  Reprinted from \cite{Kadowaki_etal:2015} by permission of the
AAS. }
\label{figAGN-BHB}
\end{figure}

More recently, \cite{Kadowaki_etal:2015} revisited the aforementioned model and extended the
study to explore also the gamma-ray flare emission of these sources.  The
current detectors of high energy gamma-ray emission, particularly at TeVs (e.g.,
the FERMI-LAT satellite and the ground observatories HESS, VERITAS and MAGIC)
have too poor resolution to determine whether this emission is produced in the
core or along the jets of these sources. This study confirmed the earlier trend
found in GL05 and \cite{deGouveiaDalPino_etal:2010a} and verified that if fast
reconnection is driven by turbulence, there is a correlation between the
calculated  fast magnetic reconnection power and the  BH mass spanning $10^{10}$
orders of magnitude. This can explain not only the observed radio,  but also the
gamma-ray emission from GBHs and low luminous AGNs (LLAGNs). This match has been
found for an extensive sample of more than 230 sources which include those of  the so called
$fundamental$ $plane$ $of$  $black$ $hole$ $activity$
(\citealt{Merloni_etal:2003}) as shown in Figure~\ref{figAGN-BHB}. This figure
also shows that the observed emission from blazars (i.e., high luminous AGNs
whose jet points to the line of sight) and GRBs does not follow the same trend
as that of the low luminous AGNs and GBHs, suggesting that the observed radio
and gamma-ray emission in these cases is $not$ produced in the core of these
sources. This result  is actually  exactly what one should expect because the
jet in these systems points to the line of sight thus screening the nuclear
emission, so that in these sources the emission is expected to be produced by
another population of particles accelerated  along the jet.

In another concomitant  work, \cite{Singh_etal:2015} explored the same
mechanism, but instead of considering a standard thin, optically thick accretion
disk as in the  works above, adopted a  magnetically-dominated advective
accretion flow (M-ADAF; \cite{NarayanYi:1995, Meier:2012})  around the BH, which
is suitable for sub-Eddington sources. The results   obtained are very similar
to those of \cite{Kadowaki_etal:2015} depicted in Figure~\ref{figAGN-BHB}
ensuring that the details of the accretion physics are not relevant for the
turbulent magnetic reconnection process which actually occurs in the corona
around the BH and the disk.

\begin{figure}[t]
\centering
\includegraphics[width=0.80\textwidth]{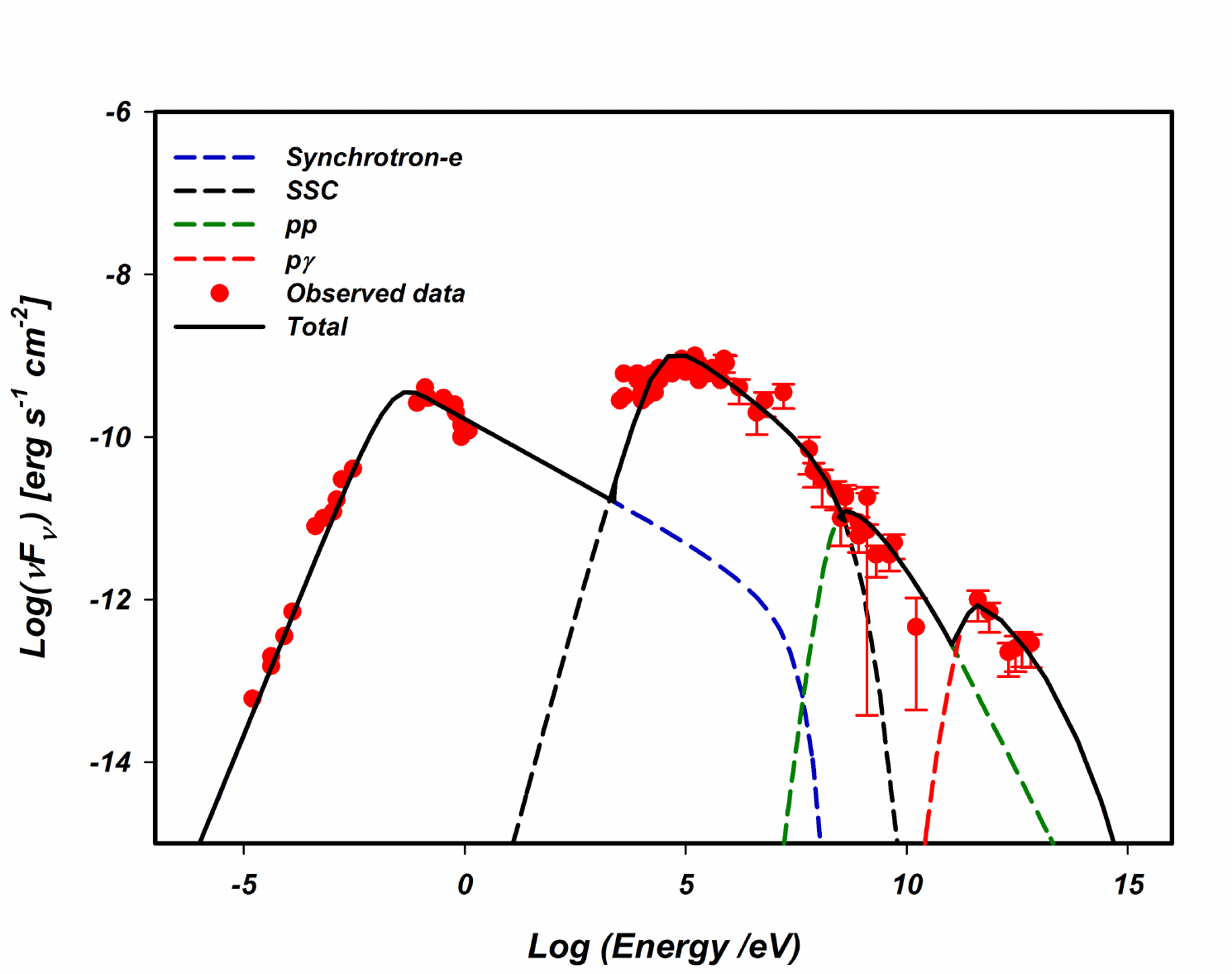}
\caption{Calculated spectral energy distribution (SED) for the AGN Cen A
employing  the turbulent magnetic reconnection acceleration model in the core
region. The data depicted in the radio to optical energy range
($10^{-5}\rm{eV}- 1 eV$) are from ISO and SCUBA and   in the  hard x-rays range
from \textit{Swift-BAT},  \textit{OSSE}  and \textit{COMPTEL}.  The data
observed in the energies $10^8-10^{10}\rm{eV}$ are taken by \textit{EGRET}  and
in the energies $10^8-10^{10}\rm{eV}$ by \textit{Fermi-LAT}. The \rm{TeV}\  data
 are taken by \textit{HESS}.  Reprinted from \cite{Khiali_etal:2015b}.
\label{SED-CenA}}
\end{figure}

The correlations found  in Figure \ref{figAGN-BHB} \citep{Kadowaki_etal:2015,
Singh_etal:2015} have motivated further investigation.  Employing  the
particle acceleration induced by turbulent reconnection and
considering the relevant non-thermal  loss processes  of the accelerated
particles  (namely, Synchrotron, inverse Compton, proton-proton and
proton-photon processes), \cite{Khiali_etal:2015} and \cite{Khiali_etal:2015b}
have computed the spectral energy distribution (SED) of several GBHs and LLAGNs
and found that these match very well with the observations (see for instance
Figure \ref{SED-CenA}, which depicts the SED of the radio galaxy Cen A),
especially at the gamma-ray tail, thus strengthening the conclusions above in
favor of a core emission origin for the very high energy emission of these
sources. The model also naturally  explains  the observed very fast variability
of the emission. The same model has been also recently applied to  explain the
high energy neutrinos observed by the IceCube as due to LLAGNs
\citep{KhialideGouveiaDalPino:2015}.

\section{Comparison of approaches to magnetic reconnection}

\subsection{Turbulent reconnection and numerical simulations}

Whether MHD numerical simulations reflect the astrophysical reality depends on
whether magnetic reconnection is correctly presented within these simulations.
The problem is far from trivial. With the Lundquist number being sometimes more
than $10^{15}$ orders different, direct numerical simulation may potentially be
very misleading. To deal with the issue Large Eddy Simulations (LES) approach
may look promising \cite[see][]{Miesch_etal:2015}.  The catch here is that LES
requires the explicit parametrization of reconnection rates. For instance, assume
that following the ideas of tearing research we adopt a particular maximal value
of reconnection speed, e.g. $0.01 V_A$. This means that the motions where the
fluids are moving with velocities larger than this chosen reconnection speed
will be constrained. In MHD transAlfv\'enic turbulence, this would predict
constraining the motions of eddies on the scales $[10^{-6} L, L]$ if we
adopt the usual Kolmogorov $v\sim l^{1/3}$ scaling. This means that for this
range of scales our results obtained with MHD turbulence theory are not
applicable and the physics of many related processes is radically different.
We believe that if wired this into LES, this will provide erroneous unphysical
results.

From the point of view of the turbulent reconnection theory a normal MHD code
reproduces magnetic reconnection correctly for turbulent regions, as for
turbulent volumes the reconnection rate does not depend on resistivity and
varies with the level of turbulence. As turbulence is the generic state of
astrophysical fluids, the regions that are turbulent within numerical studies
are correctly represented in terms of magnetic reconnection. On the contrary,
the regions where the turbulence is damped in simulations due to numerical
diffusivity do not represent magnetic reconnection correctly. Situations where
the initial set up is laminar requires following the development of turbulent
reconnection and the prescriptions based on the corresponding simulation may be
useful for parameterizing the process.

\subsection{Turbulent reconnection versus tearing reconnection}

It has been known for quite a while that Sweet-Parker current sheet is unstable
to tearing and this can affect the reconnection rate
\cite[see][LV99]{Syrovatskii:1981}. What was a more recent development is that
the 2D current sheet starting with a particular Lundquist number larger than
$10^4$ develops fast reconnection \cite[see][]{Loureiro_etal:2007,
Uzdensky_etal:2010}, i.e. reconnection that does not depend on the fluid resistivity. The study of
tearing momentarily eclipsed the earlier mainstream research of the reconnection
community, which attempted to explain fast reconnection appealing to the
collisionless plasma effects that were invoked to stabilize the Petschek-type X
point configuration for reconnection \citep{Shay_etal:1998, Drake:2001,
Drake_etal:2006}. We view this as a right step in abandoning the
artificial extended X point configurations the stability of which in the situation of realistic
astrophysical forcing was very doubtful (see discussion in LV99). However, we
 believe that tearing by itself does not provide a generic solution for the
astrophysical reconnection.

To what extend tearing is important for the onset of 3D turbulent reconnection
should be clarified by the future research. Here we can provide arguments
suggesting that tearing inevitably transfers to turbulent reconnection for
sufficiently large Lundquist numbers $S$. Indeed, from the mass conservation
constraint requirement in order to have fast reconnection one has to increase
the outflow region thickness in proportion to $L_x$, which means the
proportionality to the Lundquist number $S$. The Reynolds number $Re$ of the
outflow is $V_A \Delta/\nu$, where $\nu$ is viscosity, grows also as $S$. The
outflow gets turbulent for sufficiently large $Re$. It is natural to assume that
once the shearing rate introduced by eddies is larger than the rate of the
tearing instability growth, the instability should get suppressed.

If one assumes that tearing is the necessary requirement for fast reconnection,
this entails the conclusion that tearing should proceed at the critically damped
rate, which implies that the $Re$ number and therefore $\Delta$ should not
increase. This entails, however, the decrease of reconnection rate driven by
tearing in proportion $L_x\sim S$ as a result of mass conservation. As a result,
the reconnection should stop being fast. Fortunately, we know that turbulence
itself provides fast reconnection irrespectively whether tearing is involved or
not.

We also note that tearing reconnection in numerical simulations provides the
reconnection rate around  $0.01V_A$ for collisional and somewhat larger rates
for collisionless reconnection. These limitations are incompatible with the
requirements of astrophysical reconnection, which, for instance, requires
reconnection of the order of $V_A$ for large scale eddies in transAlfv\'enic
turbulence. In addition, fixed reconnection rates do not explain why observed
reconnection may sometimes be slow and sometimes fast.

\subsection{Turbulent reconnection versus turbulent resistivity and mean field approach}

Attempts to describe turbulent reconnection introducing some sort of turbulent
resistivity are futile and misleading. It is possible to show that
``turbulent/eddy resistivity'' description has fatal problems of inaccuracy and
unreliability, due to its poor physical foundations for turbulent flow. It is
true that coarse-graining the MHD equations by eliminating modes at scales
smaller than some length $l$ will introduce a ``turbulent electric field'', i.e.
an effective field acting on the large scales induced by motions of magnetized
eddies at smaller scales. However, it is well-known in the fluid dynamics
community that the resulting turbulent transport is not ``down-gradient'' and
not well-represented by an enhanced diffusivity. Indeed, turbulence lacks the
separation in scales to justify a simple ``eddy-resistivity'' description. As a
consequence, energy is often not absorbed by the smaller eddies, but supplied by
them, a phenomenon called ``backscatter''. The turbulent electric field often
creates magnetic flux rather than destroys it.

If we know the reconnection rate, e.g. from LV99, then an eddy-resistivity can
always be tuned by hand to achieve that rate. But this is not science. While the
tuned reconnection rate will be correct by construction, other predictions will
be wrong. The required large eddy-resistivity will smooth out all turbulence
magnetic structure below the coarse-graining scale $l$. In reality, the
turbulence will produce strong small-scale inhomogeneities, such as current
sheets, from the scale $l$ down to the micro-scale. In addition, field-lines in
the flow smoothed by eddy-resistivity will not show the explosive,
super-diffusive Richardson-type separation at scales below $l$. These are just
examples of effects that will be lost if the wrong concept of ``eddy
resistivity'' is adopted. Note, that the aforementioned are important for
understanding particle transport/scattering/acceleration in the turbulent
reconnection zone. We can also point out that in the case of relativistic
reconnection that we also deal with in this review, turbulent resistivities will
introduce non-causal, faster than light propagation effects.
Nevertheless, the worst feature of the crude ``eddy-resistivity''
parametrization is its unreliability: because it has no scientific
justification whatsoever, it cannot be applied with any confidence to
astrophysical problems.

We would like to stress that the fast turbulent reconnection concept is
definitely not equivalent to the dissipation of magnetic field by resistivity.
While the parametrization of some particular effects of turbulent fluid may be
achieved in models with different physics, e.g. of fluids with enormously
enhanced resistivity, the difference in physics will inevitably result in other
effects being wrongly represented by this effect. For instance, turbulence with
fluid having resistivity corresponding to the value of ``turbulent resistivity''
must have magnetic field and fluid decoupled on most of its inertia range
turbulent scale, i.e. the turbulence should not be affected by magnetic field in
gross contradiction with theory, observations and numerical simulations.
Magnetic helicity conservation which is essential for astrophysical dynamo
should also be grossly violated\footnote{Increasing the resistivity to the values
required to account for the diffusivities that arise from turbulent reconnection
would make astrophysical dynamos resistive or slow, in gross contradiction
to the fact that astrophysical dynamos operate in fluids with engligible
resistivity and therefore can be only modeled by "fast dynamo" (see Parker
1979).}.

The approach that we advocate is quite different. It is not based on coarse-graining. The
stochasticity of magnetic field-lines is a real, verified physical phenomenon in
turbulent fluids. Whereas ``eddy-resistivity'' ideas predict that magnetic flux
is destroyed by turbulence, we argue that turbulent motions constantly
change connectivity of magnetic field lines without dissipating magnetic
fields. Being moved by fluid motions the stochastic world-lines in relativistic
turbulence do remain within the light-cone and no non-causal effects
such as produced by ``eddy-resistivity'' being entailed (see also ELV11).

Understanding that ``resistivity arising from turbulence'' is not a real plasma
non-ideality ``created'' by the turbulence is essential for understanding why
mean field approach fails in dealing with reconnection in turbulent fluids.
Indeed, turbulence induced apparent non-ideality is dependent on the length and
timescales of the averaging and it emerges only as a consequence of observing
the plasma dynamics at a low resolution, so that the observed coarse-grained
velocity and magnetic field do not satisfy the true microscopic equations of
motion. It is obvious, that coarse-graining or averaging is a purely passive
operation which does not change the actual plasma dynamics. The non-ideality in a
turbulent plasma observed at length-scales in the inertial-range or larger is a
valid representation of the effects of turbulent eddies at smaller scales.
However, such apparent non-ideality cannot be represented by an effective
``resistivity'', a representation which in the fluid turbulence literature has
been labeled the ``gradient-transport fallacy'' \citep{TennekesLumley:1972}.

A recent paper that attempts to address turbulent reconnection using mean field
approach is \cite{Guo_etal:2012}, where ideas  originally proposed in
\cite{KimDiamond:2001} were modified and extended. Thus while the study in
\cite{KimDiamond:2001} concluded that turbulence cannot accelerate reconnection,
the more recent study came to the opposite conclusions. The expressions for
reconnection rates in \cite{Guo_etal:2012} are different from those in LV99 and
grossly contradict the results of numerical testing of turbulent reconnection in
\cite{Kowal_etal:2009}. Another model of turbulent reconnection based on the
mean field approach is presented in \cite{HigashimoriHoshino:2012}.

The mean field approach invoked in the aforementioned studies is plagued by poor
foundations and conceptual inconsistencies (ELV11).
In such an
approach effects of turbulence are described using parameters such as
anisotropic turbulent magnetic diffusivity experienced by the fields once
averaged over ensembles. The problem is that it is the lines of the full
magnetic field that must be rapidly reconnected, not just the lines of the mean
field. ELV11 stress that the former implies the latter, but not conversely. No
mean-field approach can claim to have explained the observed rapid pace of
magnetic reconnection unless it is shown that the reconnection rates obtained in
the theory are strictly independent of the length and timescales of the
averaging.

Other attempts to get fast magnetic reconnection from turbulence are related to
the so-called hyper-resistivity concept \citep{Strauss:1986,
BhattacharjeeHameiri:1986, HameiriBhattacharjee:1987, DiamondMalkov:2003}, which
is another attempt to derive fast reconnection from turbulence within the
context of mean-field resistive MHD. Apart from the generic problems of using
the mean field approach,  the derivation of the hyper-resistivity is
questionable from yet another point of view. The form of the parallel electric
field is derived from magnetic helicity conservation. Integrating by parts one
obtains a term which looks like an effective resistivity proportional to the
magnetic helicity current. There are several assumptions implicit in this
derivation, however. Fundamental to the hyper-resistive approach is the
assumption that the magnetic helicity of mean fields and of small scale,
statistically stationary turbulent fields are separately conserved, up to tiny
resistivity effects. However, this ignores magnetic helicity fluxes through open
boundaries, essential for stationary reconnection, that vitiate the conservation
constraint \cite[see more discussion in LV99, ELV11 and][]{Lazarian_etal:2015b}.

\subsection{Turbulent reconnection: 3D versus 2D}

A lot of physical phenomena are different in 3D and 2D with hydrodynamic
turbulence being a striking example. However, due to numerical constraints many
numerical studies of the physical phenomena are initially attempted in
the systems of reduced dimensions. Whether the physics in the system of reduced
dimensions is representative of the physics of the 3D system in such situations
must be theoretically justified and  tested. For magnetic reconnection the justification
of similarity of systems of reduced dimention is difficult, as crucial differences between 2D and 3D
magnetic reconnection are stressed e.g. in Priest (this volume) and also in
publications by \cite{Boozer:2012, Boozer:2013}. There it is shown that an
extrapolation from reconnection physics obtained in 2D to 3D is poorly justified. In what follows, we
add additional points why we do not believe that 2D turbulent reconnection can be a
guide for our understanding of the 3D astrophysical reconnection.

\cite{MatthaeusLamkin:1985, MatthaeusLamkin:1986} explored numerically turbulent
reconnection in 2D.  As a theoretical motivation the authors emphasized
analogies between the magnetic reconnection layer at high Lundquist numbers and
homogeneous MHD turbulence. They also pointed out various turbulence mechanisms
that would enhance reconnection rates, including multiple X-points as
reconnection sites, compressibility effects, motional electromotive force (EMF) of magnetic bubbles
advecting out of the reconnection zone. However, the authors did not realize the
importance of stochastic magnetic field wandering and they did not arrive at an
analytical prediction for the reconnection speed. Although an enhancement of the
reconnection rate was reported in their numerical study, but the setup precluded
the calculation of a long-term average reconnection rate.

We would like to stress the importance of this study in terms of attracting the
attention of the community to the influence of turbulence on reconnection.
However, the relation of this study with LV99 is not clear, as the nature of
turbulence in 2D is different. In particular, shear-Alfv\'en waves that play the
dominant role in 3D MHD turbulence according to GS95 are entirely lacking in 2D,
where only pseudo-Alfv\'en wave modes exist.

We believe that the question whether turbulent reconnection is fast in 2D has
not been resolved yet if we judge from the available publications. For instance,
in a more recent study along the lines of the approach in
\cite{MatthaeusLamkin:1985}, i.e. in \cite{Watson_etal:2007}, the effects of
small-scale turbulence on 2D reconnection were studied and no significant
effects of turbulence on reconnection was reported. \cite{Servidio_etal:2010}
have more recently made a study of Ohmic electric fields at X-points in
homogeneous, decaying 2D MHD turbulence. However, they studied a case of
small-scale magnetic reconnection and their results are not directly relevant to
the issue of reconnection of large-scale flux tubes that we deal with in this review.
The study by \cite{Loureiro_etal:2009} and that by \cite{KulpaDybel_etal:2010}
came to different conclusions on whether 2D turbulent reconnection is fast in
2D. Irrespectively of the solution of this particular controversy, we believe that
 2D turbulent reconnection is radically
different from that in 3D, as both the nature of MHD turbulence that drives the
reconnection  and the nature of 2D and 3D reconnection processes are very different
(see our discussions above).

\subsection{Turbulent reconnection and small scale reconnection events}

Magnetic reconnection is frequently presented as a particular example of a problem where
the resolving processes from the smallest to the largest scales is vital. We argue that this is not
the case of turbulent reconnection. Indeed, turbulence is a phenomenon in which the dynamics of large eddies is not affected by the microphysics of the dissipation at small scales. Similarly,
it was shown in LV99 that the rate of turbulent reconnection does not depend on the rate of local reconnection events. As a result, the rate of turbulent reconnection is expected to be the same in collisionless and collisional media. Therefore the arguments that tearing decreases the length of reconnection layers transferring reconnection in collisionless regime and this way makes the reconnection faster are not applicable to turbulent reconnection.

We also stress that the small scale plasma turbulence that can change the local resistivity and local reconnection rates (see Karimabadi \& Lazarian 2014 for a review) does not affect the global rate of turbulent reconnection. The latter is a MHD phenomenon that depends on the
properties of turbulence only.

\section{Final remarks}

\subsection{Suggestive evidence}

There are pieces of evidence that are consistent with turbulent reconnection and
can be interpreted as indirect suggestive evidence. For instance,
\cite{MininniPouquet:2009} showed that {\it fast dissipation} takes place in 3D
MHD turbulence.  This phenomenon is consistent with the idea of fast
reconnection, but naturally cannot be treated as any proof. Obviously, fast
dissipation and fast magnetic reconnection are rather different physical
processes, dealing with decrease of energy on the one hand and decrease of
magnetic flux on the other.

Similarly, work by \cite{GalsgaardNordlund:1997} could also be considered as
being in agreement with fast reconnection idea.  The authors showed that in
their simulations they could not produce highly twisted magnetic fields. These
configurations are subject to kink instability and the instability can produce
turbulence and induce reconnection. However, in view of many uncertainties of
the numerical studies, this relation is unclear. In fact, with low resolution
involved in the simulations the Reynolds numbers could not allow a turbulent
inertial-range. It is more likely that numerical finding in
\cite{LapentaBettarini:2011}  which showed that reconnecting magnetic
configurations spontaneously get chaotic and dissipate are  related to LV99
reconnection. This connection is discussed in \cite{LapentaLazarian:2012}.

\subsection{Interrelation of different concepts}

The concept of turbulent magnetic field wandering is a well-known concept that
was long adopted and widely used in the cosmic ray literature to explain
observational evidence for the fast diffusion of cosmic rays perpendicular to
the mean galactic magnetic field\footnote{It was shown in
\cite{LazarianYan:2014} that the mathematical formulation of the field wandering
has an error in the classical papers, but this does not diminish the importance
of the idea. In fact, learning about the efficient diffusion perpendicular to the mean magnetic field by charged particles provided an important impetus for one of
the authors of this review towards the idea of turbulent reconnection.}. Although this concept
existed in cosmic ray literature decades before the theory of turbulent
reconnection was formulated, it is easy to understand that it is impossible to
account for the magnetic field wandering if magnetic field lines do not
reconnect. As we discussed, magnetic wandering arises from the Richardson dispersion
process and implies continuous reconnection and changing of connectivity of magnetic field lines.

The concept of transport of heat by turbulent eddies in magnetized plasmas
\citep{Cho_etal:2002, Maron_etal:2004,Lazarian:2006} can only be understood
within the paradigm of fast turbulent reconnection that allows mixing motions
perpendicular to the local magnetic field. The same is applicable to the
turbulent transport of metals in interstellar gas. Both phenomena are related to
the reconnection diffusion. At scales smaller than the scale of injection scale,
reconnection diffusion follows the superdiffusive law dictated by the Richardson
dispersion, which on the scales larger than the injection scale, the diffusive
behavior is restored.

At scales smaller than the injection scale the magnetic field wandering quantified
in LV99 represents the Richardson dispersion in space. Last, but not the least,
fast turbulent reconnection makes the GS95 theory of MHD turbulence
self-consistent \cite[LV99, see also][]{Lazarian_etal:2015b}. All in all,
magnetic turbulence and turbulent reconnection are intrinsically related.

\subsection{Relation to other Chapters of the volume}

This review deals with turbulent reconnection in MHD regime and explains
how 3D magnetic turbulence makes reconnection fast for both collisional
and collisionless fluids\footnote{Turbulence is known to make other processes, e.g.
diffusion of passive impurities independent of microphysics of diffusion in hydro.
Thus the theory of turbulent reconnection just adds to the list of the transport processes
which turbulence makes universal, i.e. independent of the detailed physics at the
micro level.}
 MHD reconnection in
laminar regime is addressed in the contribution by Priest (this volume). An
interesting overlap in terms of conclusions is that the 3D and 2D reconnection
processes are very different. Plasma effects related to reconnection and their
laboratory studies are covered by Yamada et al. (this volume). The corresponding
Reynolds numbers of reconnecting plasma outflows is insufficient for observing
the regime of turbulent reconnection, but corresponds well to the present day
two fluid simulations. The fact that thickness of the reconnection layers in the
experiments is comparable with the ion inertial length makes this experiments
very relevant to magnetospheric reconnection (see Cassak \& Fuselier;
Petrukovich et al., Raymond et al. this volume). We hope that in future
experiments the reconnection will be studied in the conditions closer to those
that we discuss in the review and turbulent driving will be employed to test
our predictions.

A case of interlaced, but not turbulent fields is considered by Parker \&
Rappazzo (this volume). In terms of turbulent reconnection, such configurations
generating flares of reconnection may be important for inducing turbulence in
the system.

Within turbulent reconnection the microphysics of individual local small scale
reconnection events is not relevant for determining the global reconnection
rate. However, many processes that accompany turbulent reconnection events, e.g.
energization/acceleration of particles from the thermal pool do depend on the
detailed microphysics, e.g. collisionless plasma physics (see Yamada et al.;
Scudder et al., this volume), electric field at the separatrix of local
reconnection events (Lapenta et al., this volume).

Turbulent reconnection theory that we described does not deal with complex
radiative processes that frequently manifest the reconnection events
observationally. Their effects are described in detail by Uzdensky (this volume).
In the case of relativistic reconnection we have shown that the effects of
compressibility can play an important role changing the reconnection rate.
Similarly, heating and cooling of media considered by Uzdensky (this volume)
is expected to affect the rate of turbulent reconnection through the density of
the outflow. At the same time we do not expect the change of resistivity that
is related to heating to affect the turbulent reconnection rates.

The closest in spirit review is that by Shibata and Takasao (this volume).
Fractal structure is the accepted feature of turbulence and transition from
laminar to fractal reconnection that the authors describe has the important
overlap with the physical processes that we describe in this review. The
authors, however, more focused on reconnection mediated by plasmoid generation,
which may also be viewed as a complementary approach. As we discuss above we
view tearing as a transient process that for systems of sufficiently high
Reynolds numbers leads to turbulent reconnection. Unlike tearing, the final stage
of turbulent reconnection does not depend on plasma microphysics, but only on
the level of turbulence.

We feel that the beauty of turbulent
reconnection as opposed to other suggestions is that it makes magnetic reconnection fast independently of detailed
properties of plasmas, making magnetic reconnection really universal. This corresponds both to the
principal of parsimony and to observations that do not see much difference between the two
media as far as the rate of magnetic reconnection is concerned. However, the detailed microphysics of reconnection
should not be disregarded. While we claim that it may not determine the resulting reconnection rates within
sufficiently turbulent medium, the microphysics is likely to be important to other processes that
accompany magnetic reconnection. In addition, particular intensively studied systems like Earth magnetosphere as well as many cases of plasmas presented in laboratory devices the thickness of reconnection layers is comparable with the plasma scales, making plasma effects important
for reconnection in these systems. Thus magnetic reconnection presents a multi-facet problem
where different approaches can be complementary.

\begin{acknowledgement}
AL acknowledges the NSF grant AST 1212096, a distinguished visitor PVE/CAPES
appointment at the Physics Graduate Program of the Federal University of Rio
Grande do Norte and thanks the INCT INEspao and Physics Graduate Program/UFRN
for hospitality. Final work on the review were done in a stimulating atmosphere
of Bochum University during the visit supported by the Humboldt Foundation. GK
acknowledges support from FAPESP (grants no. 2013/04073-2 and 2013/18815-0).
EMGDP:  Brazilian agencies FAPESP (2013/10559-5) and CNPq (306598/2009-4). JC
acknowledges support from the National Research Foundation of Korea
(NRF-2013R1A1A2064475).
\end{acknowledgement}

{\small
\bibliography{ms_long}
}

\end{document}